\documentclass[12pt]{article}
\usepackage{a4wide}
\usepackage{latexsym}
\usepackage{cite}
\usepackage{axodraw}

\usepackage{pslatex}
\usepackage[latin1]{inputenc}
\usepackage[T1]{fontenc}

\def\bq{\begin{eqnarray}}
\def\eq{\end{eqnarray}}
\def\l{\langle}
\def\r{\rangle} 
\def\eps{\varepsilon}

\begin{document}

\thispagestyle{empty}

\begin{flushright}
  MZ-TH/09-12
\end{flushright}

\vspace{1.5cm}

\begin{center}
  {\Large\bf The infrared structure of $e^+ e^- \rightarrow \mbox{3 jets}$ at NNLO reloaded\\
  }
  \vspace{1cm}
  {\large Stefan Weinzierl\\
\vspace{2mm}
      {\small \em Institut f{\"u}r Physik, Universit{\"a}t Mainz,}\\
      {\small \em D - 55099 Mainz, Germany}\\
  } 
\end{center}

\vspace{2cm}

\begin{abstract}\noindent
  {
This paper gives detailed information on the structure of the infrared singularities for the process
$e^+ e^- \rightarrow \mbox{3 jets}$ at next-to-next-to-leading order in perturbation theory.
Particular emphasis is put on singularities associated to soft gluons.
The knowledge of the singularity structure allows the construction of appropriate subtraction terms,
which in turn can be implemented into a numerical Monte Carlo program. 
   }
\end{abstract}

\vspace*{\fill}

\newpage

\section{Introduction}
\label{sec:intro}

The process $e^+ e^- \rightarrow \mbox{3 jets}$ is of particular interest for the extraction of the
strong coupling $\alpha_s$.
Observables related to three-jet events are well suited for this task
because the leading term in a perturbative calculation 
of three-jet observables is already proportional to the strong coupling.
An accurate and precise extraction of $\alpha_s$ from the experimental measured observables requires
also a precise theoretical prediction for these observables.
Infrared-safe observables can be computed in perturbation theory and higher precision
is obtained by including the next-higher term in the perturbative series.
Infrared-safe observables related to $e^+ e^- \rightarrow \mbox{3 jets}$ have 
in the past been calculated at leading-order (LO) 
and next-to-leading order (NLO) \cite{Ellis:1980nc,Ellis:1981wv,Fabricius:1981sx,Kunszt:1989km,Giele:1992vf,Catani:1996jh} in QCD. The NLO electro-weak corrections have been calculated recently in \cite{CarloniCalame:2008qn}.
This paper reports on the next-to-next-to-leading order (NNLO) QCD calculation for
$e^+ e^- \rightarrow \mbox{3 jets}$.
This is a highly non-trivial calculation which triggered research in many directions.
The relevant tree-, one- and two-loop-amplitudes have been calculated in 
refs.~\cite{Berends:1989yn,Hagiwara:1989pp,Falck:1989uz,Schuler:1987ej,Korner:1990sj,Giele:1992vf,Bern:1997ka,Bern:1997sc,Campbell:1997tv,Glover:1997eh,Garland:2001tf,Garland:2002ak,Moch:2002hm}.
For the computation of the two-loop amplitudes new integration techniques have been 
developed \cite{Gehrmann:1999as,Gehrmann:2000zt,Gehrmann:2001ck,Moch:2001zr}
together with numerical routines for the evaluation 
of polylogarithms \cite{Gehrmann:2001pz,Gehrmann:2001jv,Vollinga:2004sn},
which occur in the integral functions.
The inclusion of the third term of the perturbative expansion into a numerical
program, which allows the calculation of an arbitrary infrared-safe three-jet observable is very
challenging.
Most of the complications are related to infrared singularities, which occur in intermediate steps.
Methods to handle these singularities at NNLO have been discussed in 
refs.~\cite{Kosower:2002su,Kosower:2003cz,Weinzierl:2003fx,Weinzierl:2003ra,Kilgore:2004ty,Frixione:2004is,Gehrmann-DeRidder:2003bm,Gehrmann-DeRidder:2004tv,Gehrmann-DeRidder:2005hi,Gehrmann-DeRidder:2005aw,Gehrmann-DeRidder:2005cm,GehrmannDeRidder:2007jk,Somogyi:2005xz,Somogyi:2006da,Somogyi:2006db,Catani:2007vq,Somogyi:2008fc,Aglietti:2008fe},
generalising ideas from NLO
\cite{Giele:1992vf,Giele:1993dj,Keller:1998tf,Frixione:1996ms,Catani:1997vz,Dittmaier:1999mb,Phaf:2001gc,Catani:2002hc}.
Unlike other processes like $e^+ e^- \rightarrow \mbox{2 jets}$, $p p \rightarrow H$ or Drell-Yan, which have been calculated
previously to next-to-next-to-leading 
order \cite{Anastasiou:2004qd,Weinzierl:2006ij,Weinzierl:2006yt,Anastasiou:2002yz,Anastasiou:2003yy,Anastasiou:2003ds,Anastasiou:2004xq,Anastasiou:2005qj,Anastasiou:2007mz,Melnikov:2006di,Anastasiou:2005pn,Catani:2001ic,Catani:2001cr,Grazzini:2008tf,Harlander:2001is,Harlander:2002wh,Harlander:2003ai}
and which involve only two hard coloured partons, the process $e^+ e^- \rightarrow \mbox{3 jets}$ is the first process
calculated at NNLO with three hard coloured partons.

Here new complications related to soft gluons occur.
This paper documents the singularity structure of $e^+ e^- \rightarrow \mbox{3 jets}$
and gives a detailed discussion of the singularities related to soft gluons.
Another group published results on the infrared structure of $e^+ e^- \rightarrow \mbox{3 jets}$
at NNLO earlier on in \cite{GehrmannDeRidder:2007hr},
but omitted certain subtraction terms related to soft gluons.
In the calculation presented here the methods used are in many parts similar to the ones used 
in \cite{GehrmannDeRidder:2007hr} and it should be mentioned that 
the authors of \cite{GehrmannDeRidder:2007hr} made
major contributions to the development of these methods \cite{Gehrmann-DeRidder:2003bm,Gehrmann-DeRidder:2004tv,Gehrmann-DeRidder:2005hi,Gehrmann-DeRidder:2005aw,Gehrmann-DeRidder:2005cm}.
The purpose of this paper is to present a complete set of subtraction terms for the process $e^+ e^- \rightarrow \mbox{3 jets}$.
Meanwhile the other group has added the missing subtraction terms in the colour factors $N_c^2$ and $N_c^0$.

In general the subtraction terms are not unique, since arbitrary finite terms can be added to them.
There are a few minor points, where the subtraction terms of this paper differ from the corrected
subtraction terms of ref.~\cite{GehrmannDeRidder:2007hr}.
First of all, ref.~\cite{GehrmannDeRidder:2007hr} introduces ``dipole'' subtraction terms by dividing a
three-parton antenna into two dipoles.
The subtraction terms in this paper are based entirely on complete antennas.
Secondly, ref.~\cite{GehrmannDeRidder:2007hr} symmetrises in the quark momenta, thus eliminating contributions anti-symmetric in the
quark momenta.
In this paper I give subtraction terms which approximate the double-real emission contribution in all singular limits
without the need for symmetrisation in the quark momenta.
Thirdly, ref.~\cite{GehrmannDeRidder:2007hr} uses in the colour factor $N_f N_c$ in the NLO subtraction terms
for a collinear quark pair always an antenna with a gluon as spectator.
As a consequence no soft subtraction term is needed with this choice.
In this paper the NLO subtraction terms are the canonical ones, obtained from the colour flow of the amplitudes.
In this case a soft subtraction term occurs also in the colour factor $N_f N_c$.
In general the occurrence of soft subtraction terms is related to the specific choice of the NLO subtraction terms.
In this work the subtraction terms are developed from the starting point, that the NLO subtraction terms
are given by the canonical choice.
A second possibility, which completely avoids soft subtraction terms, is to modify the NLO subtraction terms
appropriately. This possibility has been discussed recently in ref.~\cite{Somogyi:2009ri}.

The subtraction terms used in this work are based on spin-averaged antenna functions.
Therefore they do not capture phase-space singularities associated to spin-correlations 
and are therefore not local subtraction terms.
These remaining singularities are rather mild and treated by a slicing procedure.
Therefore the complete treatment of all infrared singularities for the process $e^+ e^- \rightarrow \mbox{3 jets}$
is based on a hybrid method of subtraction and slicing.

This paper is necessarily rather technical. It was written for the following reason:
Understanding the infrared singularity structure of the process $e^+ e^- \rightarrow \mbox{3 jets}$ at NNLO
is not only relevant to three-jet observables in electron-positron annihilation,
but to all processes with three or more hard partons, if calculated at NNLO.
The knowledge of the singularity structure allows the construction of appropriate subtraction terms
and I give a full account of the subtraction terms used in the 
numerical program ``Mercutio2'' \cite{Weinzierl:2008iv,Weinzierl:1999yf}.
Numerical results for jet rates and event shapes related to the process $e^+ e^- \rightarrow \mbox{3 jets}$
have been presented in \cite{Weinzierl:2008iv,GehrmannDeRidder:2007bj,GehrmannDeRidder:2007hr,GehrmannDeRidder:2008ug,Ridder:2009dp}.
The numerical results from the program ``Mercutio2'' are presented in a companion paper \cite{Weinzierl:2009bbb}.

This paper is organised as follows:
Section~\ref{sect:notation} introduces the notation and recalls the most important facts of the relevant amplitudes.
Section~\ref{sect:irdiv} discusses the general features of the method used to handle the infrared divergences.
The explicit subtraction terms are given in section~\ref{sect:subtraction_terms}.
Section~\ref{sect_finiteness} shows explicitly the cancellation of all singularities.
Section~\ref{sect_conclusions} contains a summary and the conclusions.
In appendix~\ref{sect:alternative_soft_term} an alternative form for the soft subtraction term is given.
Appendix~\ref{sect:limits} lists the explicit forms of the various terms in the singular limits.

\section{Notation and conventions}
\label{sect:notation}

\subsection{Cross sections and amplitudes}

The master formula to calculate an observable at an collider with no 
initial-state hadrons (e.g. an electron-positron collider) is given by
\bq
\label{master_formula}
\l {\cal O}^{(j)} \r & = & \frac{1}{2 K(s)}
             \frac{1}{\left( 2 J_1+1 \right)}
             \frac{1}{\left( 2 J_2+1 \right)}
             \sum\limits_n
             \int d\phi_{n}
             {\cal O}^{(j)}_n\left(p_1,...,p_n,q_1,q_2\right)
             \sum\limits_{helicity} 
             \left| {\cal A}_{n} \right|^2,
\eq
where $q_1$ and $q_2$ are the momenta of the initial-state particles, 
$2K(s)=2s$ is the flux factor and $s=(q_1+q_2)^2$ is the center-of-mass energy squared.
Kinematical invariants will be denoted by
\bq 
 s_{ij} = \left( p_i + p_j \right)^2,
 \;\;\;\;
 s_{ijk} = \left( p_i + p_j + p_k \right)^2,
 \;\;\;\;
 \mbox{etc.}
\eq
The factors $1/(2J_1+1)$ and $1/(2J_2+1)$ correspond
to an averaging over the initial helicities. $d\phi_{n}$ is the invariant phase space measure
for $n$ final state particles and
${\cal O}^{(j)}_n\left(p_1,...,p_n,q_1,q_2\right)$ is the observable, evaluated with a configuration
depending on $n$ final state partons and two initial state particles.
The index $j$ indicates that the leading order contribution depends on $j$ final-state partons.
The observable has to be infrared safe, in particular this implies that in single and double
unresolved limits we must have
\bq
{\cal O}^{(j)}_{n+1}(p_1,...,p_{n+1},q_1,q_2) & \rightarrow & {\cal O}^{(j)}_n(p_1',...,p_n',q_1',q_2')
 \;\;\;\;\;\;\mbox{for single unresolved limits},
 \nonumber \\
{\cal O}^{(j)}_{n+2}(p_1,...,p_{n+2},q_1,q_2) & \rightarrow & {\cal O}^{(j)}_n(p_1',...,p_n',q_1',q_2')
 \;\;\;\;\;\;\mbox{for double unresolved limits}.
\eq
${\cal A}_n$ is the amplitude with $n$ final-state partons.
At NNLO we need the following perturbative expansions of the amplitudes:
\bq
& & 
 \left| {\cal A}_n \right|^2
 =  
   \left. {\cal A}_n^{(0)} \right.^\ast {\cal A}_n^{(0)} 
 + 
   \left(
             \left. {\cal A}_n^{(0)} \right.^\ast {\cal A}_n^{(1)} 
           + \left. {\cal A}_n^{(1)} \right.^\ast {\cal A}_n^{(0)} 
       \right)
 + 
   \left(
             \left. {\cal A}_n^{(0)} \right.^\ast {\cal A}_n^{(2)} 
           + \left. {\cal A}_n^{(2)} \right.^\ast {\cal A}_n^{(0)}  
           + \left. {\cal A}_n^{(1)} \right.^\ast {\cal A}_n^{(1)} 
       \right),
 \nonumber \\
& &
  \left| {\cal A}_{n+1} \right|^2
 =  
   \left. {\cal A}_{n+1}^{(0)} \right.^\ast {\cal A}_{n+1}^{(0)}
 + 
   \left(
          \left. {\cal A}_{n+1}^{(0)} \right.^\ast {\cal A}_{n+1}^{(1)} 
        + \left. {\cal A}_{n+1}^{(1)} \right.^\ast {\cal A}_{n+1}^{(0)}
 \right),
 \nonumber \\ 
& &
  \left| {\cal A}_{n+2} \right|^2
 = 
   \left. {\cal A}_{n+2}^{(0)} \right.^\ast {\cal A}_{n+2}^{(0)}. 
\eq
Here ${\cal A}_n^{(l)}$ denotes an amplitude with $n$ final-state partons and $l$ loops.
We rewrite the master formula eq.~(\ref{master_formula})
symbolically as 
\bq
\l {\cal O}^{(j)} \r & = & 
             \sum\limits_n \int {\cal O}^{(j)}_{n} \; d\sigma_{n}
\eq
and the LO, NLO and NNLO contribution as 
\bq
\label{def_LO_NLO_NNLO}
\l {\cal O}^{(j)} \r^{LO} & = & \int {\cal O}^{(j)}_{n} \; d\sigma_{n}^{(0)},
 \nonumber \\
\l {\cal O}^{(j)} \r^{NLO} & = & \int {\cal O}^{(j)}_{n+1} \; d\sigma_{n+1}^{(0)} + \int {\cal O}^{(j)}_{n} \; d\sigma_{n}^{(1)},
 \nonumber \\
\l {\cal O}^{(j)} \r^{NNLO} & = & \int {\cal O}^{(j)}_{n+2} \; d\sigma_{n+2}^{(0)} 
                   + \int {\cal O}^{(j)}_{n+1} \; d\sigma_{n+1}^{(1)} 
                   + \int {\cal O}^{(j)}_{n} \; d\sigma_{n}^{(2)}.
\eq
The LO, NLO and NNLO contributions may be decomposed into gauge-invariant colour pieces.
For the process $e^+ e^- \rightarrow \mbox{3 jets}$ the colour decomposition reads
\bq
\label{colour_decomposition}
\l {\cal O}^{(3)} \r^{LO} & = & \frac{1}{2} \left(N_c^2-1\right) \l {\cal O}^{(3)} \r^{LO}_{lc}
 \nonumber \\
\l {\cal O}^{(3)} \r^{NLO} & = & \frac{1}{4} \left(N_c^2-1\right) 
       \left[ N_c \l {\cal O}^{(3)} \r^{NLO}_{lc} 
            + N_f \l {\cal O}^{(3)} \r^{NLO}_{nf}
            + \frac{1}{N_c} \l {\cal O}^{(3)} \r^{NLO}_{sc} \right]
 \nonumber \\
\l {\cal O}^{(3)} \r^{NNLO} & = & \frac{1}{8} \left(N_c^2-1\right) 
       \left[ N_c^2 \l {\cal O}^{(3)} \r^{NNLO}_{lc} 
            + \l {\cal O}^{(3)} \r^{NNLO}_{sc}
            + \frac{1}{N_c^2} \l {\cal O}^{(3)} \r^{NNLO}_{ssc}
\right. \nonumber \\
 & & \left.
            + N_f N_c \l {\cal O}^{(3)} \r^{NNLO}_{nf} 
            + \frac{N_f}{N_c} \l {\cal O}^{(3)} \r^{NNLO}_{nfsc} 
            + N_f^2 \l {\cal O}^{(3)} \r^{NNLO}_{nfnf} \right],
\eq
where $N_c$ denotes the number of colours and $N_f$ the number of light quark flavours.
I use the following normalisation for the colour matrices:
\bq
\mbox{Tr} \; T^a T^b = \frac{1}{2} \delta^{ab}.
\eq
The factors $1/2$, $1/4$ and $1/8$ in eq.~(\ref{colour_decomposition}) are a consequence of this normalisation.
In addition to the terms shown in eq.~(\ref{colour_decomposition}), there are in the colour decomposition 
singlet contributions, which arise from interference terms of amplitudes, where
the electro-weak boson couples to two different fermion lines.
These singlet contributions are expected to be numerically 
small \cite{Dixon:1997th,vanderBij:1988ac,Garland:2002ak}
and neglected in the present calculation.

The computation of the NNLO correction for the process $e^+ e^- \rightarrow \mbox{3 jets}$ 
requires the knowledge of the amplitudes
for the three-parton final state 
$e^+ e^- \rightarrow \bar{q} q g$ up to two-loops \cite{Garland:2002ak,Moch:2002hm},
the amplitudes of the four-parton final states
$e^+ e^- \rightarrow \bar{q} q g g$
and
$e^+ e^- \rightarrow \bar{q} q \bar{q}' q'$
up to one-loop \cite{Bern:1997ka,Bern:1997sc,Campbell:1997tv,Glover:1997eh}
and the five-parton final states
$e^+ e^- \rightarrow \bar{q} q g g g$
and
$e^+ e^- \rightarrow \bar{q} q \bar{q}' q' g$
at tree level \cite{Berends:1989yn,Hagiwara:1989pp,Nagy:1998bb}.
The one-loop amplitude $e^+ e^- \rightarrow g g g$
enters only the NNLO term as a quark singlet contribution and is therefore neglected.
I will label the momenta of the particles of the amplitudes as follows:
\bq
\label{collection_amplitudes}
 {\cal A}_{3,\bar{q}gq}^{(l)}(0,1,2) 
 & = & {\cal A}_{3,\bar{q}gq}^{(l)}(\bar{q}_0,g_1,q_2,e^-_3,e^+_4),
 \nonumber \\
 {\cal A}_{4,\bar{q}ggq}^{(l)}(0,1,2,3) 
 & = & {\cal A}_{4,\bar{q}ggq}^{(l)}(\bar{q}_0,g_1,g_2,q_3,e^-_4,e^+_5),
 \nonumber \\
 {\cal A}_{4,\bar{q}\bar{q}'q'q}^{(l)}(0,1,2,3) 
 & = & {\cal A}_{4,\bar{q}\bar{q}'q'q}^{(l)}(\bar{q}_0,\bar{q}_1',q_2',q_3,e^-_4,e^+_5),
 \nonumber \\
 {\cal A}_{5,\bar{q}gggq}^{(l)}(0,1,2,3,4) 
 & = & {\cal A}_{5,qggg\bar{q}}^{(l)}(\bar{q}_0,g_1,g_2,g_3,q_4,e^-_5,e^+_6),
 \nonumber \\
 {\cal A}_{5,\bar{q}\bar{q}'gq'q}^{(l)}(0,1,2,3,4) 
 & = & {\cal A}_{5,\bar{q}\bar{q}'gq'q}^{(l)}(\bar{q}_0,\bar{q}_1',g_2,q_3',q_4,e^-_5,e^+_6).
\eq
The ordering of the external particles follows the conventions of \cite{Weinzierl:2005dd}.
I use the convention that all particle momenta are outgoing. Therefore $e^-_i$ in
eq.~(\ref{collection_amplitudes}) corresponds
to an outgoing electron with momentum $p_i$,
or equivalently to an incoming positron with momentum $-p_i$. 
Not all amplitudes contribute to all colour factors.
Table \ref{table_colour_decomposition} shows the contributions at NNLO to the individual 
colour factors.
\begin{table}[t]
\begin{center}
\begin{tabular}{|l|c|c|c|c|c|c|}
\hline
  & $N_c^2$ & $N_c^0$ & $N_c^{-2}$ & $N_f N_c$ & $N_f/N_c$ & $N_f^2$ \\
\hline 
 ${\cal A}_{3,\bar{q}gq}^{(0)\ast} {\cal A}_{3,\bar{q}gq}^{(2)}$           & x & x & x & x & x & x \\
 ${\cal A}_{3,\bar{q}gq}^{(1)\ast} {\cal A}_{3,\bar{q}gq}^{(1)}$           & x & x & x & x & x & x \\
 ${\cal A}_{4,\bar{q}ggq}^{(0)\ast} {\cal A}_{4,\bar{q}ggq}^{(1)}$          & x & x & x & x & x & - \\
 ${\cal A}_{4,\bar{q}\bar{q}'q'q}^{(0)\ast} {\cal A}_{4,\bar{q}\bar{q}'q'q}^{(1)}$  & - & x & x & x & x & x \\
 ${\cal A}_{5,\bar{q}gggq}^{(0)\ast} {\cal A}_{5,\bar{q}gggq}^{(0)}$         & x & x & x & - & - & - \\
 ${\cal A}_{5,\bar{q}\bar{q}'gq'q}^{(0)\ast} {\cal A}_{5,\bar{q}\bar{q}'gq'q}^{(0)}$ & - & x & x & x & x & - \\
\hline
\end{tabular}
\caption{
Contributions at NNLO of the various amplitudes to the individual colour factors.
}
\label{table_colour_decomposition}
\end{center}
\end{table}

\subsection{Colour decomposition of the amplitudes}
\label{subsect:colour_decomp}

For the discussion of intermediate results it is useful to have the colour decomposition
of all the relevant amplitudes.
We start with the amplitudes for the three-parton final state $e^+ e^- \rightarrow \bar{q} q g$.
The colour structure of the tree amplitude, the one-loop amplitude and the two-loop amplitude is
\bq
\label{eq_colour_decomp_3_parton}
{\cal A}^{(0)}_3(\bar{q}_0,g_1,q_2)  & = &
 e^2 g c_0 
 \left( T^1 \right)_{20} A^{(0)}_3(\bar{q}_0,g_1,q_2),
 \nonumber \\
{\cal A}^{(1)}_3(\bar{q}_0,g_1,q_2)  & = &
 e^2 g c_0 
        \left( T^1 \right)_{20} \left( \frac{\alpha_s}{2\pi} \right)
        A^{(1)}_{3}(\bar{q}_0,g_1,q_2),
 \nonumber \\
{\cal A}^{(2)}_3(\bar{q}_0,g_1,q_2)  & = &
 e^2 g c_0 
        \left( T^1 \right)_{20} \left( \frac{\alpha_s}{2\pi} \right)^2
        A^{(2)}_{3}(\bar{q}_0,g_1,q_2).
\eq
Here, $e$ denotes the electro-magnetic coupling, $g$ the strong coupling and as usual
we set
$\alpha=e^2/(4\pi)$ and $\alpha_s=g^2/(4\pi)$.
$c_0$ denotes a factor from the electro-magnetic coupling:
\bq
c_0 = -Q^q + v^e v^q {\cal P}_Z(s), 
 & &
{\cal P}_Z(s) = \frac{s}{s-m_Z^2+ i \Gamma_Z m_Z}.
\eq
The electron -- positron pair can either annihilate into a photon or a $Z$-boson. The first term in the expression
for $c_0$ corresponds to an intermediate photon, whereas the last term corresponds to a $Z$-boson.
The left- and right-handed couplings of the $Z$-boson to fermions are given by
\bq
v^f_L = \frac{I^f_3 - Q^f \sin^2 \theta_W}{\sin \theta_W \cos \theta_W}, & &
v^f_R = \frac{- Q^f \sin \theta_W}{\cos \theta_W},
\eq
where $Q^f$ and $I^f_3$ are the charge and the third component of the weak isospin of the fermion. 
If the anti-quark $\bar{q}_0$ in eq.~(\ref{eq_colour_decomp_3_parton}) has colour index $j_0$, the quark $q_2$ has colour index
$i_2$ and the gluon $g_1$ has colour index $a_1$, we abbreviate the notation for the colour matrix by
\bq
 \left( T^1 \right)_{20} & = & T^{a_1}_{i_2 j_0}.
\eq
The one-loop amplitude ${\cal A}^{(1)}_3$ has the colour decomposition
\bq
{\cal A}^{(1)}_3 & = & \frac{N_c}{2} {\cal A}^{(1)}_{3,lc} 
               + \frac{1}{2 N_c} {\cal A}^{(1)}_{3,sc}
               + \frac{N_f }{2} {\cal A}^{(1)}_{3,nf}
               + \frac{1}{2} {\cal A}^{(1)}_{3,vec}
               + \frac{1}{2} {\cal A}^{(1)}_{3,ax}.
\eq
${\cal A}^{(1)}_{3,vec}$ and ${\cal A}^{(1)}_{3,ax}$ represent contributions, where the electro-weak gauge boson
couples to a closed fermion loop with a vector or axial-vector coupling, respectively. 
Due to Furry's theorem (i.e. charge conjugation)
the amplitudes ${\cal A}^{(1)}_{3,vec}$ vanish.
Furthermore, since the $u$, $d$, $s$, $c$ quark masses may be neglected only the $t$, $b$ quark pair
survives an isodoublet cancellation in the fermion loop and contributes to 
${\cal A}^{(1)}_{3,ax}$.
Explicit results for the amplitude ${\cal A}^{(1)}_{3}$ 
can be found in refs.~\cite{Giele:1992vf,Bern:1997sc}.
The two-loop amplitude ${\cal A}^{(2)}_3$ has the colour decomposition
\bq
{\cal A}^{(2)}_3 & = & \frac{N_c^2}{4} {\cal A}^{(2)}_{3,lc}
               + \frac{1}{4} {\cal A}^{(2)}_{3,sc}
               + \frac{1}{4 N_c^2} {\cal A}^{(2)}_{3,ssc}
               + \frac{N_c N_f}{4} {\cal A}^{(2)}_{3,nf}
               + \frac{N_f}{4 N_c} {\cal A}^{(2)}_{3,nfsc}
               + \frac{N_f^2}{4} {\cal A}^{(2)}_{3,nfnf}
 \nonumber \\
 & &
               + \frac{N_c}{4} {\cal A}^{(2)}_{3,vec,lc}
               + \frac{N_c}{4} {\cal A}^{(2)}_{3,ax,lc}
               + \frac{1}{4 N_c} {\cal A}^{(2)}_{3,vec,sc}
               + \frac{1}{4 N_c} {\cal A}^{(2)}_{3,ax,sc}.
\eq
${\cal A}^{(2)}_{3,vec,lc}$, ${\cal A}^{(2)}_{3,vec,sc}$, ${\cal A}^{(2)}_{3,ax,lc}$ and ${\cal A}^{(2)}_{3,ax,sc}$ represent again 
contributions, where the electro-weak gauge boson
couples to a closed fermion loop with a vector or axial-vector coupling.
These contributions are expected to be small and neglected in the present analysis.
Explicit results for the amplitude ${\cal A}^{(2)}_{3}$ 
can be found in refs.~\cite{Garland:2002ak,Moch:2002hm}.

We now turn to the colour decomposition  of the amplitudes of the four-parton final states.
We first consider the decomposition of the amplitudes for the sub-process
$e^+ e^- \rightarrow \bar{q} q g g$.
The colour decomposition of the Born amplitude is
\bq
{\cal A}^{(0)}_4(\bar{q}_0,g_1,g_2,q_3)  = 
   e^2 g^2 c_0 \left[  \left( T^2 T^1 \right)_{30} A^{(0)}_4(\bar{q}_0,g_1,g_2,q_3)
                     + \left( T^1 T^2 \right)_{30} A^{(0)}_4(\bar{q}_0,g_2,g_1,q_3) \right].
\eq
The one-loop amplitude is first decomposed into partial amplitudes
\bq
\lefteqn{
{\cal A}^{(1)}_4(\bar{q}_0,g_1,g_2,q_3)  = 
 e^2 g^2 c_0 \left( \frac{\alpha_s}{2\pi} \right)
 } \nonumber \\
 & & 
 \left[ N_c \left( (T^2 T^1)_{30} A^{(1)}_{4,1}(\bar{q}_0,g_1,g_2,q_3) 
                 + (T^1 T^2)_{30} A^{(1)}_{4,1}(\bar{q}_0,g_2,g_1,q_3) \right) 
\right . \nonumber \\
 & & \left.
                    + \frac{1}{2} \delta^{12} \delta_{30} A^{(1)}_{4,3}(\bar{q}_0,g_1,g_2,q_3)
+ \left( (T^2 T^1)_{30} + (T^1 T^2)_{30} - \frac{1}{N_c} \delta^{12} \delta_{30} \right) A^{(1)}_{4,vec}(\bar{q}_0,g_1,g_2,q_3)
\right. \nonumber \\
 & & \left.
+ \left( (T^2 T^1)_{30} - \frac{1}{2N_c} \delta^{12} \delta_{30} \right) A^{(1)}_{4,4,ax}(\bar{q}_0,g_1,g_2,q_3) 
\right. \nonumber \\
 & & \left.
+ \left( (T^1 T^2)_{30} - \frac{1}{2N_c} \delta^{12} \delta_{30} \right) A^{(1)}_{4,4,ax}(\bar{q}_0,g_2,g_1,q_3) 
+ \frac{1}{2N_c} \delta^{12} \delta_{30} A^{(1)}_{4,5,ax}(\bar{q}_0,g_1,g_2,q_3)
             \right].
\eq
In this formula the one-loop leading-colour partial amplitude is denoted by
$A^{(1)}_{4,1}$ and $A^{(1)}_{4,3}$ denotes a sub-leading-colour contribution.
The partial amplitudes $A^{(1)}_{4,vec}$, $A^{(1)}_{4,4,ax}$ and $A^{(1)}_{4,5,ax}$  represent 
the contributions from a photon
or $Z$ coupling to a fermion loop through a vector or axial-vector coupling.
These contributions are small and neglected in the further analysis.
The partial amplitude $A^{(1)}_{4,1}$ may be further 
decomposed: 
\bq
\lefteqn{
A^{(1)}_{4,1}(\bar{q}_0,g_1,g_2,q_3)
 = } & & 
\nonumber \\
 & & 
 A^{(1)}_{4,1,lc}(\bar{q}_0,g_1,g_2,q_3)
 + \frac{N_f}{N_c} A^{(1)}_{4,1,nf}(\bar{q}_0,g_1,g_2,q_3)
 + \frac{1}{N_c^2} A^{(1)}_{4,1,sc}(\bar{q}_0,g_1,g_2,q_3).
\eq
Explicit results for the amplitude ${\cal A}^{(1)}_{4}(\bar{q}_0,g_1,g_2,q_3)$ 
can be found in ref.~\cite{Bern:1997sc}.

We now consider the colour decomposition of the amplitudes for the four-parton final state
$\bar{q}q\bar{q}'q'$.
The colour decomposition of the tree amplitude reads
\bq
{\cal A}^{(0)}_4(\bar{q}_0,\bar{q}_1',q_2',q_3)
 & = &
 e^2 g^2
 \left\{ 
        \frac{1}{2} \left( \delta_{31} \delta_{20} - \frac{1}{N_c} \delta_{30} \delta_{21} \right) 
        \chi^{(0)}_4(\bar{q}_0,\bar{q}_1',q_2',q_3) 
 \right.
 \nonumber \\
 & & \left.
        - \delta_{flav} 
        \frac{1}{2} \left( \delta_{30} \delta_{21} - \frac{1}{N_c} \delta_{31} \delta_{20} \right) 
        \chi^{(0)}_4(\bar{q}_1',\bar{q}_0,q_2',q_3) \right\}, 
 \nonumber \\
\chi^{(0)}_4(\bar{q}_0,\bar{q}_1',q_2',q_3) 
 & = & 
 c_0(3) A^{(0)}_4(\bar{q}_0,\bar{q}_1',q_2',q_3) + c_0(2) A^{(0}_4(\bar{q}_1',\bar{q}_0,q_3,q_2').
\eq
Here $c_0(q)$ denotes the electro-weak coupling for quark $q$.
The decomposition of the one-loop amplitude is
\bq
{\cal A}^{(1)}_4(\bar{q}_0,\bar{q}_1',q_2',q_3)
 & = & 
 \chi^{(1)}_4(\bar{q}_0,\bar{q}_1',q_2',q_3) - \delta_{flav} \chi^{(1)}_4(\bar{q}_1',\bar{q}_0,q_2',q_3), 
 \nonumber \\
\chi^{(1)}_4(\bar{q}_0,\bar{q}_1',q_2',q_3)
 & = & 
 e^2 g^2 \left( \frac{\alpha_s}{2\pi} \right)
 \frac{1}{2}
 \left( 
        N_c \delta_{31} \delta_{20} \chi^{(1)}_{4,1}(\bar{q}_0,\bar{q}_1',q_2',q_3) 
        + \delta_{30} \delta_{21} \chi^{(1)}_{4,2}(\bar{q}_0,\bar{q}_1',q_2',q_3) 
 \right. \nonumber \\
 & & \left. 
        + \left( \delta_{31} \delta_{20} 
                 - \frac{1}{N_c} \delta_{30} \delta_{21} \right) \chi^{(1)}_{4,ax}(\bar{q}_0,\bar{q}_1',q_2',q_3) \right), 
 \nonumber \\
\chi^{(1)}_{4,1}(\bar{q}_0,\bar{q}_1',q_2',q_3) & = & c_0(3) A^{(1)}_{4,1}(\bar{q}_0,\bar{q}_1',q_2',q_3) + c_0(2) A^{(1)}_{4,1}(\bar{q}_1',\bar{q}_0,q_3,q_2'), 
 \nonumber \\
\chi^{(1)}_{4,2}(\bar{q}_0,\bar{q}_1',q_2',q_3) & = & c_0(3) A^{(1)}_{4,2}(\bar{q}_0,\bar{q}_1',q_2',q_3) + c_0(2) A^{(1)}_{4,2}(\bar{q}_1',\bar{q}_0,q_3,q_2').
\eq
The function $\chi^{(1)}_{4,ax}$ arises from a fermion triangle graph with an axial coupling of the $Z$
and is neglected in the further analysis.
The partial amplitudes $A^{(1)}_{4,1}$ and $A^{(1)}_{4,2}$ can be further decomposed according to
\bq
\lefteqn{
A^{(1)}_{4,1}(\bar{q}_0,\bar{q}_1',q_2',q_3)
 = } & & \nonumber \\
 & &
 A^{(1)}_{4,1,lc}(\bar{q}_0,\bar{q}_1',q_2',q_3)
 + \frac{N_f}{N_c} A^{(1)}_{4,1,nf}(\bar{q}_0,\bar{q}_1',q_2',q_3)
 + \frac{1}{N_c^2} A^{(1)}_{4,1,sc}(\bar{q}_0,\bar{q}_1',q_2',q_3),
 \nonumber \\
\lefteqn{
A^{(1)}_{4,2}(\bar{q}_0,\bar{q}_1',q_2',q_3)
 = } & & \nonumber \\
 & &
 A^{(1)}_{4,2,lc}(\bar{q}_0,\bar{q}_1',q_2',q_3)
 + \frac{N_f}{N_c} A^{(1)}_{4,2,nf}(\bar{q}_0,\bar{q}_1',q_2',q_3)
 + \frac{1}{N_c^2} A^{(1)}_{4,2,sc}(\bar{q}_0,\bar{q}_1',q_2',q_3).
\eq
Explicit results for the amplitude ${\cal A}^{(1)}_{4}(\bar{q}_0,\bar{q}_1',q_2',q_3)$ 
can be found in ref.~\cite{Bern:1997ka}.

Finally let us discuss the colour decomposition of the amplitudes related to the five-parton final state.
The colour decomposition of the tree-level amplitude for $e^+ e^- \rightarrow \bar{q} q g g g$ is
\bq
\lefteqn{
{\cal A}_5^{(0)}(\bar{q}_0,g_1,g_2,g_3,q_4) 
 = } \nonumber & & \\
& &
          ( T^3 T^2 T^1 )_{40} A_5^{(0)}(\bar{q}_0,g_1,g_2,g_3,q_4)
        + ( T^2 T^3 T^1 )_{40} A_5^{(0)}(\bar{q}_0,g_1,g_3,g_2,q_4) 
\nonumber \\ & &
        + ( T^1 T^3 T^2 )_{40} A_5^{(0)}(\bar{q}_0,g_2,g_3,g_1,q_4)
        + ( T^3 T^1 T^2 )_{40} A_5^{(0)}(\bar{q}_0,g_2,g_1,g_3,q_4) 
\nonumber \\ & &
        + ( T^2 T^1 T^3 )_{40} A_5^{(0)}(\bar{q}_0,g_3,g_1,g_2,q_4)
        + ( T^1 T^2 T^3 )_{40} A_5^{(0)}(\bar{q}_0,g_3,g_2,g_1,q_4).
\eq
Explicit results for the amplitude ${\cal A}^{(0)}_{5}(\bar{q}_0,g_1,g_2,g_3,q_4)$ 
can be found in ref.~\cite{Berends:1989yn,Nagy:1998bb}.

The colour structure of the tree-level amplitude for $e^+ e^- \rightarrow \bar{q} q \bar{q}' q' g$
is given by
\bq
{\cal A}_{5}^{(0)}(\bar{q}_0,\bar{q}_1',g_2,q_3',q_4)
 & = & \frac{1}{2} \delta_{40} T^2_{31} D_1(\bar{q}_0,\bar{q}_1',g_2,q_3',q_4) 
     + \frac{1}{2} \delta_{31} T^2_{40} D_2(\bar{q}_0,\bar{q}_1',g_2,q_3',q_4) 
 \\
 & &
     - \frac{1}{2} \delta_{41} T^2_{30} D_3(\bar{q}_0,\bar{q}_1',g_2,q_3',q_4)
     - \frac{1}{2} \delta_{30} T^2_{41} D_4(\bar{q}_0,\bar{q}_1',g_2,q_3',q_4).
 \nonumber 
\eq
The functions $D_j(\bar{q}_0,\bar{q}_1',g_2,q_3',q_4)$ can be further decomposed according to
\bq
D_1(\bar{q}_0,\bar{q}_1',g_2,q_3',q_4) 
 & = & 
 c_0(4) B_1(\bar{q}_0,\bar{q}_1',g_2,q_3',q_4) + c_0(3) B_2(\bar{q}_1',\bar{q}_0,g_2,q_4,q_3') 
 \\ & & 
 + \delta_{flav} \frac{1}{N_c} 
       \left( c_0(4) B_3(\bar{q}_1',\bar{q}_0,g_2,q_3',q_4) + c_0(3) B_4(\bar{q}_0,\bar{q}_1',g_2,q_4,q_3') \right), 
 \nonumber \\
D_2(\bar{q}_0,\bar{q}_1',g_2,q_3',q_4) 
 & = & 
 c_0(4) B_2(\bar{q}_0,\bar{q}_1',g_2,q_3',q_4) + c_0(3) B_1(\bar{q}_1',\bar{q}_0,g_2,q_4,q_3') 
 \nonumber \\ & &
 + \delta_{flav} \frac{1}{N_c} 
       \left( c_0(4) B_4(\bar{q}_1',\bar{q}_0,g_2,q_3',q_4) + c_0(3) B_3(\bar{q}_0,\bar{q}_1',g_2,q_4,q_3') \right),
 \nonumber \\
D_3(\bar{q}_0,\bar{q}_1',g_2,q_3',q_4) 
 & = & 
 c_0(4) \frac{1}{N_c} B_3(\bar{q}_0,\bar{q}_1',g_2,q_3',q_4) + c_0(3) \frac{1}{N_c} B_4(\bar{q}_1',\bar{q}_0,g_2,q_4,q_3') 
 \nonumber \\ & &
 + \delta_{flav} 
       \left( c_0(4) B_1(\bar{q}_1',\bar{q}_0,g_2,q_3',q_4) + c_0(3) B_2(\bar{q}_0,\bar{q}_1',g_2,q_4,q_3') \right),
 \nonumber \\
D_4(\bar{q}_0,\bar{q}_1',g_2,q_3',q_4) 
 & = & 
 c_0(4) \frac{1}{N_c} B_4(\bar{q}_0,\bar{q}_1',g_2,q_3',q_4) + c_0(3) \frac{1}{N_c} B_3(\bar{q}_1',\bar{q}_0,g_2,q_4,q_3') 
 \nonumber \\ & &
 + \delta_{flav} 
       \left( c_0(4) B_2(\bar{q}_1',\bar{q}_0,g_2,q_3',q_4) + c_0(3) B_1(\bar{q}_0,\bar{q}_1',g_2,q_4,q_3') \right).
 \nonumber
\eq
Explicit results for the amplitude ${\cal A}^{(0)}_{5}(\bar{q}_0,\bar{q}_1',g_2,q_3',q_4)$ 
can be found in ref.~\cite{Berends:1989yn,Nagy:1998bb}.

\subsection{Ultraviolet renormalisation of loop amplitudes}

The loop amplitudes have explicit divergences. It is common practise to regulate these divergences
by dimensional regularisation.
The origin of these divergences are either related to ultraviolet or to infrared singularities.
I first discuss ultraviolet divergences. These are removed by redefining the parameters of the theory.
For the QCD corrections to the process $e^+ e^- \rightarrow \mbox{3 jets}$ with massless quarks it is 
sufficient to renormalise the strong coupling constant.
In the $\overline{\mbox{MS}}$ scheme the relation between
the bare coupling $\alpha_0$ and the renormalised coupling $\alpha_s(\mu^2)$ evaluated
at the renormalisation scale $\mu^2$ reads:
\begin{eqnarray}
\alpha_0 & = & \alpha_s S_\eps^{-1} \mu^{2\eps} \left[ 1 
 -\frac{\beta_0}{2\eps} \left( \frac{\alpha_s}{2\pi} \right)
 + \left( \frac{\beta_0^2}{4\eps^2} - \frac{\beta_1}{8\eps} \right) 
   \left( \frac{\alpha_s}{2\pi} \right)^2
 + {\cal O}(\alpha_s^3) \right],
\end{eqnarray}
where 
\begin{eqnarray}
S_\eps & = & \left( 4 \pi \right)^\eps e^{-\eps\gamma_E} \, ,
\end{eqnarray}
is the typical phase-space volume factor in $D =4-2\eps$ dimensions, 
$\gamma_E$ is Euler's constant,
and $\beta_0$ and $\beta_1$ are the first two coefficients of the QCD $\beta$-function:
\begin{eqnarray}
\beta_0 = \frac{11}{3} C_A - \frac{4}{3} T_R N_f,
&\;\;\;&
\beta_1 = \frac{34}{3} C_A^2 - \frac{20}{3} C_A T_R N_f - 4 C_F T_R N_f,
\end{eqnarray}
with the colour factors
\begin{eqnarray}
C_A = N_c, \;\;\; C_F = \frac{N_c^2-1}{2N_c}, \;\;\; T_R = \frac{1}{2}.
\end{eqnarray}
Let the expansion in the strong coupling 
of the unrenormalised amplitude be
\bq
{\cal A}_{n,bare} & = & 
 \left( 4 \pi \alpha_0 \right)^{\frac{n-2}{2}}
 \left[
        \hat{\cal A}_{n,bare}^{(0)} + \left( \frac{\alpha_0}{2\pi} \right) \hat{\cal A}_{n,bare}^{(1)}
                    + \left( \frac{\alpha_0}{2\pi} \right)^2 \hat{\cal A}_{n,bare}^{(2)}
        + O(\alpha_s^3)
 \right].
\eq
Then, the renormalised two-loop amplitude can be expressed as
\bq
{\cal A}_{n,ren} & = & 
 \left( 4 \pi \alpha_s \right)^{\frac{n-2}{2}}
 \left( S_\eps^{-1} \mu^{2\eps} \right)^{\frac{n-2}{2}}
 \left[
        \hat{\cal A}_{n,ren}^{(0)} + \left( \frac{\alpha_s}{2\pi} \right) \hat{\cal A}_{n,ren}^{(1)}
                    + \left( \frac{\alpha_s}{2\pi} \right)^2 \hat{\cal A}_{n,ren}^{(2)}
        + O(\alpha_s^3)
 \right].
\eq
The relations between
the renormalised and the bare amplitudes are
given by
\bq
\hat{\cal A}_{n,ren}^{(0)} & = & \hat{\cal A}_{n,bare}^{(0)},
 \nonumber \\
\hat{\cal A}_{n,ren}^{(1)} & = & S_\eps^{-1} \mu^{2\eps} \hat{\cal A}_{n,bare}^{(1)}
                      - (n-2) \frac{\beta_0}{4\eps} \hat{\cal A}_{n,bare}^{(0)},
 \nonumber \\
\hat{\cal A}_{n,ren}^{(2)} & = & S_\eps^{-2} \mu^{4\eps} \hat{\cal A}_{n,bare}^{(2)}
                      - n \frac{\beta_0}{4\eps} S_\eps^{-1} \mu^{2\eps} \hat{\cal A}_{n,bare}^{(1)}
                      + \frac{(n-2)}{16} 
                        \left( n \frac{\beta_0^2}{2\eps^2} - \frac{\beta_1}{\eps} \right) \hat{\cal A}_{n,bare}^{(0)}.
\eq

\subsection{Infrared structure of loop amplitudes}
\label{sec:infrared}

Let us now turn to the infrared singularities of loop amplitudes.
The infrared pole structure of loop amplitudes in the dimensional regularisation parameter $\eps$
is well understood \cite{Catani:1998bh,Sterman:2002qn,Mitov:2006xs,Becher:2009cu,Gardi:2009qi}
and can be stated explicitly for one- and two-loop amplitudes.
If we regard a loop amplitude as a vector in colour space, then the infrared poles
of the loop amplitude can be expressed through an operator acting on this vector
in colour space. 
We start with the one-loop amplitude, which can be written as
\bq
{\cal A}^{(1)}_n & = & 
 {\bf I}^{(1)}(\eps) {\cal A}^{(0)}_n 
 + {\cal A}^{(1)}_{n,fin}.
\eq
Here ${\bf I}^{(1)}(\eps)$ contains all infrared double and single poles in $1/\eps$ and
${\cal A}^{(1)}_{n,fin}$ is a finite remainder.
At two-loops, the corresponding formula reads:
\bq
{\cal A}^{(2)}_n & = & 
 {\bf I}^{(1)}(\eps) {\cal A}^{(1)}_n 
 + {\bf I}^{(2)}(\eps) {\cal A}^{(0)}_n 
 + {\cal A}^{(2)}_{n,fin}.
\eq
The one-loop insertion operator ${\bf I}^{(1)}$ is given by
\bq
{\bf I}^{(1)}(\eps) & = & 
 \frac{1}{2} \frac{1}{\Gamma(1-\eps)} e^{\eps \gamma_E} 
 \sum\limits_{i} \frac{1}{ {\bf T}_i^2} {\cal V}_i(\eps)
 \sum\limits_{j \neq i} {\bf T}_i {\bf T}_j
 \left( \frac{\mu^2}{-s_{ij}} \right)^\eps,
\eq
where 
\bq
 {\cal V}_i(\eps) & = &
  {\bf T}_i^2 \frac{1}{\eps^2} + \gamma_i \frac{1}{\eps} \, ,
\eq
and the coefficients ${\bf T}_i^2$ and $\gamma_i$ are
\bq
{\bf T}_q^2 = {\bf T}_{\bar{q}}^2 = C_F,
 \;\;\;
{\bf T}_g^2 = C_A, 
 \;\;\;
\gamma_q = \gamma_{\bar{q}} = \frac{3}{2} C_F,
  \;\;\;
\gamma_g = \frac{\beta_0}{2}.
\eq
In general, the colour operators ${\bf T}_i {\bf T}_j$ give rise to colour
correlations. 
However, the colour structure for the Born amplitude  $e^+e^- \rightarrow \bar{q} q g$ 
is rather trivial and the colour operators are in this case proportional to the identity matrix
in colour space:
\bq
{\bf T}_q {\bf T}_{\bar{q}} {\cal A}^{(0)}_3 = T_R {1\over N_c} {\cal A}^{(0)}_3, & &
{\bf T}_q {\bf T}_{g} {\cal A}^{(0)}_3 = {\bf T}_g {\bf T}_{\bar{q}} {\cal A}^{(0)}_3
 = - T_R N_c {\cal A}^{(0)}_3.
\eq
The two-loop insertion operator has the form 
\bq
{\bf I}^{(2)}(\eps) & = &
 - \frac{1}{2} {\bf I}^{(1)}(\eps) \left( {\bf I}^{(1)}(\eps) + \frac{\beta_0}{\eps} \right)
 + e^{-\eps \gamma_E} \frac{\Gamma(1-2\eps)}{\Gamma(1-\eps)}
   \left( \frac{\beta_0}{2 \eps} + K \right) {\bf I}^{(1)}(2 \eps)
 + {\bf H}^{(2)},
\eq
where
\bq
K & = & \left( \frac{67}{18} - \frac{\pi^2}{6} \right) C_A - \frac{10}{9} T_R N_f.
\eq
The operator ${\bf H}^{(2)}$ is process- and scheme-dependent and for 
$e^+e^- \rightarrow q \bar{q} g$, it is given by \cite{Garland:2001tf,Bern:2002tk}
\bq
 {\bf H}^{(2)} {\cal A}^{(0)}_3
 & = & \frac{1}{4} \frac{1}{\Gamma(1-\eps)} e^{\eps \gamma_E}
   \frac{1}{\eps}
   \left( H_q^{(2)} + H_g^{(2)} + H_{\bar{q}}^{(2)} \right) {\cal A}^{(0)}_3,
\eq
where $H_q^{(2)} = H_{\bar{q}}^{(2)}$ and
\bq
H_q^{(2)} & = & 
\left( \frac{7}{4} \zeta_3 - \frac{11}{96} \pi^2 + \frac{409}{864} \right) N_c^2
+ \left( - \frac{1}{4} \zeta_3 - \frac{\pi^2 }{96} - \frac{41}{108} \right)
+ \left( - \frac{3}{2} \zeta_3 + \frac{\pi^2 }{8} - \frac{3}{32} \right) 
   \frac{1}{ N_c^2}
\nonumber \\
 & & 
 + \left( \frac{\pi^2}{48} - \frac{25}{216} \right) \frac{N_c^2 -1}{N_c} N_f,
 \nonumber \\
H_g^{(2)} & = & 
\left( \frac{\zeta_3}{2} + \frac{11}{144} \pi^2 + \frac{5}{12} \right) N_c^2
+ \left( - \frac{\pi^2}{72} - \frac{89}{108} \right) N_c N_f
- \frac{N_f}{4N_c}
+ \frac{5}{27} \, N_f^2. 
\eq

\section{Cancellation of infrared divergences}
\label{sect:irdiv}

\subsection{Singular behaviour in phase space}

QCD amplitudes become singular in phase space, when the momenta of one or more external particles become degenerate.
In perturbative calculations this phenomen occurs first in next-to-leading order calculations
and is related to the term $d\sigma_{n+1}^{(0)}$.
Singularities occur if either two particles become collinear or if one particle becomes soft.
The factorisation properties are most conveniently discussed through decomposing QCD amplitudes into
primitive amplitudes, as it was done in section~\ref{subsect:colour_decomp}.
Primitive amplitudes are defined by
a fixed cyclic ordering of the QCD partons,
a definite routing of the external fermion lines through the diagram
and the particle content circulating in the loop.
In the soft limit one has the factorisation
\bq
\lim\limits_{j \; soft}
A_{n+1}^{(0)}(...,p_i,p_j,p_k,...) & = & \mbox{Eik}_3^{(0)}(p_i,p_j,p_k) A_{n}^{(0)}(...,p_i,p_k,...),
\eq
where the eikonal factor is given by
\bq
\mbox{Eik}_3^{(0)}(p_i,p_j,p_k)
 & = & 
 \frac{2 p_i \cdot \eps(p_j)}{s_{ij}} - \frac{2 p_k \cdot \eps(p_j)}{s_{jk}}.
\eq
The square of the eikonal factor is given by
\bq
\label{sqreikonal}
\sum\limits_{\lambda_j}
\left. \; \mbox{Eik}_3^{(0)}(p_i,p_j,p_k) \right.^\ast 
       \mbox{Eik}_3^{(0)}(p_i,p_j,p_k) & = & 
 4 \frac{s_{ik}}{s_{ij}s_{jk}}.
\eq
Soft singularities lead to colour correlations among the primitive amplitudes $A_{n}^{(0)}$.
\\
\\
In the collinear limit one has the factorisation
\bq
\label{factcollinearlimit}
A_{n+1}^{(0)}(...,p_i,p_j,...) & = & 
 \sum\limits_{\lambda} \; \mbox{Split}_3^{(0)}(p_i,p_j) \; A_{n}^{(0)}(...,p,...).
\eq
where $p_i$ and $p_j$ are the momenta of two adjacent legs and
the sum is over all polarisations.
Squaring the splitting amplitudes one obtains
\bq
\label{PLO}
P_{3,quark}^{(0)}(\lambda,\lambda') & = &  \sum\limits_{\lambda_i, \lambda_j}
 u(p,\lambda) \left. \; \mbox{Split}_3^{(0)}(p_i,p_j) \right.^\ast \mbox{Split}_3^{(0)}(p_i,p_j) \;\bar{u}(p,\lambda'),
 \nonumber \\
P_{3,gluon}^{(0)}(\lambda,\lambda') & = &  \sum\limits_{\lambda_i, \lambda_j}
 \left. \eps^\mu(p,\lambda) \right.^\ast \;
                \left. \mbox{Split}_3^{(0)}(p_i,p_j) \right.^\ast \mbox{Split}_3^{(0)}(p_i,p_j) \; \eps^\nu(p,\lambda').
\eq
$P^{(0)}(\lambda,\lambda')$ is a tensor in spin space. Collinear singularities lead therefore to correlations
in spin space.
The spin-averaged splitting functions are obtained by
\bq
 \langle P_{3,quark}^{(0)} \rangle & = & \frac{1}{2} \sum\limits_\lambda P_{3,quark}^{(0)}(\lambda,\lambda),
 \nonumber \\
 \langle P_{3,gluon}^{(0)} \rangle & = & \frac{1}{2(1-\eps)} \sum\limits_\lambda P_{3,gluon}^{(0)}(\lambda,\lambda).
\eq
One-loop amplitudes factorise in single unresolved limits (i.e. soft or collinear limits) as
\cite{Bern:1994zx,Bern:1998sc,Kosower:1999xi,Kosower:1999rx,Bern:1999ry,Catani:2000pi,Kosower:2003cz}
\bq
\label{oneloopfactformula}
\lim \; A^{(1)}_{n+1}
  & = &
 \sum
 \left(
  \mbox{Sing}_3^{(0)} 
  \cdot A^{(1)}_{n} +
  \mbox{Sing}_3^{(1)} \cdot A^{(0)}_{n}
 \right).
\eq
In the soft limit, the sum involves only one term, whereas in the collinear limit the sum is over the polarisations of the
intermediate parton.
As in the previous case, the sum over the polarisations leads to correlations in spin space.
The one-loop splitting amplitudes have a decomposition into primitive splitting amplitudes:
\bq
 \mbox{Sing}_3^{(1)} & = & \mbox{Sing}_{3,lc}^{(1)} + \frac{N_f}{N_c} \mbox{Sing}_{3,nf}^{(1)} + \frac{1}{N_c^2} \mbox{Sing}_{3,sc}^{(1)}.
\eq
The one-loop splitting amplitudes have an ultraviolet divergence and are renormalised according to
\bq
\mbox{Sing}_{3,ren}^{(1)} & = & S_\eps^{-1} \mu^{2\eps} \; \mbox{Sing}_{3,bare}^{(1)}
                      - \frac{\beta_0}{2 N_c \eps} \; \mbox{Sing}_{3,bare}^{(0)},
\eq
where $\mu^2$ is the renormalisation scale.
At NNLO we have to consider in addition double unresolved limits related to the term $d\sigma_{n+2}^{(0)}$.
Generic double unresolved limits are the double soft limit, the soft-collinear limit and the triple collinear limit.
Tree amplitudes factorise in the double unresolved limits as
\cite{Berends:1989zn,Gehrmann-DeRidder:1998gf,Campbell:1998hg,Catani:1998nv,Catani:1999ss,DelDuca:1999ha,Kosower:2002su}
\bq
\label{factsing}
\lim \; A^{(0)}_{n+2}
  & = &
 \sum
  \mbox{Sing}_4^{(0)} \cdot A^{(0)}_{n}.
\eq
In the double soft limit the sum involves only one term, while in the soft-collinear limit 
and in the triple collinear limit the sum is over the polarisations of the intermediate parton.
Again, the sum over the polarisations leads to correlations in spin space.
The double unresolved limit of two collinear pairs factorises in a product of two splitting functions.

\subsection{The subtraction method}

The individual contributions on the r.h.s. of eq.~(\ref{def_LO_NLO_NNLO})
to $\l {\cal O}^{(j)} \r^{NLO}$ and $\l {\cal O}^{(j)} \r^{NNLO}$ are in general infrared divergent, only the sum is finite.
However, these contributions live on different phase spaces, which prevents a naive Monte Carlo integration approach.
To render the individual contributions finite, one adds and subtracts suitable chosen terms.
The NLO contribution is given by
\bq
\l {\cal O}^{(j)} \r^{NLO} & = & 
   \int \left( {\cal O}^{(j)}_{n+1} \; d\sigma_{n+1}^{(0)} - {\cal O}^{(j)}_{n} \circ d\alpha^{(0,1)}_{n} \right)
 + \int \left( {\cal O}^{(j)}_{n} \; d\sigma_{n}^{(1)} + {\cal O}^{(j)}_{n} \circ d\alpha^{(0,1)}_{n} \right).
\eq
The notation ${\cal O}^{(j)}_{n} \circ d\alpha^{(0,1)}_{n}$ is a reminder, that
in general the approximation is a sum of terms
\bq
{\cal O}^{(j)}_{n} \circ d\alpha^{(0,1)}_{n} & = & \sum {\cal O}^{(j)}_{n} \; d\alpha^{(0,1)}_{n}
\eq
and the mapping used to relate the $n+1$ parton configuration to a $n$ parton configuration
differs in general for each summand.
\\
\\
In a similar way, the NNLO contribution is written as
\bq
\label{nnlo_subtraction}
\l {\cal O}^{(j)} \r^{NNLO} & = &
 \int \left( {\cal O}^{(j)}_{n+2} \; d\sigma_{n+2}^{(0)} 
             - {\cal O}^{(j)}_{n+1} \circ d\alpha^{(0,1)}_{n+1}
             - {\cal O}^{(j)}_{n} \circ d\alpha^{(0,2)}_{n} 
      \right) \nonumber \\
& &
 + \int \left( {\cal O}^{(j)}_{n+1} \; d\sigma_{n+1}^{(1)} 
               + {\cal O}^{(j)}_{n+1} \circ d\alpha^{(0,1)}_{n+1}
               - {\cal O}^{(j)}_{n} \circ d\alpha^{(1,1)}_{n}
        \right) \nonumber \\
& & 
 + \int \left( {\cal O}^{(j)}_{n} \; d\sigma_n^{(2)} 
               + {\cal O}^{(j)}_{n} \circ d\alpha^{(0,2)}_{n}
               + {\cal O}^{(j)}_{n} \circ d\alpha^{(1,1)}_{n}
        \right).
\eq
$d\alpha^{(0,1)}_{n+1}$ is the NLO subtraction term for $(n+1)$-parton configurations,
$d\alpha^{(0,2)}_{n}$ and $d\alpha^{(1,1)}_{n}$ are generic NNLO subtraction terms.
It is convenient to rewrite these terms for a process like $e^+ e^- \rightarrow \mbox{3 jets}$ as
\bq
\label{decomp_subtr_terms}
 d\alpha^{(0,1)}_{n+1} & = & d\alpha^{single}_{n+1}, 
 \nonumber \\
 d\alpha^{(0,2)}_{n} & = & d\alpha^{double}_{n} + d\alpha^{almost}_{n} + d\alpha^{soft}_{n} 
                           - d\alpha^{iterated}_{n},
 \nonumber \\
 d\alpha^{(1,1)}_{n} & = & d\alpha^{loop}_{n} + d\alpha^{product}_{n}
                         - d\alpha^{almost}_{n} - d\alpha^{soft}_{n} + d\alpha^{iterated}_{n}.
\eq
Each term has an interpretation.
$d\alpha^{single}_{n+1}$ is the NLO subtraction term for $(n+1)$-jet observables, containing only
three parton tree-level antenna functions.
At NNLO there are several new subtraction terms required, each of them with a specific structure.
The term $d\alpha^{double}$ contains the four-parton tree-level antenna functions.
The term $d\alpha^{loop}$ contains three-parton one-loop antenna functions together with tree-level
matrix elements and three-parton tree-level antenna functions together with one-loop matrix elements.
The remaining terms $d\alpha^{iterated}$, $d\alpha^{almost}$, $d\alpha^{product}$ and 
$d\alpha^{soft}$ all contain a product of two three-parton tree-level antenna functions. 
In $d\alpha^{iterated}$ and $d\alpha^{almost}$ one antenna function has five-parton kinematics, 
while the other antenna has four-parton kinematics. The former subtraction term is an approximation
to $d\alpha^{single}$, while the latter approximates $d\sigma_{n+2}^{(0)}$ in almost colour-correlated double
unresolved configurations. 
Almost colour-correlated double unresolved configurations correspond to two three-parton tree-level antenna functions
sharing one common hard radiator.
In $d\alpha^{product}$ both antennas have four-parton kinematics.
The term $d\alpha^{soft}$ approximates soft gluon singularities.
The decomposition of eq.~(\ref{decomp_subtr_terms}) applies to the process
$e^+ e^- \rightarrow \mbox{3 jets}$.
For processes with more than three hard partons one has in addition products of 
three-parton tree-level antenna functions corresponding to colour disconnected contributions.

\subsection{Antenna functions}

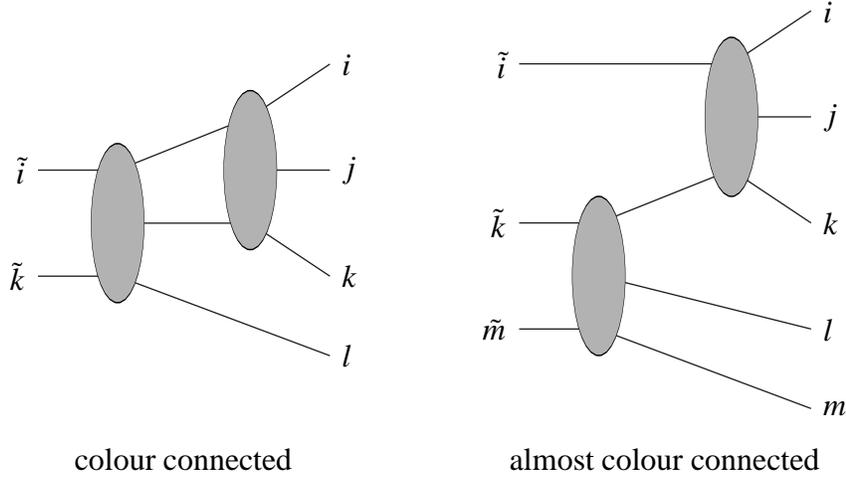
\begin{figure}[t]
\begin{center}
\begin{tabular}{ll}
\begin{picture}(170,170)(0,-30)
 \Line(40,70)(90,90)
 \Line(40,30)(120,0)
 \Line(40,50)(90,50)
 \Line(40,70)(10,70)
 \Line(40,30)(10,30)
 \Line(90,90)(120,110)
 \Line(90,70)(120,70)
 \Line(90,50)(120,30)
 \GOval(40,50)(30,10)(0){0.7}
 \GOval(90,70)(30,10)(0){0.7}
 \Text(125,110)[l]{$i$}
 \Text(125,70)[l]{$j$}
 \Text(125,30)[l]{$k$}
 \Text(125,0)[l]{$l$}
 \Text(5,70)[r]{$\tilde{i}$}
 \Text(5,30)[r]{$\tilde{k}$}
 \Text(65,-35)[t]{colour connected}
\end{picture}
&
\begin{picture}(140,170)(0,-10)
 \Line(40,70)(90,90)
 \Line(40,30)(120,0)
 \Line(40,50)(120,30)
 \Line(40,70)(10,70)
 \Line(40,30)(10,30)
 \Line(90,130)(120,150)
 \Line(90,110)(120,110)
 \Line(90,90)(120,70)
 \Line(90,130)(10,130)
 \GOval(40,50)(30,10)(0){0.7}
 \GOval(90,110)(30,10)(0){0.7}
 \Text(125,150)[l]{$i$}
 \Text(125,110)[l]{$j$}
 \Text(125,70)[l]{$k$}
 \Text(125,30)[l]{$l$}
 \Text(125,0)[l]{$m$}
 \Text(5,130)[r]{$\tilde{i}$}
 \Text(5,70)[r]{$\tilde{k}$}
 \Text(5,30)[r]{$\tilde{m}$}
 \Text(65,-15)[t]{almost colour connected}
\end{picture}
\\
\end{tabular}
\end{center}
\caption{
At NNLO also the iteration of
the three parton tree-level antenna functions $X_3^0$ occur.
}
\label{fig_2}
\end{figure}

For the subtraction terms we need functions which interpolate between the various singular limits, such that
each limit is exactly approximated once and not over-subtracted.
A possibility are antenna functions \cite{Kosower:1998zr,Kosower:2002su,Kosower:2003cz,Gehrmann-DeRidder:2003bm,Gehrmann-DeRidder:2004tv,Gehrmann-DeRidder:2005hi,Gehrmann-DeRidder:2005aw,Gehrmann-DeRidder:2005cm,GehrmannDeRidder:2007jk,Weinzierl:2006ij}.
I will largely follow the notation of ref.~\cite{Gehrmann-DeRidder:2005cm}. 
In a few smaller points the notation differs and
this paragraph defines the notation used in this article 
for the antenna functions.
The small modifications simplify the presentation of the subtraction terms.
Generically the unintegrated antenna functions are denoted by
\bq
 X_3^0(i,j,k),
 \;\;\;
 X_3^1(i,j,k),
 \;\;\;
 X_4^0(i,j,k,l),
\eq
where $X_3^0$ denotes a three-parton tree-level antenna function,
$X_3^1$ denotes a three-parton one-loop antenna function
and $X_4^0$ denotes a four-parton tree-level antenna function.
\begin{figure}[t]
\begin{center}
\begin{tabular}{lll}
\begin{picture}(130,100)(0,0)
 \Line(40,70)(70,90)
 \Line(40,30)(70,10)
 \Line(40,50)(70,50)
 \Line(40,70)(10,70)
 \Line(40,30)(10,30)
 \GOval(40,50)(30,10)(0){0.7}
 \Text(75,90)[l]{$i$}
 \Text(75,50)[l]{$j$}
 \Text(75,10)[l]{$k$}
 \Text(5,70)[r]{$\tilde{i}$}
 \Text(5,30)[r]{$\tilde{k}$}
 \Text(40,-5)[t]{single unresolved}
\end{picture}
&
\begin{picture}(130,100)(0,0)
 \Line(40,70)(70,90)
 \Line(40,30)(70,10)
 \Line(40,50)(70,50)
 \Line(40,70)(10,70)
 \Line(40,30)(10,30)
 \GOval(40,50)(30,10)(0){0.7}
 \GOval(40,50)(15,5)(0){1.0}
 \Text(75,90)[l]{$i$}
 \Text(75,50)[l]{$j$}
 \Text(75,10)[l]{$k$}
 \Text(5,70)[r]{$\tilde{i}$}
 \Text(5,30)[r]{$\tilde{k}$}
 \Text(40,-5)[t]{one-loop unresolved}
\end{picture}
&
\begin{picture}(100,100)(0,0)
 \Line(40,70)(70,90)
 \Line(40,30)(70,10)
 \Line(40,43)(70,37)
 \Line(40,57)(70,63)
 \Line(40,70)(10,70)
 \Line(40,30)(10,30)
 \GOval(40,50)(30,10)(0){0.7}
 \Text(75,90)[l]{$i$}
 \Text(75,63)[l]{$j$}
 \Text(75,37)[l]{$k$}
 \Text(75,10)[l]{$l$}
 \Text(5,70)[r]{$\tilde{i}$}
 \Text(5,30)[r]{$\tilde{l}$}
 \Text(40,-5)[t]{double unresolved}
\end{picture}
\\
\end{tabular}
\end{center}
\caption{
The basic antenna functions occurring in an NNLO calculation: 
The three parton tree-level antenna function $X_3^0$ (left),
the three parton one-loop antenna function $X_3^1$ (middle) and
the four-parton tree-level antenna functions $X_4^0$ (right).
}
\label{fig_1}
\end{figure}
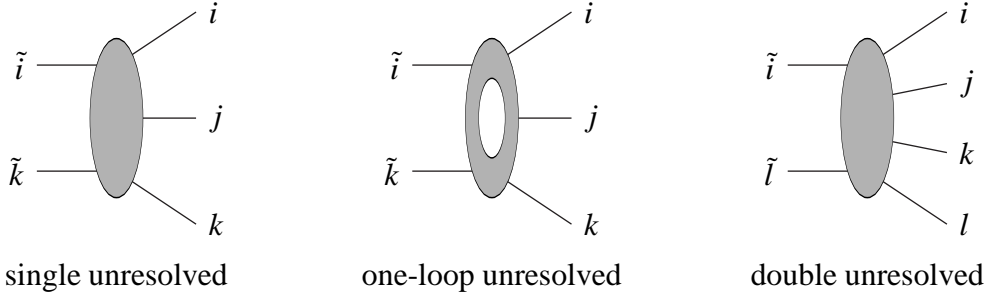
A pictorial representation for these basic antenna functions is shown in fig.~\ref{fig_1}.
At NNLO we also have iterations of the three-parton tree-level antenna functions. For three hard coloured partons
the two generic cases shown in fig.~\ref{fig_2} occur.
They are labelled 'colour-connected' and 'almost colour-connected'.
For four or more hard partons also colour disconnected configurations occur. These are rather trivial and not shown in 
fig.~\ref{fig_2}.
\\
Letters from the beginning of the alphabet are used to denote specific antenna functions:
\bq
 A_3^0(i_q,j_g,k_{\bar{q}}), \; A_3^1(i_q,j_g,k_{\bar{q}}): & & q_{\tilde{i}} + \bar{q}_{\tilde{k}} \rightarrow q_i + g_j + \bar{q}_k,
 \nonumber \\
 D_3^0(i_q,j_g,k_g), \; D_3^1(i_q,j_g,k_g): & & q_{\tilde{i}} + g_{\tilde{k}} \rightarrow q_i + g_j + g_k,
 \nonumber \\
 E_3^0(i_q,j_q,k_{\bar{q}}), \; E_3^1(i_q,j_q,k_{\bar{q}}): & & q_{\tilde{i}} + g_{\tilde{k}} \rightarrow q_i + q_j + \bar{q}_k,
 \nonumber \\
 F_3^0(i_g,j_g,k_g), \; F_3^1(i_g,j_g,k_g): & & g_{\tilde{i}} + g_{\tilde{k}} \rightarrow g_i + g_j + g_k,
 \nonumber \\
 G_3^0(i_g,j_q,k_{\bar{q}}), \; G_3^1(i_g,j_q,k_{\bar{q}}): & & g_{\tilde{i}} + g_{\tilde{k}} \rightarrow g_i + q_j + \bar{q}_k.
\eq
The relevant four-parton antenna functions are
\bq
 A_4^0(i_q,j_g,k_g,l_{\bar{q}}): & & q_{\tilde{i}} + \bar{q}_{\tilde{l}} \rightarrow q_i + g_j + g_k + \bar{q}_l,
 \nonumber \\
 B_4^0(i_q,j_q,k_{\bar{q}},l_{\bar{q}}): & & q_{\tilde{i}} + \bar{q}_{\tilde{l}} \rightarrow q_i + q_j' + \bar{q}_k' + \bar{q}_l,
 \nonumber \\
 C_4^0(i_q,j_q,k_{\bar{q}},l_{\bar{q}}): & & q_{\tilde{i}} + \bar{q}_{\tilde{l}} \rightarrow q_i + q_j + q_k + \bar{q}_l,
 \nonumber \\
 D_4^0(i_q,j_g,k_g,l_g): & & q_{\tilde{i}} + g_{\tilde{l}} \rightarrow q_i + g_j + g_k + g_l,
 \nonumber \\
 E_4^0(i_q,j_q,k_{\bar{q}},l_g): & & q_{\tilde{i}} + g_{\tilde{l}} \rightarrow q_i + q_j + \bar{q}_k + g_l.
\eq
A subscript ``sc'' or ``nf'' is used to indicate contributions
sub-leading in colour or proportional to the number of light flavours $N_f$.
(In ref.~\cite{Gehrmann-DeRidder:2005cm} a tilde and a hat are used to denote these contributions.)
The leading colour one-loop three-parton antenna function is denoted with a subscript ``lc''.
For the four-parton tree-level antenna functions no extra subscript is used for the contribution leading in colour.
For the three-parton one-loop antenna functions we therefore have
\bq
 & &
 A_{3,lc}^1(i_q,j_g,k_{\bar{q}}), \; A_{3,nf}^1(i_q,j_g,k_{\bar{q}}), \; A_{3,sc}^1(i_q,j_g,k_{\bar{q}}),
 \nonumber \\
 & &
 D_{3,lc}^1(i_q,j_g,k_g), \; D_{3,nf}^1(i_q,j_g,k_g),
 \nonumber \\
 & &
 E_{3,lc}^1(i_q,j_q,k_{\bar{q}}), \; E_{3,nf}^1(i_q,j_q,k_{\bar{q}}), \; E_{3,sc}^1(i_q,j_q,k_{\bar{q}}),
 \nonumber \\
 & &
 F_{3,lc}^1(i_g,j_g,k_g), \; F_{3,nf}^1(i_g,j_g,k_g), 
 \nonumber \\
 & &
 G_{3,lc}^1(i_g,j_q,k_{\bar{q}}), \; G_{3,nf}^1(i_g,j_q,k_{\bar{q}}), \; G_{3,sc}^1(i_g,j_q,k_{\bar{q}}).
\eq
We denote by $X_3^1$ the combination
\bq
 X_3^1 & = & X_{3,lc}^1 + \frac{N_f}{N_c} X_{3,nf}^1 - \frac{1}{N_c^2} X_{3,sc}^1.
\eq
The minus sign in the sub-leading colour term is convention.
For the four-parton tree-level antenna functions we have
\bq
 & &
 A_4^0(i_q,j_g,k_g,l_{\bar{q}}), \; A_{4,sc}^0(i_q,j_g,k_g,l_{\bar{q}}),
 \nonumber \\
 & &
 E_4^0(i_q,j_q,k_{\bar{q}},l_g), \; E_{4,sc}^0(i_q,j_q,k_g,l_{\bar{q}}).
\eq
The arguments of the antenna function $E_{4,sc}^0(i_q,j_q,k_g,l_{\bar{q}})$
are labelled according to the singularity structure. 
The four-parton tree-level antenna functions $B_4^0$, $C_4^0$ and $D_4^0$ have only one colour contribution.
I will assume that the antenna functions are already ordered.
Therefore the antenna functions contain only singularities corresponding to the order of the arguments.
In particular the first and the last argument of the ordered antenna functions
indicate the hard partons.
This is the major difference with the notation of ref.~\cite{Gehrmann-DeRidder:2005cm}.
This concerns the three-parton antenna functions
\bq
 D_3^l(i_q,j_g,k_g), \;\;\;
 F_3^l(i_g,j_g,k_g), \;\;\;\;\;\;
 l=1,2 
\eq
and the four-parton antenna functions
\bq
 A_{4,sc}^0(i_q,j_g,k_g,l_{\bar{q}}),
 \;
 D_4^0(i_q,j_g,k_g,l_g),
 \;
 E_4^0(i_q,j_q,k_{\bar{q}},l_g).
\eq
We have the relations
\bq
 D_{3,unordered}^l(i_q,j_g,k_g) 
 & = &
 \frac{1}{2} \left( D_3^l(i_q,j_g,k_g) + D_3^l(i_q,k_g,j_g) \right),
 \nonumber \\
 F_{3,unordered}^l(i_g,j_g,k_g) 
 & = &
 \frac{1}{3} \left( F_3^l(i_g,j_g,k_g) + F_3^l(j_g,k_g,i_g) + F_3^l(k_g,i_g,j_g) \right),
\eq
as well as
\bq
 A_{4,sc,unordered}^0(i_q,j_g,k_g,l_{\bar{q}})
 & = & 
 \frac{1}{2} \left(
 A_{4,sc}^0(i_q,j_g,k_g,l_{\bar{q}})
 +
 A_{4,sc}^0(i_q,k_g,j_g,l_{\bar{q}}) \right),
 \nonumber \\
 D_{4,unordered}^0(i_q,j_g,k_g,l_g)
 & = &
 \frac{1}{4} \left(
 D_4^0(i_q,j_g,k_g,l_g)
 +
 D_4^0(i_q,l_g,k_g,j_g)
 +
 D_4^0(i_q,j_g,l_g,k_g)
 \right. \nonumber \\
 & & \left.
 +
 D_4^0(i_q,l_g,j_g,k_g)
 \right),
 \nonumber \\
 E_{4,unordered}^0(i_q,j_q,k_{\bar{q}},l_g)
 & = &
 E_{4,qqqg}^0(i_q,j_q,k_{\bar{q}},l_g) + E_{4,qgqq}^0(i_q,l_g,j_q,k_{\bar{q}}).
\eq
The integrated antenna functions are denoted by a calligraphic letter, 
the generic ones are
\bq
 {\cal X}_3^0(s_{ijk}),
 \;\;\;
 {\cal X}_3^1(s_{ijk}),
 \;\;\;
 {\cal X}_4^0(s_{ijkl}).
\eq
The integrated antenna functions are related to the unintegrated antenna functions by
\bq
\label{more_orderings}
 {\cal X}_3^l = S_\eps^{-1} \mu^{2\eps} \sum\limits_{orderings} \int d\phi_{unres} \; X_3^l,
 & &
 {\cal X}_4^0 = S_\eps^{-2} \mu^{4\eps} \sum\limits_{orderings} \int d\phi_{unres} \; X_4^0.
\eq
Each antenna is given as a product of a prefactor $P$ or ${\cal P}$, a symmetry factor $S$,
a colour factor $C$ and a kinematical factor $K$ or ${\cal K}$:
\bq
 X = P S C K,
 & &
 {\cal X} = {\cal P} S C {\cal K}.
\eq
The prefactors are given for the unintegrated antenna functions by
\bq
 P & = & \left\{ \begin{array}{ll}
 8 \pi \alpha_s, & \mbox{for} \; X_3^0 \\
 8 \pi \alpha_s \left( \frac{\alpha_s}{\pi} \right), & \mbox{for} \; X_3^1 \\
 \left( 8 \pi \alpha_s \right)^2, & \mbox{for} \; X_4^0 \\
 \end{array} \right.
\eq
For the integrated antenna functions the prefactor is given by
\bq
 {\cal P} & = & \left\{ \begin{array}{ll}
 \frac{\alpha_s}{\pi}, & \mbox{for} \; {\cal X}_3^0 \\
 \left( \frac{\alpha_s}{\pi} \right)^2, & \mbox{for} \; {\cal X}_3^1 \\
 \left( \frac{\alpha_s}{\pi} \right)^2, & \mbox{for} \; {\cal X}_4^0 \\
 \end{array} \right.
\eq
The symmetry and colour factors are equal for the unintegrated and integrated antenna functions and are given 
in table~\ref{table_symmetry_and_colour_factors}.
\begin{table}[t]
\begin{center}
\begin{tabular}{l|lllll|lllll|lllll}
           & $A_3^0$ &       $D_3^0$ &           $E_3^0$ &        $F_3^0$ &           $G_3^0$ 
           & $A_3^1$ &       $D_3^1$ &           $E_3^1$ &        $F_3^1$ &           $G_3^1$ 
           & $A_4^0$ &       $B_4^0$ &           $C_4^0$ &        $D_4^0$ &           $E_4^0$ \\
 \hline
 &&&&& &&&&& &&&&& \\
$S$        &    $1$ & $\frac{1}{2}$ &              $1$ & $\frac{1}{3}$ &               $1$ 
           &    $1$ & $\frac{1}{2}$ &              $1$ & $\frac{1}{3}$ &               $1$ 
           &    $1$ &           $1$ &    $\frac{1}{2}$ & $\frac{1}{2}$ &               $1$ \\
$C$        &    $1$ &           $1$ & $\frac{T_R}{C_A}$ &           $1$ & $\frac{T_R}{C_A}$ 
           &    $1$ &           $1$ & $\frac{T_R}{C_A}$ &           $1$ & $\frac{T_R}{C_A}$ 
           &    $1$ &           $1$ &              $1$ &           $1$ &               $1$ \\ 
\end{tabular}
\caption{
The symmetry and colour factors for the antenna functions.
}
\label{table_symmetry_and_colour_factors}
\end{center}
\end{table}
The normalisation of the kinematical factors $K$ and ${\cal K}$ agrees with the antenna functions 
listed in ref.~\cite{Gehrmann-DeRidder:2005cm}.
The introduction of the symmetry factor ensures for example 
that the antenna functions $A_3^0$, $D_3^0$ and $F_3^0$ or
$A_4^0$ and $D_4^0$ have the same soft singularities.
As an example I list here the $qg\bar{q}$-antenna function in the integrated
and unintegrated form:
\bq
 A_3^0(i_q,j_g,k_{\bar{q}}) & = & 
 8\pi \alpha_s 
 \frac{1}{s_{ijk}} 
 \left( 2 \frac{s_{ik}s_{ijk}}{s_{ij}s_{jk}} 
        + (1-\eps) \frac{s_{jk}}{s_{ij}} 
        + (1-\eps) \frac{s_{ij}}{s_{jk}} 
        - 2 \eps \right),
 \nonumber \\
 {\cal A}_3^0(s_{\tilde{i}\tilde{k}})
 & = & 
 \frac{\alpha_s}{\pi}
 \left( \frac{s_{\tilde{i}\tilde{k}}}{\mu^2} \right)^{-\eps}
 \left[ \frac{1}{\eps^2} + \frac{3}{2\eps} + \frac{19}{4} - \frac{7}{12} \pi^2 \right] + {\cal O}(\eps).
\eq
The one-loop three-parton antenna functions $X_3^1(i,j,k)$ are ultraviolet divergent and need to be renormalised.
I will assume that these functions are all renormalised
at a scale $\mu$. The relation between the renormalised and bare one-loop three-parton antenna functions is given by
\bq
 K_{3,ren}^1 & = & S_\eps^{-1} \mu^{2\eps} K_{3,bare}^1
                      - \frac{\beta_0}{4\eps} K_{3,bare}^0,
\eq
and an identical relation holds between ${\cal K}_{3,ren}^1$ and ${\cal K}_{3,bare}^1$.
It is also useful to know the infrared poles of the one-loop antenna functions.
They are given by
\bq
 \left. A_3^1(i,j,k) \right|_{IR-Poles} & = & 
 \left.
 \left\{
        {\cal A}_3^0(s_{ijk}) - {\cal D}_3^0(s_{ij}) - N_f {\cal E}_3^0(s_{ij}) 
                              - {\cal D}_3^0(s_{jk}) - N_f {\cal E}_3^0(s_{jk}) 
 \right. \right. \nonumber \\
 & & \left. \left.
 - \frac{1}{N_c^2}
 \left[ {\cal A}_3^0(s_{ijk}) - {\cal A}_3^0(s_{ik}) \right]
 \right\}
 A_3^0(i,j,k)
 \right|_{IR-Poles},
\nonumber \\
 \left. D_3^1(i,j,k) \right|_{IR-Poles} & = & 
 \left.
 \left\{
        2 {\cal D}_3^0(s_{ijk}) + 2 N_f {\cal E}_3^0(s_{ijk}) 
      - {\cal D}_3^0(s_{ij}) - N_f {\cal E}_3^0(s_{ij}) 
      - {\cal D}_3^0(s_{ik}) - N_f {\cal E}_3^0(s_{ik}) 
 \right. \right. \nonumber \\
 & & \left. \left.
      - {\cal F}_3^0(s_{jk}) - 2 N_f {\cal G}_3^0(s_{jk}) 
 \right\}
 D_3^0(i,j,k)
 \right|_{IR-Poles},
 \nonumber \\
 \left. E_3^1(i,j,k) \right|_{IR-Poles} & = & 
 \left.
 \left\{
        2 {\cal D}_3^0(s_{ijk}) + 2 N_f {\cal E}_3^0(s_{ijk}) 
        - {\cal A}_3^0(s_{ij}) - {\cal A}_3^0(s_{ik}) 
 \right. \right. \nonumber \\
 & & \left. \left.
 + \frac{1}{N_c^2} {\cal A}_3^0(s_{jk}) 
 \right\}
 E_3^0(i,j,k)
 \right|_{IR-Poles}.
\eq

\subsection{Phase space mappings}

The antenna functions in the subtraction terms are accompanied by amplitudes
and observable functions, which are evaluated with configurations where
one or more partons have been removed.
The momenta of these reduced configurations are related by a map of momenta to the original
configuration. The momentum map has to satisfy momentum conservation and the on-shell
conditions.
We first consider the case where a single particle is removed.
Let $(p_i,p_j,p_k)$ be a set of momenta corresponding to an antenna, 
such that particle $j$ is emitted by the antenna formed
by particles $i$ and $k$
and denote the mapped momenta of the reduced configuration by $(\tilde{p}_i,\tilde{p}_k)$.
There are several possibilities to relate the mapped momenta $(\tilde{p}_i,\tilde{p}_k)$
to the original momenta $(p_i,p_j,p_k)$.
One such possibility is \cite{Kosower:1998zr}:
\bq
\label{momenta_map}
\tilde{p}_i & = & \frac{(1+\rho)s_{ijk} - 2 r s_{jk}}{2(s_{ijk}-s_{jk})} p_i 
          + r p_j
          + \frac{(1-\rho)s_{ijk} - 2 r s_{ij}}{2(s_{ijk}-s_{ij})} p_k,
 \nonumber \\
\tilde{p}_k & = & \frac{(1-\rho)s_{ijk} - 2 (1-r) s_{jk}}{2(s_{ijk}-s_{jk})} p_i 
          + (1-r) p_j
          + \frac{(1+\rho)s_{ijk} - 2 (1-r) s_{ij}}{2(s_{ijk}-s_{ij})} p_k,
\eq
where
\bq
 \rho = \sqrt{1 + 4 r (1-r) \frac{s_{ij}s_{jk}}{s_{ijk} s_{ik}} },
 & &
 r= \frac{s_{jk}}{s_{ij}+s_{jk}}.
\eq
This choice interpolates smoothly between the singular limits $p_i||p_j$ and $p_j||p_k$.
Setting however $r=1$ yields the Catani-Seymour choice \cite{Catani:1997vz}
\bq
\label{catani_ij_k}
\tilde{p}_i = 
 p_i + p_j - \frac{s_{ij}}{s_{ijk}-s_{ij}} p_k,
 & &
\tilde{p}_k = \frac{s_{ijk}}{s_{ijk}-s_{ij}} p_k,
\eq
which can be used in the $p_i||p_j$ limit, but not in the $p_j||p_k$ limit.
This mapping is symmetric in $(i,j)$.
On the other hand, setting $r=0$ yields
\bq
\label{catani_i_jk}
\tilde{p}_i = \frac{s_{ijk}}{s_{ijk}-s_{jk}} p_i,
 & &
\tilde{p}_k = - \frac{s_{jk}}{s_{ijk}-s_{jk}} p_i 
          + p_j
          + p_k,
\eq
the Catani-Seymour mapping for the opposite case, which can be used in the
$p_j||p_k$ limit, but not in the $p_i||p_j$ limit.
This mapping is symmetric in $(j,k)$.
For the three-parton antenna functions, which have singularities in $p_i||p_j$ and 
$p_j||p_k$ we use the mapping eq.~(\ref{momenta_map}).
This concerns the antenna functions
\bq
 A_3^l(i_q,j_g,k_{\bar{q}}), 
 \;\;\;
 D_3^l(i_q,j_g,k_g),
 \;\;\;
 F_3^l(i_g,j_g,k_g).
\eq
The antenna functions $E_3^0$ and $G_3^0$ have on the other hand only singularities
for $p_j||p_k$ and are symmetric in $(j,k)$.
We therefore use the mapping eq.~(\ref{catani_i_jk}), which respects the symmetry
in $(j,k)$ for the antenna functions
\bq
 E_3^l(i_q,j_q,k_{\bar{q}}),
 \;\;\;
 G_3^l(i_g,j_q,k_{\bar{q}}).
\eq
For the double unresolved subtraction terms we have to relate four momenta
$(p_i,p_j,p_k,p_l)$ to two hard momenta $(\tilde{p}_i,\tilde{p}_l)$.
The $4\rightarrow 2$ mapping can be realised through two $3\rightarrow 2$ mappings.
There are two possibilities for four-parton antennas with an ordered singularity structure.
The first possibility is to combine first $(p_i,p_j,p_k)$ into $(\hat{p}_i,\hat{p}_k)$
and then to combine $(\hat{p}_i,\hat{p}_k,p_l)$ into $(\tilde{p}_i,\tilde{p}_l)$:
\bq
\label{mapping_a}
 (p_i,p_j,p_k,p_l) \rightarrow (\hat{p}_i,\hat{p}_k,p_l) \rightarrow (\tilde{p}_i,\tilde{p}_l).
\eq
The second possibility is to combine first $(p_j,p_k,p_l)$ into $(\hat{p}_j,\hat{p}_l)$
and then to combine $(p_i,\hat{p}_j,\hat{p}_l)$ into $(\tilde{p}_i,\tilde{p}_l)$:
\bq
\label{mapping_b}
 (p_i,p_j,p_k,p_l) \rightarrow (p_i,\hat{p}_j,\hat{p}_l) \rightarrow (\tilde{p}_i,\tilde{p}_l).
\eq
As a criteria which mapping should be used we take that the smallest antenna should be recombined
first.
As a measure for the smallness of an antenna $X_3^l(i,j,k)$ we take the product
\bq
 s_{ij} s_{jk}.
\eq
In the limit where these invariants are small this quantity is proportional to the transverse
momentum of the emitted particle $j$.
For the order $(i,j,k,l)$ we therefore compare $s_{ij} s_{jk}$ with $s_{jk} s_{kl}$, or 
equivalently $s_{ij}$ with $s_{kl}$.
If $s_{ij}$ is smaller, one uses eq.~(\ref{mapping_a}),
if on the other hand $s_{kl}$ is smaller one uses eq.~(\ref{mapping_b}).
This procedure is used in combination with eq.~(\ref{momenta_map}) for the antenna functions
\bq
 A_4^0(i_q,j_g,k_g,l_{\bar{q}}),
 \;\;\;
 A_{4,sc}^0(i_q,j_g,k_g,l_{\bar{q}}),
 \;\;\;
 D_4^0(i_q,j_g,k_g,l_g).
\eq
For the antenna functions
\bq
 B_4^0(i_q,j_q,k_{\bar{q}},l_{\bar{q}}),
 \;\;\;
 C_4^0(i_q,j_q,k_{\bar{q}},l_{\bar{q}}),
 \;\;\;
 E_{4,qqqg}^0(i_q,j_q,k_{\bar{q}},l_g).
\eq
we use eq.~(\ref{catani_ij_k}) or eq.~(\ref{catani_i_jk})
for the first mapping.
For the antenna function
\bq
 E_{4,sc}^0(i_q,j_q,k_g,l_{\bar{q}})
\eq
we always map $(j,k,l)$ first with eq.~(\ref{momenta_map}) followed by a mapping
$(p_i,\hat{p}_j,\hat{p}_l)$ with eq.~(\ref{catani_i_jk}).
For the antenna function
\bq
 E_{4,qgqq}^0(i_q,j_g,k_q,l_{\bar{q}})
\eq
we map in the case $s_{ij}<s_{kl}$ first $(i,j,k)$ with eq.~(\ref{momenta_map})
followed by a map $(\hat{p}_i,\hat{p}_k,p_l)$ with eq.~(\ref{catani_i_jk}).
In the case $s_{ij}>s_{kl}$ the momenta $(j,k,l)$ are mapped first with eq.~(\ref{catani_i_jk}) 
followed by a map $(p_i,\hat{p}_j,\hat{p}_l)$ with eq.~(\ref{momenta_map}).
These mappings basically follow the singularity structure of the antenna functions.

A final remark on the notation for the arguments of the amplitudes: The particles in an amplitude
are usually labelled by consecutive numbers.
We have denoted the mapped momenta with a tilde, keeping the original numbering untouched.
The mapped event has one or two partons less and leaves therefore a gap in the 
consecutive numbering. It is convenient to restore the consecutive numbering by relabelling all
particles in the event. The momenta after this relabelling will be denoted with a prime.
As an example consider the five-parton amplitude
${\cal A}_{5,qggg\bar{q}}^{(0)}(p_0,p_1,p_2,p_3,p_4)$ together with the mapping
$(p_1,p_2,p_3) \rightarrow (\tilde{p}_1,\tilde{p}_3)$.
This yields
\bq
 {\cal A}_{4,qgg\bar{q}}^{(0)}(p_0,\tilde{p}_1,\tilde{p}_3,p_4)
\eq
which will be denoted by
\bq
 {\cal A}_{4,qgg\bar{q}}^{(0)}(p_0',p_1',p_2',p_3').
\eq
Therefore
\bq
 p_0'=p_0,
 \;\;\;
 p_1'=\tilde{p}_1,
 \;\;\;
 p_2'=\tilde{p}_3,
 \;\;\;
 p_4'=p_3.
\eq
The subtraction terms to the amplitude 
${\cal A}_{5,\bar{q}\bar{q}'gq'q}^{(0)}(p_0,p_1,p_2,p_3,p_4)$
proportional to $E_3^0$ correspond to a $g \rightarrow q \bar{q}$ splitting and we obtain as the
reduced amplitude ${\cal A}_{4,qgg\bar{q}}^{(0)}$.
By convention we place the merged gluon immediately after the anti-quarks, if the spectator
was an anti-quark.
We place the merged gluon immediately before the quarks, if the spectator was a quark.
Therefore for $E_3^0(0,1,3)$ we have
\bq
 {\cal A}_{4,qgg\bar{q}}^{(0)}(p_0',p_1',p_2',p_3')
 & = &
 {\cal A}_{4,qgg\bar{q}}^{(0)}(\tilde{p}_0,\tilde{p}_3,p_2,p_4),
\eq
while for $E_3^0(4,3,1)$ we have
\bq
 {\cal A}_{4,qgg\bar{q}}^{(0)}(p_0',p_1',p_2',p_3')
 & = &
 {\cal A}_{4,qgg\bar{q}}^{(0)}(p_0,p_2,\tilde{p}_1,\tilde{p}_4).
\eq

\subsection{The hybrid method of subtraction and slicing}

Let us recall eq.~(\ref{nnlo_subtraction}) for the subtraction method
at NNLO:
The NNLO contribution is written as
\bq
\l {\cal O}^{(3)} \r^{NNLO} & = &
 \l {\cal O}^{(3)} \r^{NNLO}_3 + \l {\cal O}^{(3)} \r^{NNLO}_4 + \l {\cal O}^{(3)} \r^{NNLO}_5,
\eq
with
\bq
\l {\cal O}^{(3)} \r^{NNLO}_3 & = &
   \int \left( {\cal O}^{(3)}_{3} \; d\sigma_3^{(2)} 
               + {\cal O}^{(3)}_{3} \circ d\alpha^{(0,2)}_{3}
               + {\cal O}^{(3)}_{3} \circ d\alpha^{(1,1)}_{3}
        \right),
 \nonumber \\
\l {\cal O}^{(3)} \r^{NNLO}_4 & = &
   \int \left( {\cal O}^{(3)}_{4} \; d\sigma_{4}^{(1)} 
               + {\cal O}^{(3)}_{4} \circ d\alpha^{(0,1)}_{4}
               - {\cal O}^{(3)}_{3} \circ d\alpha^{(1,1)}_{3}
        \right), \nonumber \\
\l {\cal O}^{(3)} \r^{NNLO}_5 & = &
 \int \left( {\cal O}^{(3)}_{5} \; d\sigma_{5}^{(0)} 
             - {\cal O}^{(3)}_{4} \circ d\alpha^{(0,1)}_{4}
             - {\cal O}^{(3)}_{3} \circ d\alpha^{(0,2)}_{3} 
      \right).
\eq
The strongest conditions, which we can put on the form of the subtraction terms, are as follows:
\begin{description}
\item{(1)} The explicit poles in the dimensional regularisation parameter $\eps$
of the three-parton contribution $\l {\cal O}^{(3)} \r^{NNLO}_3$ cancel
for each point of the three-particle phase space.
\item{(2a)} The explicit poles in the dimensional regularisation parameter $\eps$
of the four-parton contribution $\l {\cal O}^{(3)} \r^{NNLO}_4$ cancel
for each point of the four-particle phase space.
\item{(2b)} The phase space singularities 
of the four-parton contribution $\l {\cal O}^{(3)} \r^{NNLO}_4$ cancel
point-by-point in the four-particle phase space.
\item{(3)} The phase space singularities 
of the five-parton contribution $\l {\cal O}^{(3)} \r^{NNLO}_5$ cancel
point-by-point in the five-particle phase space.
\end{description}
Due to spin-correlations in the collinear limit the spin-averaged antenna function
do not lead to a point-by-point cancellation of the phase space singularities.
Therefore we have to relax the requirements on the subtraction terms. 
In addition we relax also the requirement on the cancellation of the explicit poles 
in the four-particle phase space.
In a hybrid scheme
of subtraction and slicing the last three items above are replaced by
\begin{description}
\item{(2a')} The explicit poles in the dimensional regularisation parameter $\eps$
of the four-parton contribution $\l {\cal O}^{(3)} \r^{NNLO}_4$ cancel
after integration over unresolved phase spaces
for each point of the resolved phase space.
\item{(2b')} The phase space singularities 
of the four-parton contribution $\l {\cal O}^{(3)} \r^{NNLO}_4$ cancel
after azimuthal averaging has been performed.
\item{(3')} The phase space singularities 
of the five-parton contribution $\l {\cal O}^{(3)} \r^{NNLO}_5$ cancel
after azimuthal averaging has been performed.
\end{description}
The azimuthal average is not performed in the Monte Carlo integration.
Instead a slicing parameter $\eta$ is introduced to regulate the phase space singularities
related to spin-dependent terms.
It is important to note that in this scheme no numerically large contributions proportional
to a power of $\ln \eta$ are cancelled between $\l {\cal O}^{(3)} \r^{NNLO}_5$,
$\l {\cal O}^{(3)} \r^{NNLO}_4$ or $\l {\cal O}^{(3)} \r^{NNLO}_3$.
Each contribution itself is independent of $\eta$ in the limit $\eta\rightarrow 0$.

Let me discuss the cancellation of the explicit poles 
in the four-particle phase space (item 2a') in more detail: We consider contributions of the form
\bq
\label{discussion_2ap}
 \int d\phi_3 {\cal O}_3^{(3)} \left| {\cal A}_3^{(0)} \right|^2
 \; \left[ 
 \int d\phi_{unres,ijk} X_3^0(i,j,k) \sum {\cal Y}_3^0(i,j,k,...) \right].
\eq
The functions ${\cal Y}_3^0(i,j,k,...)$ contain the explicit poles in the dimensional regularisation parameter $\eps$.
The dots indicate that these functions may depend on partons other than partons $i$, $j$ or $k$.
Item 2a' states that we do not require that the explicit poles cancel point by point in the four-parton phase space, 
which is given by $d\phi_4 = d\phi_3 d\phi_{unres,ijk}$. 
Instead we only require that these poles vanish after integration over the unresolved phase-space $d\phi_{unres,ijk}$.
This is sufficient to ensure that the expression in the square bracket is finite and the integration over the three-parton
phase space $d\phi_3$ can be performed in four dimensions.
The observable ${\cal O}_3^{(3)}$  and the matrix element $| {\cal A}_3^{(0)} |^2$
can also be evaluated in four dimensions.

Now we can distinguish two important cases. In the first case the expression in the square bracket vanishes after integration
over the $D$-dimensional unresolved phase space $d\phi_{unres,ijk}$ to all order in $\eps$. This case is relevant for 
the subtraction terms in the main part of this paper. 
Obviously, since the square bracket vanishes to all orders in $\eps$ we have in particular that the
term of order $\eps^0$ vanishes. Therefore we do not obtain any finite terms in this case.

In the second case only the pole terms vanish after integration over the $D$-dimensional unresolved phase space $d\phi_{unres,ijk}$.
In this case eq.~(\ref{discussion_2ap}) will lead to finite terms. This case is discussed in 
appendix~\ref{sect:alternative_soft_term} and only relevant to the alternative subtraction terms presented there.


\section{The subtraction terms}
\label{sect:subtraction_terms}

\subsection{The tree-level single unresolved subtraction terms}
\label{sect:subtraction_terms_nlo}

We first give the NLO subtraction terms $d\alpha_3^{single}$ and $d\alpha_4^{single}$.
The first one gives the complete subtraction term for the NLO coefficient, the second one
enters the NNLO coefficient.
We have
\bq
 d\alpha_{3,\bar{q}ggq}^{single} & = &
 \frac{1}{2} \left\{ 
  \frac{N_c}{2} \left[ D_3^0(0,1,2) + D_3^0(0,2,1) + D_3^0(3,1,2) + D_3^0(3,2,1) \right]
 \right.
 \\
 & & \left.
  - \frac{1}{2N_c} \left[ A_3^0(0,1,3) + A_3^0(0,2,3) \right]
 \right\} \circ \left| {\cal A}_3^{(0)} \right|^2 d\phi_4,
 \nonumber \\
 d\alpha_{3,\bar{q}\bar{q}'q'q}^{single} & = &
 \left\{
 \left( \frac{1}{4} + \frac{N_f-1}{2} \right)
  \frac{N_c}{2} \left[ E_3^0(0,1,2) + E_3^0(1,0,3) + E_3^0(2,3,0) + E_3^0(3,2,1) \right]
 \right. 
 \nonumber \\
 & &
 \left.
 + \frac{1}{4} 
  \frac{N_c}{2} \left[ E_3^0(0,1,3) + E_3^0(1,0,2) + E_3^0(3,2,0) + E_3^0(2,3,1) \right]
 \right\} \circ \left| {\cal A}_3^{(0)} \right|^2 d\phi_4.
 \nonumber 
\eq
The amplitude ${\cal A}_3^{(0)}$ is evaluated with momenta obtained from eq.~(\ref{momenta_map}),
which depend on the partons, which form the antenna.
Integration over an one-parton phase space yields
\bq
\lefteqn{
 \int\limits_1 d\alpha_3^{single} = 
 \int\limits_1 \left( d\alpha_{3,\bar{q}ggq}^{single} + d\alpha_{3,\bar{q}\bar{q}'q'q}^{single} \right)
 = } & & \\
 & &
 \left\{
  \frac{N_c}{2} \left[ {\cal D}_3^0(s_{01}) + N_f {\cal E}_3^0(s_{01}) 
                   + {\cal D}_3^0(s_{12}) + N_f {\cal E}_3^0(s_{12}) 
              \right]
  - \frac{1}{2N_c} {\cal A}_3^0(s_{02}) 
 \right\} \left| {\cal A}_3^{(0)} \right|^2 d\phi_3.
 \nonumber
\eq
For the subtraction term $d\alpha_4^{single}$ we have to take into account colour correlations.
For the sub-process $e^+ e^- \rightarrow \bar{q} g g g q$ we first define a two-component vector in colour space obtained
from the colour-ordered amplitudes of $e^+ e^- \rightarrow \bar{q} g g q$:
\bq
 \vec{\cal A}_4^{(0)}(\bar{q}_0,g_1,g_2,q_3)
 & = & 
 e^2 g^2 c_0
 \left( \begin{array}{c}
 A^{(0)}_4(\bar{q}_0,g_1,g_2,q_3) \\
 A^{(0)}_4(\bar{q}_0,g_2,g_1,q_3) \\
        \end{array}
 \right).
\eq
For the subtraction term we have
\bq
\lefteqn{
 d\alpha_{4,\bar{q}gggq}^{single} = 
 \frac{1}{3} \frac{N_c^2-1}{8} 
} & & \nonumber \\
 & &
\left\{
 \left[ D_3^0(0,1,2) + D_3^0(0,2,1) + D_3^0(0,3,1) \right]
 \circ
 \vec{\cal A}_4^{(0)\dagger} \left( \begin{array}{cc}
          N_c^2-1 & -1 \\
               -1 & -1 \\
 \end{array} \right) \vec{\cal A}_4^{(0)}
 \right. \nonumber \\
 & & \left.
+
 \left[ D_3^0(0,2,3) + D_3^0(0,3,2) + D_3^0(0,1,3) \right]
 \circ
 \vec{\cal A}_4^{(0)\dagger} \left( \begin{array}{cc}
          -1 &      -1 \\
          -1 & N_c^2-1 \\
 \end{array} \right) \vec{\cal A}_4^{(0)}
 \right. \nonumber \\
 & & \left.
+
 \left[ F_3^0(1,2,3) + F_3^0(1,3,2) + F_3^0(2,1,3) \right]
 \circ
 \vec{\cal A}_4^{(0)\dagger} \left( \begin{array}{cc}
          N_c^2 & 0     \\
              0 & N_c^2 \\
 \end{array} \right) \vec{\cal A}_4^{(0)}
 \right. \nonumber \\
 & & \left.
+
 \left[ D_3^0(4,2,1) + D_3^0(4,3,1) + D_3^0(4,1,2) \right]
 \circ
 \vec{\cal A}_4^{(0)\dagger} \left( \begin{array}{cc}
          -1 &      -1 \\
          -1 & N_c^2-1 \\
 \end{array} \right) \vec{\cal A}_4^{(0)}
 \right. \nonumber \\
 & & \left.
+
 \left[ D_3^0(4,3,2) + D_3^0(4,1,3) + D_3^0(4,2,3) \right]
 \circ
 \vec{\cal A}_4^{(0)\dagger} \left( \begin{array}{cc}
          N_c^2-1 & -1 \\
               -1 & -1 \\
 \end{array} \right) \vec{\cal A}_4^{(0)}
 \right. \nonumber \\
 & & \left.
+
 \left[ A_3^0(0,1,4) + A_3^0(0,2,4) + A_3^0(0,3,4) \right]
 \circ
 \vec{\cal A}_4^{(0)\dagger} \left( \begin{array}{cc}
          \frac{1}{N_c^2} & 1 + \frac{1}{N_c^2} \\
          1 + \frac{1}{N_c^2} & \frac{1}{N_c^2} \\ 
 \end{array} \right) \vec{\cal A}_4^{(0)}
 \right\} d\phi_5.
\eq
For the sub-process $e^+ e^- \rightarrow \bar{q} q \bar{q}' q' g$ we have two subtractions terms,
one involving the two-quark two-gluon amplitudes of $e^+ e^- \rightarrow \bar{q} g g q$,
the other one involving the four-quark amplitudes of $e^+ e^- \rightarrow \bar{q} q \bar{q}' q'$:
\bq
 d\alpha_{4,\bar{q}\bar{q}'gq'q}^{single} & = & 
  d\alpha_{4,\bar{q}\bar{q}'gq'q \rightarrow \bar{q}ggq}^{single} 
 + d\alpha_{4,\bar{q}\bar{q}'gq'q \rightarrow \bar{q}\bar{q}'q'q}^{single}.
\eq
We have
\bq
\lefteqn{
 d\alpha_{4,\bar{q}\bar{q}'gq'q \rightarrow \bar{q}ggq}^{single} = 
 \frac{N_c^2-1}{8}
 } & & \nonumber \\
 & &
 \left\{
 \left[ 
        2 \left( \frac{1}{4} + \frac{N_f-1}{2} \right)
        \left( E_3^0(0,1,3) + E_3^0(1,0,4) + E_3^0(3,4,0) + E_3^0(4,3,1) \right)
 \right. \right. \nonumber \\
 & & \left. \left.
       +
        2 \frac{1}{4}
        \left( E_3^0(0,1,4) + E_3^0(1,0,3) + E_3^0(4,3,0) + E_3^0(3,4,1) \right)       
 \right]
 \circ
 \vec{\cal A}_4^{(0)\dagger} \left( \begin{array}{cc}
          -1 &      -1 \\
          -1 & N_c^2-1 \\
 \end{array} \right) \vec{\cal A}_4^{(0)}
 \right. \nonumber \\
 & & \left.
+
 \left[ 
        2 \left( \frac{1}{4} + \frac{N_f-1}{2} \right)
        \left( G_3^0(2,1,3) + G_3^0(2,0,4) \right)
       +
        2 \frac{1}{4}
        \left( G_3^0(2,0,3) + G_3^0(2,1,4) \right)       
 \right]
 \right. \nonumber \\
 & & \left.
 \circ
 \vec{\cal A}_4^{(0)\dagger} \left( \begin{array}{cc}
          N_c^2 & 0     \\
              0 & N_c^2 \\
 \end{array} \right) \vec{\cal A}_4^{(0)}
 \right\} d\phi_5.
\eq
The asymmetry in the colour factors for the $E$-terms is just related to way the partons
are labelled in the amplitudes.
For the subtraction term $d\alpha_{4,\bar{q}\bar{q}'gq'q \rightarrow \bar{q}\bar{q}'q'q}^{single}$
we introduce a two-component vector in colour-space obtained from the colour-ordered amplitudes
of $e^+ e^- \rightarrow \bar{q} q \bar{q}' q'$:
\bq
 \vec{\cal \chi}_4^{(0)}(\bar{q}_0,\bar{q}_1',q_2',q_3)
 & = & 
 e^2 g^2 
 \left( \begin{array}{c}
 \chi^{(0)}_4(\bar{q}_0,\bar{q}_1',q_2',q_3) \\
 \delta_{flav} \chi^{(0)}_4(\bar{q}_1',\bar{q}_0,q_2',q_3) \\
        \end{array}
 \right).
\eq
We then have
\bq
 d\alpha_{4,\bar{q}\bar{q}'gq'q \rightarrow \bar{q}\bar{q}'q'q}^{single} & = &
 \frac{N_c^2-1}{8}
 \left\{
 \left[ A_3^0(0,2,3) + A_3^0(1,2,4) \right]
 \circ
 \vec{\cal \chi}_4^{(0)\dagger} \left( \begin{array}{cc}
   N_c - \frac{2}{N_c} & - \frac{1}{N_c^2} \\
   - \frac{1}{N_c^2} & - \frac{1}{N_c} \\
 \end{array} \right) \vec{\cal \chi}_4^{(0)}
 \right. \nonumber \\
 & & \left.
+
 \left[ A_3^0(0,2,4) + A_3^0(1,2,3) \right]
 \circ
 \vec{\cal \chi}_4^{(0)\dagger} \left( \begin{array}{cc}
   - \frac{1}{N_c}   & - \frac{1}{N_c^2} \\
   - \frac{1}{N_c^2} & N_c - \frac{2}{N_c} \\
 \end{array} \right) \vec{\cal \chi}_4^{(0)}
 \right. \\
 & & \left.
+
 \left[ A_3^0(0,2,1) + A_3^0(3,2,4) \right]
 \circ
 \vec{\cal \chi}_4^{(0)\dagger} \left( \begin{array}{cc}
 \frac{2}{N_c} & 1 + \frac{1}{N_c^2} \\
 1 + \frac{1}{N_c^2} & \frac{2}{N_c} \\
 \end{array} \right) \vec{\cal \chi}_4^{(0)}
 \right\} d\phi_5.
 \nonumber 
\eq
Integration of these subtraction terms yields
\bq
\lefteqn{
 \int\limits_1 \left( d\alpha_{4,\bar{q}gggq}^{single} + d\alpha_{4,\bar{q}\bar{q}'gq'q \rightarrow \bar{q}ggq}^{single} \right)
 = 
 \frac{N_c^2-1}{8}
 } & & \nonumber \\
 & &
 \left\{
 \left[ {\cal D}_3^0(s_{01}) + {\cal D}_3^0(s_{23}) + N_f {\cal E}_3^0(s_{01}) + N_f {\cal E}_3^0(s_{23}) \right]
 \vec{\cal A}_4^{(0)\dagger} \left( \begin{array}{cc}
          N_c^2-1 & -1 \\
               -1 & -1 \\
 \end{array} \right) \vec{\cal A}_4^{(0)}
 \right. \nonumber \\
 & & \left.
+
 \left[ {\cal D}_3^0(s_{02}) + {\cal D}_3^0(s_{13}) + N_f {\cal E}_3^0(s_{02}) + N_f {\cal E}_3^0(s_{13}) \right]
 \vec{\cal A}_4^{(0)\dagger} \left( \begin{array}{cc}
          -1 &      -1 \\
          -1 & N_c^2-1 \\
 \end{array} \right) \vec{\cal A}_4^{(0)}
 \right. \nonumber \\
 & & \left.
+
 \left[ {\cal F}_3^0(s_{12}) + 2 N_f {\cal G}_3^0(s_{12}) \right]
 \vec{\cal A}_4^{(0)\dagger} \left( \begin{array}{cc}
          N_c^2 & 0     \\
              0 & N_c^2 \\
 \end{array} \right) \vec{\cal A}_4^{(0)}
 \right. \nonumber \\
 & & \left.
+
 \left[ {\cal A}_3^0(s_{03}) \right]
 \vec{\cal A}_4^{(0)\dagger} \left( \begin{array}{cc}
          \frac{1}{N_c^2} & 1 + \frac{1}{N_c^2} \\
          1 + \frac{1}{N_c^2} & \frac{1}{N_c^2} \\ 
 \end{array} \right) \vec{\cal A}_4^{(0)}
 \right\} d\phi_4,
\eq
\bq
 \int\limits_1 d\alpha_{4,\bar{q}\bar{q}'gq'q \rightarrow \bar{q}\bar{q}'q'q}^{single} & = & 
 \frac{N_c^2-1}{8}
 \left\{
 \left[ {\cal A}_3^0(s_{02}) + {\cal A}_3^0(s_{13}) \right]
 \vec{\cal \chi}_4^{(0)\dagger} \left( \begin{array}{cc}
   N_c - \frac{2}{N_c} & - \frac{1}{N_c^2} \\
   - \frac{1}{N_c^2} & - \frac{1}{N_c} \\
 \end{array} \right) \vec{\cal \chi}_4^{(0)}
 \right. \nonumber \\
 & & \left.
+
 \left[ {\cal A}_3^0(s_{03}) + {\cal A}_3^0(s_{12}) \right]
 \vec{\cal \chi}_4^{(0)\dagger} \left( \begin{array}{cc}
   - \frac{1}{N_c}   & - \frac{1}{N_c^2} \\
   - \frac{1}{N_c^2} & N_c - \frac{2}{N_c} \\
 \end{array} \right) \vec{\cal \chi}_4^{(0)}
 \right. \nonumber \\
 & & \left.
+
 \left[ {\cal A}_3^0(s_{01}) + {\cal A}_3^0(s_{23}) \right]
 \vec{\cal \chi}_4^{(0)\dagger} \left( \begin{array}{cc}
 \frac{2}{N_c} & 1 + \frac{1}{N_c^2} \\
 1 + \frac{1}{N_c^2} & \frac{2}{N_c} \\
 \end{array} \right) \vec{\cal \chi}_4^{(0)}
 \right\} d\phi_4.
\nonumber
\eq
In the first equation
terms proportional to ${\cal E}_3^0$ have been equally distributed between the two colour matrices.

\subsection{The one-loop single unresolved subtraction terms}
\label{sect:subtraction_terms_loop}

For the subtraction term $d\alpha_3^{loop}$ we have two contributions corresponding to the sub-processes
$e^+ e^- \rightarrow \bar{q} g g q$ and $e^+ e^- \rightarrow \bar{q} q \bar{q}' q'$.
Each of these contributions will have terms proportional to the one-loop interference term 
$2\mbox{Re}\;{\cal A}_3^{(0)\dagger} {\cal A}_3^{(1)}$ and the tree-level antennas $X_3^0$ as well as terms
proportional to the Born matrix element $| {\cal A}_3^{(0)} |^2$ and one-loop antenna functions $X_3^1$.
We therefore write
\bq
 d\alpha_{3,\bar{q}ggq}^{loop} & = & d\alpha_{3,\bar{q}ggq,a}^{loop} + d\alpha_{3,\bar{q}ggq,b}^{loop},
 \nonumber \\
 d\alpha_{3,\bar{q}\bar{q}'q'q}^{loop} & = & d\alpha_{3,\bar{q}\bar{q}'q'q,a}^{loop} + d\alpha_{3,\bar{q}\bar{q}'q'q,b}^{loop}.
\eq
We have
\bq
 d\alpha_{3,\bar{q}ggq,a}^{loop} & = & 
 \frac{1}{2} \left\{ 
  \frac{N_c}{2} \left[ D_3^0(0,1,2) + D_3^0(0,2,1) + D_3^0(3,1,2) + D_3^0(3,2,1) \right]
 \right.
 \\
 & & \left.
  - \frac{1}{2N_c} \left[ A_3^0(0,1,3) + A_3^0(0,2,3) \right]
 \right\} \circ \left( {\cal A}_3^{(0)\dagger} {\cal A}_3^{(1)} + {\cal A}_3^{(1)\dagger} {\cal A}_3^{(0)} \right) d\phi_4,
 \nonumber \\
d\alpha_{3,\bar{q}ggq,b}^{loop} & = & 
 \frac{1}{2} \left\{ 
  \frac{N_c^2}{4} \left[ 
    D_3^1(0,1,2) + D_3^1(0,2,1)
  + D_3^1(3,1,2) + D_3^1(3,2,1)
    \right]
 \right.
 \nonumber \\
 & & \left.
  - \frac{1}{4} \left[ 
    A_3^1(0,1,3) 
    + \left( N_c^2 + 1 \right) A_{3,sc}^1(0,1,3)
 \right. \right. \nonumber \\
 & & \left. \left.
  + A_3^1(0,2,3) 
    + \left( N_c^2 + 1 \right) A_{3,sc}^1(0,2,3)
  \right]
 \right\} \circ \left| {\cal A}_3^{(0)} \right|^2 d\phi_4,
 \nonumber \\
 d\alpha_{3,\bar{q}\bar{q}'q'q,a}^{loop} & = &
 \left\{
 \left( \frac{1}{4} + \frac{N_f-1}{2} \right)
  \frac{N_c}{2} \left[ E_3^0(0,1,2) + E_3^0(1,0,3) + E_3^0(2,3,0) + E_3^0(3,2,1) \right]
 \right. 
 \nonumber \\
 & &
 \left.
 + \frac{1}{4} 
  \frac{N_c}{2} \left[ E_3^0(0,1,3) + E_3^0(1,0,2) + E_3^0(3,2,0) + E_3^0(2,3,1) \right]
 \right\}
 \nonumber \\
 & & 
 \circ \left( {\cal A}_3^{(0)\dagger} {\cal A}_3^{(1)} + {\cal A}_3^{(1)\dagger} {\cal A}_3^{(0)} \right) d\phi_4,
 \nonumber \\
d\alpha_{3,\bar{q}\bar{q}'q'q,b}^{loop} & = &
 \left\{
 \left( \frac{1}{4} + \frac{N_f-1}{2} \right)
  \frac{N_c^2}{4} \left[ 
    E_3^1(0,1,2)
  + E_3^1(1,0,3)
  + E_3^1(2,3,0)
  + E_3^1(3,2,1) 
  \right]
 \right. 
 \nonumber \\
 & &
 \left.
 + \frac{1}{4} 
  \frac{N_c^2}{4} \left[ 
    E_3^1(0,1,3)
  + E_3^1(1,0,2)
  + E_3^1(3,2,0)
  + E_3^1(2,3,1)
  \right]
 \right\} \circ \left| {\cal A}_3^{(0)} \right|^2 d\phi_4.
 \nonumber 
\eq
Integration over an one-parton phase space yields
\bq
\lefteqn{
 \int\limits_1 d\alpha_3^{loop} = 
 \int\limits_1 \left( d\alpha_{3,\bar{q}ggq}^{loop} + d\alpha_{3,\bar{q}\bar{q}'q'q}^{loop} \right)
 = } & & \nonumber \\
 & &
 \left\{
  \frac{N_c}{2} \left[ {\cal D}_3^0(s_{01}) + N_f {\cal E}_3^0(s_{01}) 
                   + {\cal D}_3^0(s_{12}) + N_f {\cal E}_3^0(s_{12}) 
              \right]
  - \frac{1}{2N_c} {\cal A}_3^0(s_{02}) 
 \right\} 
 \nonumber \\
 & &
 \cdot
\left( {\cal A}_3^{(0)\dagger} {\cal A}_3^{(1)} + {\cal A}_3^{(1)\dagger} {\cal A}_3^{(0)} \right) d\phi_3
\nonumber \\
 & &
 +
 \left\{
  \frac{N_c^2}{4} \left[ {\cal D}_3^1(s_{01}) + N_f {\cal E}_3^1(s_{01}) 
                   + {\cal D}_3^1(s_{12}) + N_f {\cal E}_3^1(s_{12}) 
              \right]
  - \frac{1}{4} \left[ {\cal A}_3^1(s_{02}) + (N_c^2+1) {\cal A}_{3,sc}^1(s_{02}) \right] 
 \right\} 
 \nonumber \\
 & & 
 \cdot
 \left| {\cal A}_3^{(0)} \right|^2 d\phi_3.
\eq

\subsection{The iterated single unresolved subtraction terms}
\label{sect:subtraction_terms_product}

Whereas the subtraction term $d\alpha_3^{loop}$ approximates the one-loop matrix element in singular phase space regions, there are
on the four-parton phase space additional terms, descending from the five-parton phase space:
\bq
 d\alpha_{4}^{single} + d\alpha_3^{almost} + d\alpha_3^{iterated} + d\alpha_3^{soft}.
\eq
These terms themselves are singular in single unresolved phase space regions, except for the term
$d\alpha_3^{soft}$, which is integrable in all limits.
The term $d\alpha_3^{product}$ is a subtraction term for these contributions.
In the unintegrated version $d\alpha_3^{product}$ lives on the four-parton phase space and involves one
unintegrated three-parton tree-level antenna function together with one integrated three-parton tree-level antenna function,
which however does only depend on the hard three-parton kinematics.
Therefore this integrated antenna function factorises from the integration over the unresolved phase space
and the integrated version is just a product of two integrated three-parton tree-level antenna functions
with three-parton kinematics. We have
\bq
\lefteqn{
 d\alpha_{3,\bar{q}ggq}^{product} = 
 \frac{1}{2} \left\{ 
  \frac{N_c^2}{4} \left[ D_3^0(0,1,2) + D_3^0(0,2,1) - D_3^0(3,1,2) - D_3^0(3,2,1) \right]
\right. } & & \\
 & & \left.
 \circ
 \left( 
        - \frac{1}{2} {\cal D}_3^0(s_{0'1'}) - \frac{N_f}{2} {\cal E}_3^0(s_{0'1'})
        + \frac{1}{2} {\cal D}_3^0(s_{1'2'}) + \frac{N_f}{2} {\cal E}_3^0(s_{1'2'})
 \right)
 \right. \nonumber \\
 & & \left.
- \left( \frac{N_c^2}{4} \frac{1}{2}  + \frac{1}{4} \right)
  \left[ D_3^0(0,1,2) + D_3^0(0,2,1) + D_3^0(3,1,2) + D_3^0(3,2,1) \right]
 \circ
        {\cal A}_3^0(s_{0'2'})
 \right.
 \nonumber \\
 & & \left.
  + \frac{N_c^2}{4} \left[ A_3^0(0,1,3) + A_3^0(0,2,3) \right]
 \circ
 \left( 
          \frac{1}{2} {\cal D}_3^0(s_{0'1'}) + \frac{N_f}{2} {\cal E}_3^0(s_{0'1'})
        + \frac{1}{2} {\cal D}_3^0(s_{1'2'}) + \frac{N_f}{2} {\cal E}_3^0(s_{1'2'})
 \right)
 \right.
 \nonumber \\
 & & \left.
  + 
 \left( \frac{N_c^2}{4} \frac{1}{2}  + \frac{1}{4} \right) \left[ A_3^0(0,1,3) + A_3^0(0,2,3) \right]
 \circ
        {\cal A}_3^0(s_{0'2'})
  \right\} \circ \left| {\cal A}_3^{(0)} \right|^2 d\phi_4,
 \nonumber \\
\lefteqn{
 d\alpha_{3,\bar{q}\bar{q}'q'q}^{product} = 
 \left\{
 \frac{N_c^2}{4} \left[
 \left( \frac{1}{4} + \frac{N_f-1}{2} \right)
  \left[ E_3^0(0,1,2) + E_3^0(1,0,3) - E_3^0(2,3,0) - E_3^0(3,2,1) \right]
 \right. \right. } & &
 \nonumber \\
 & &
 \left. \left.
 + \frac{1}{4} 
  \left[ E_3^0(0,1,3) + E_3^0(1,0,2) - E_3^0(3,2,0) - E_3^0(2,3,1) \right]
 \right]
 \circ
 \left( 
        - \frac{1}{2} {\cal D}_3^0(s_{0'1'}) 
        + \frac{1}{2} {\cal D}_3^0(s_{1'2'}) 
 \right)
 \right.
 \nonumber \\
 & &
 \left.
- 
 \left( \frac{N_c^2}{4} \frac{1}{2}  + \frac{1}{4} \right) 
 \left[
 \left( \frac{1}{4} + \frac{N_f-1}{2} \right)
  \left[ E_3^0(0,1,2) + E_3^0(1,0,3) + E_3^0(2,3,0) + E_3^0(3,2,1) \right]
 \right. \right.
 \nonumber \\
 & &
 \left. \left.
 + \frac{1}{4} 
  \left[ E_3^0(0,1,3) + E_3^0(1,0,2) + E_3^0(3,2,0) + E_3^0(2,3,1) \right]
 \right]
 \circ {\cal A}_3^0(s_{0'2'})
 \right\} \circ \left| {\cal A}_3^{(0)} \right|^2 d\phi_4.
 \nonumber 
\eq
Integration over the one-parton phase space yields
\bq
\label{integrated_product}
\lefteqn{
 \int\limits_1 d\alpha_3^{product} 
 = 
 \left\{
  \frac{N_c^2}{4} 
  \left[ 
        - \frac{1}{2} {\cal D}_3^0(s_{01})^2
        - \frac{1}{2} {\cal D}_3^0(s_{12})^2
        + {\cal D}_3^0(s_{01}) {\cal D}_3^0(s_{12})
        + \frac{1}{2} {\cal A}_3^0(s_{02})^2
  \right]
\right.
} & & \nonumber \\
 & & \left.
  \frac{N_c^2 N_f}{4} 
  \left[
        - {\cal D}_3^0(s_{01}) {\cal E}_3^0(s_{01})
        - {\cal D}_3^0(s_{12}) {\cal E}_3^0(s_{12})
        + {\cal D}_3^0(s_{01}) {\cal E}_3^0(s_{12})
        + {\cal D}_3^0(s_{12}) {\cal E}_3^0(s_{01})
  \right]
 \right. \nonumber \\
 & & \left.
 + \frac{1}{4}
  \left[
        - {\cal D}_3^0(s_{01}) {\cal A}_3^0(s_{02})
        - {\cal D}_3^0(s_{12}) {\cal A}_3^0(s_{02})
        + {\cal A}_3^0(s_{02})^2
  \right]
 \right. \nonumber \\
 & & \left.
 + \frac{N_f}{4}
  \left[
        - {\cal E}_3^0(s_{01}) {\cal A}_3^0(s_{02})
        - {\cal E}_3^0(s_{12}) {\cal A}_3^0(s_{02})
  \right]
 \right\} \left| {\cal A}_3^{(0)} \right|^2 d\phi_3.
\eq
Remark: The term
\bq
\label{E30_squared}
  \frac{N_c^2 N_f^2}{4} 
  \left[
        2 {\cal E}_3^0(s_{01}) {\cal E}_3^0(s_{12})
        - {\cal E}_3^0(s_{01})^2
        - {\cal E}_3^0(s_{12})^2
  \right]
  \left| {\cal A}_3^{(0)} \right|^2 d\phi_3
\eq
is finite. Therefore eq.~(\ref{integrated_product}) can be re-written in a form, such that
the antenna function ${\cal E}_3^0$ occurs only in the combination ${\cal D}_3^0+N_f {\cal E}_3^0$
modulo finite terms of the form (\ref{E30_squared}).

\subsection{The colour-connected double unresolved subtraction terms}
\label{sect:subtraction_terms_double}

We consider now the subtraction terms for the double real emission contribution.
We start with the subtraction terms involving the four-parton antenna functions.
For the subtraction term $d\alpha_3^{double}$ we have two contributions corresponding to the sub-processes
$e^+ e^- \rightarrow \bar{q} g g g q$ and $e^+ e^- \rightarrow \bar{q} q \bar{q}' q' g$.
We have

\bq
 d\alpha_{3,\bar{q}gggq}^{double} & = &
 \frac{1}{6} \left\{ 
  \sum\limits_{i,j,k\in\{1,2,3\},i\neq j\neq k} \frac{N_c^2}{4} \left[ D_4^0(0,i,j,k) + D_4^0(4,i,j,k) \right]
 \right.
 \nonumber \\
 & & \left.
 + \sum\limits_{i,j\in\{1,2,3\},i\neq j}
 \frac{N_c^2}{4} \left[ - A_{4,sc}^0(0,i,j,4) \right]
 + \frac{1}{4} \left[ - A_4^0(0,i,j,4) - A_{4,sc}^0(0,i,j,4) \right]
 \right. \nonumber \\
 & & \left.
 + \frac{1}{4N_c^2} \left[ A_{4,sc}^0(0,i,j,4) \right]
 \right\} \circ \left| {\cal A}_3^{(0)} \right|^2 d\phi_5,
\eq
\bq
\lefteqn{
 d\alpha_{3,\bar{q}\bar{q}'g q'q}^{double}  =
 \left\{
 \left( \frac{1}{4} + \frac{N_f-1}{2} \right)
\right. } & & \\
 & & \left.
 \left[
 \frac{N_c}{4} 
    \left( 
           E_{4,qqqg}^0(0,1,3,2) + E_{4,qqqg}^0(1,0,4,2) 
         + E_{4,qqqg}^0(4,3,1,2) + E_{4,qqqg}^0(3,4,0,2) 
    \right)
\right. \right. \nonumber \\
 & & \left. \left.
+
 \frac{N_c}{4} 
    \left( 
           E_{4,qgqq}^0(0,2,3,1) + E_{4,qgqq}^0(1,2,4,0) 
         + E_{4,qgqq}^0(4,2,1,3) + E_{4,qgqq}^0(3,2,0,4) 
    \right)
\right. \right. \nonumber \\
 & & \left. \left.
 -\frac{1}{4N_c} \frac{1}{2}
    \left( 
           E_{4,sc}^0(0,1,2,3) + E_{4,sc}^0(1,0,2,4) 
         + E_{4,sc}^0(4,3,2,1) + E_{4,sc}^0(3,4,2,0) 
    \right)
\right. \right. \nonumber \\
 & & \left. \left.
 -\frac{1}{4N_c} \left( B_4^0(0,1,3,4) + B_4^0(1,0,4,3) \right)
 \right]
 \right. 
 \nonumber \\
 & &
 \left.
 + \frac{1}{4} 
 \left[
 \frac{N_c}{4} 
    \left( 
           E_{4,qqqg}^0(0,1,4,2) + E_{4,qqqg}^0(1,0,3,2) 
         + E_{4,qqqg}^0(4,3,0,2) + E_{4,qqqg}^0(3,4,1,2) 
    \right) 
\right. \right. \nonumber \\
 & & \left. \left.
 +
 \frac{N_c}{4} 
    \left( 
           E_{4,qgqq}^0(0,2,4,1) + E_{4,qgqq}^0(1,2,3,0) 
         + E_{4,qgqq}^0(4,2,0,3) + E_{4,qgqq}^0(3,2,1,4) 
    \right) 
\right. \right. \nonumber \\
 & & \left. \left.
 +
 -\frac{1}{4N_c} \frac{1}{2}
    \left( 
           E_{4,sc}^0(0,1,2,4) + E_{4,sc}^0(1,0,2,3) 
         + E_{4,sc}^0(4,3,2,0) + E_{4,sc}^0(3,4,2,1) 
    \right) 
\right. \right. \nonumber \\
 & & \left. \left.
 -\frac{1}{4N_c} \left( B_4^0(0,1,4,3) + B_4^0(1,0,3,4) \right) 
 \right]
\right. \nonumber \\
 & & \left.
 - \frac{1}{4} \left( 1-\frac{1}{N^2} \right) 
   \left[ C_4^0(0,1,3,4) + C_4^0(0,1,4,3) + C_4^0(1,0,4,3) + C_4^0(1,0,3,4) \right]
 \right\} \circ \left| {\cal A}_3^{(0)} \right|^2 d\phi_5.
 \nonumber
\eq
Integration over a two-parton phase space yields
\bq
\lefteqn{
 \int\limits_2 d\alpha_3^{double} = 
 \int\limits_2 \left( d\alpha_{3,\bar{q}gggq}^{double} + d\alpha_{3,\bar{q}\bar{q}'q'q}^{double} \right)
 = } & & \nonumber \\
 & &
 \left\{
 \frac{N_c^2}{4}  
 \left[ {\cal D}_4^0(s_{01})   
      + {\cal D}_4^0(s_{12}) 
      - {\cal A}_{4,sc}^0(s_{02}) \right]
 +
 \frac{N_c N_f}{4}  
 \left[ {\cal E}_4^0(s_{01}) + {\cal E}_4^0(s_{12}) \right]
 \right.
 \nonumber \\
 & &
 +
 \frac{1}{4} 
 \left[ - {\cal A}_4^0(s_{02}) - {\cal A}_{4,sc}^0(s_{02}) - 4 {\cal C}_4^0(s_{02}) \right]
 +
 \frac{N_f}{4 N_c}  
 \left[ - {\cal B}_4^0(s_{02}) - \frac{1}{2} {\cal E}_{4,sc}^0(s_{01}) 
                               - \frac{1}{2} {\cal E}_{4,sc}^0(s_{12}) \right]
 \nonumber \\
 & &
 \left.
 +
 \frac{1}{4 N_c^2} 
 \left[ {\cal A}_{4,sc}^0(s_{02}) + 4 {\cal C}_4^0(s_{02}) \right]
 \right\} \left| {\cal A}_3^{(0)} \right|^2 d\phi_3.
\eq

\subsection{The almost colour-connected double unresolved subtraction terms}
\label{sect:subtraction_terms_almost}

For the subtraction term $d\alpha_3^{almost}$ we have two contributions corresponding to the sub-processes
$e^+ e^- \rightarrow \bar{q} g g g q$ and $e^+ e^- \rightarrow \bar{q} q \bar{q}' q' g$.
For the sub-process $e^+ e^- \rightarrow \bar{q} g g g q$ we have
\bq
\lefteqn{
 d\alpha_{3,\bar{q}gggq}^{almost} = 
 \frac{1}{6}
\left\{
 \frac{N_c^2}{4} \frac{1}{2}
 \right.
} & & \nonumber \\
 & & \left.
 \left[
             \left( 
                    - D_3^0(0,1,3) - D_3^0(0,2,3) - D_3^0(0,3,2) 
                    + D_3^0(4,3,2) + D_3^0(4,1,3) + D_3^0(4,2,3) 
             \right)
             \right. \right. \nonumber \\
             & & \left. \left.
         \circ \left( 
                      D_3^0(0',1',2') - D_3^0(3',1',2')
               \right)
             \right. \right. \nonumber \\
             & & \left. \left.
     + 
             \left( 
                    - D_3^0(4,2,1) - D_3^0(4,3,1) - D_3^0(4,1,2) 
                    + D_3^0(0,1,2) + D_3^0(0,2,1) + D_3^0(0,3,1) 
             \right)
             \right. \right. \nonumber \\
             & & \left. \left.
         \circ \left(
                      D_3^0(3',2',1')- D_3^0(0',2',1')
               \right)
             \right. \right. \nonumber \\
             & & \left. \left.
     -        
             \left( 
                      A_3^0(0,1,4) + A_3^0(0,2,4) + A_3^0(0,3,4) 
             \right)
             \right. \right. \nonumber \\
             & & \left. \left.
         \circ \left(
                      D_3^0(0',1',2') + D_3^0(3',2',1')
                    + D_3^0(0',2',1') + D_3^0(3',1',2')
               \right)
             \right. \right. \nonumber \\
             & & \left. \left.
     +   
             \left( 
                      D_3^0(0,1,3) + D_3^0(0,2,3) + D_3^0(0,3,2) 
                    + D_3^0(4,3,2) + D_3^0(4,1,3) + D_3^0(4,2,3) 
             \right)
             \right. \right. \nonumber \\
             & & \left. \left.
         \circ A_3^0(0',1',3')
             \right. \right. \nonumber \\
             & & \left. \left.
     + 
             \left( 
                      D_3^0(0,1,2) + D_3^0(0,2,1) + D_3^0(0,3,1) 
                    + D_3^0(4,2,1) + D_3^0(4,3,1) + D_3^0(4,1,2) 
             \right)
             \right. \right. \nonumber \\
             & & \left. \left.
         \circ A_3^0(0',2',3')
             \right. \right. \nonumber \\
             & & \left. \left.
     +   
             \left( 
                    A_3^0(0,1,4) + A_3^0(0,2,4) + A_3^0(0,3,4) 
             \right)
         \circ \left(
                      A_3^0(0',1',3') + A_3^0(0',2',3')
               \right)
 \right]
%
             \right. \nonumber \\
             & & \left.
 + \frac{1}{4}
 \left[
                \left( 
                      A_3^0(0,1,4) + A_3^0(0,2,4) + A_3^0(0,3,4) 
                \right)
         \circ \left(
                      A_3^0(0',1',3') + A_3^0(0',2',3')
             \right. \right. \right. \nonumber \\
             & & \left. \left. \left.
                    - D_3^0(0',1',2') - D_3^0(3',2',1') 
                    - D_3^0(0',2',1') - D_3^0(3',1',2')
               \right)
 \right]
 \right\} \circ \left| {\cal A}_3^{(0)} \right|^2 d\phi_5.
\eq
For the sub-process $e^+ e^- \rightarrow \bar{q} q \bar{q}' q' g$ we have two subtractions terms,
corresponding to an intermediate $\bar{q} g g q$ or $\bar{q} q \bar{q}' q'$ state:
\bq
 d\alpha_{3,\bar{q}\bar{q}'gq'q}^{almost} & = & 
  d\alpha_{3,\bar{q}\bar{q}'gq'q \rightarrow \bar{q}ggq \rightarrow \bar{q}gq}^{almost} 
 + d\alpha_{3,\bar{q}\bar{q}'gq'q \rightarrow \bar{q}\bar{q}'q'q \rightarrow \bar{q}gq}^{almost}.
\eq
We have
\bq
\lefteqn{
 d\alpha_{3,\bar{q}\bar{q}'gq'q \rightarrow \bar{q}ggq \rightarrow \bar{q}gq}^{almost} = 
 \frac{1}{2}
 \frac{N_c^2}{4} 
 \frac{1}{2}
 } & & \\
 & &
 \left\{
             \left[ 
                    \left( \frac{1}{4} + \frac{N_f-1}{2} \right) \left( - E_3^0(0,1,3) - E_3^0(1,0,4) 
                                                                        + E_3^0(4,1,3) + E_3^0(3,0,4) \right)
 \right. \right. \nonumber \\
 & & \left. \left.
                    +      \frac{1}{4}                           \left( - E_3^0(0,1,4) - E_3^0(1,0,3) 
                                                                        + E_3^0(3,1,4) + E_3^0(4,0,3) \right) 
             \right]
 \right. \nonumber \\
 & & \left.
             \circ \left( D_3^0(0',1',2') - D_3^0(3',1',2') \right) 
 \right. \nonumber \\
 & & \left.
     + 
             \left[ 
                    \left( \frac{1}{4} + \frac{N_f-1}{2} \right) \left( - E_3^0(4,3,1) - E_3^0(3,4,0) 
                                                                        + E_3^0(0,3,1) + E_3^0(1,4,0) \right) 
 \right. \right. \nonumber \\
 & & \left. \left.
                    +      \frac{1}{4}                           \left( - E_3^0(3,4,1) - E_3^0(4,3,0) 
                                                                        + E_3^0(0,4,1) + E_3^0(1,3,0) \right) 
             \right]
 \right. \nonumber \\
 & & \left.
             \circ \left( D_3^0(3',2',1') - D_3^0(0',2',1') \right) 
 \right. \nonumber \\
 & & \left.
     + 
             \left[ 
                    \left( \frac{1}{4} + \frac{N_f-1}{2} \right) \left( E_3^0(0,1,3) + E_3^0(1,0,4) 
                                                                      + E_3^0(4,1,3) + E_3^0(3,0,4) \right) 
 \right. \right. \nonumber \\
 & & \left. \left.
                    +      \frac{1}{4}                           \left( E_3^0(0,1,4) + E_3^0(1,0,3) 
                                                                      + E_3^0(3,1,4) + E_3^0(4,0,3) \right) 
             \right] 
             \circ A_3^0(0',1',3') 
 \right. \nonumber \\
 & & \left.
     + 
             \left[ 
                    \left( \frac{1}{4} + \frac{N_f-1}{2} \right) \left( E_3^0(0,3,1) + E_3^0(1,4,0) 
                                                                      + E_3^0(4,3,1) + E_3^0(3,4,0) \right) 
 \right. \right. \nonumber \\
 & & \left. \left.
                    +      \frac{1}{4}                           \left( E_3^0(0,4,1) + E_3^0(1,3,0) 
                                                                      + E_3^0(3,4,1) + E_3^0(4,3,0) \right) 
             \right] 
             \circ A_3^0(0',2',3') 
 \right\} \circ \left| {\cal A}_3^{(0)} \right|^2 d\phi_5.
 \nonumber  \\
\lefteqn{
 d\alpha_{3,\bar{q}\bar{q}'gq'q \rightarrow \bar{q}\bar{q}'q'q \rightarrow \bar{q}gq}^{almost} 
 = 
 \left\{
 \frac{N_c^2}{4}
 \right. 
} & & \nonumber \\
 & &
 \left.
 \left[
 \left( \frac{1}{4} + \frac{N_f-1}{2} \right)
  \left(
         \left( A_3^0(1,2,4) - \frac{1}{2} A_3^0(0,2,3) - \frac{1}{2} A_3^0(3,2,4) - \frac{1}{2} A_3^0(0,2,4) \right) \circ E_3^0(0',1',2') 
         \right. \right. \right. \nonumber \\ & & \left. \left. \left.
       + \left( A_3^0(1,2,4) - \frac{1}{2} A_3^0(0,2,3) - \frac{1}{2} A_3^0(0,2,1) - \frac{1}{2} A_3^0(1,2,3) \right) \circ E_3^0(2',3',0') 
         \right. \right. \right. \nonumber \\ & & \left. \left. \left.
       + \left( A_3^0(0,2,3) - \frac{1}{2} A_3^0(1,2,4) - \frac{1}{2} A_3^0(3,2,4) - \frac{1}{2} A_3^0(1,2,3) \right) \circ E_3^0(1',0',3') 
         \right. \right. \right. \nonumber \\ & & \left. \left. \left.
       + \left( A_3^0(0,2,3) - \frac{1}{2} A_3^0(1,2,4) - \frac{1}{2} A_3^0(0,2,1) - \frac{1}{2} A_3^0(0,2,4) \right) \circ E_3^0(3',2',1') 
  \right)
 \right. \right. \nonumber \\ & & \left. \left.
 + \frac{1}{4}
  \left( 
         \left( A_3^0(1,2,3) - \frac{1}{2} A_3^0(0,2,4) - \frac{1}{2} A_3^0(3,2,4) - \frac{1}{2} A_3^0(0,2,3) \right) \circ E_3^0(0',1',3') 
         \right. \right. \right. \nonumber \\ & & \left. \left. \left.
       + \left( A_3^0(1,2,3) - \frac{1}{2} A_3^0(0,2,4) - \frac{1}{2} A_3^0(0,2,1) - \frac{1}{2} A_3^0(1,2,4) \right) \circ E_3^0(3',2',0') 
         \right. \right. \right. \nonumber \\ & & \left. \left. \left.
       + \left( A_3^0(0,2,4) - \frac{1}{2} A_3^0(1,2,3) - \frac{1}{2} A_3^0(3,2,4) - \frac{1}{2} A_3^0(1,2,4) \right) \circ E_3^0(1',0',2') 
         \right. \right. \right. \nonumber \\ & & \left. \left. \left.
       + \left( A_3^0(0,2,4) - \frac{1}{2} A_3^0(1,2,3) - \frac{1}{2} A_3^0(0,2,1) - \frac{1}{2} A_3^0(0,2,3) \right) \circ E_3^0(2',3',1') 
  \right)
 \right]
 \right. \nonumber \\
 & & \left. 
 + \frac{1}{4} 
 \left[
 \left( \frac{1}{4} + \frac{N_f-1}{2} \right)
 \left( 
     2 \left( A_3^0(0,2,1) - A_3^0(0,2,3) + A_3^0(3,2,4) - A_3^0(1,2,4) \right)
 \right. \right. \right. \nonumber \\
 & & \left. \left. \left. 
   \circ
   \left( E_3^0(0',1',2') + E_3^0(1',0',3') 
        + E_3^0(2',3',0') + E_3^0(3',2',1') \right)
 \right. \right. \right. \nonumber \\
 & & \left. \left. \left.
      - A_3^0(0,2,4) \circ \left( E_3^0(0',1',2') + E_3^0(3',2',1') \right) 
 \right. \right. \right. \nonumber \\
 & & \left. \left. \left.
      - A_3^0(1,2,3) \circ \left( E_3^0(1',0',3') + E_3^0(2',3',0') \right) 
 \right) 
 \right. \right. \nonumber \\
 & & \left. \left.
 + 
    \frac{1}{4}
   \left( 
      2 \left( A_3^0(0,2,1) - A_3^0(0,2,4) + A_3^0(3,2,4) - A_3^0(1,2,3) \right)
 \right. \right. \right. \nonumber \\
 & & \left. \left. \left. 
   \circ
   \left( E_3^0(0',1',3') + E_3^0(1',0',2') 
        + E_3^0(3',2',0') + E_3^0(2',3',1') \right)
 \right. \right. \right. \nonumber \\
 & & \left. \left. \left.
      - A_3^0(0,2,3) \circ \left( E_3^0(0',1',3') + E_3^0(2',3',1') \right)  
 \right. \right. \right. \nonumber \\
 & & \left. \left. \left.
      - A_3^0(1,2,4) \circ \left( E_3^0(1',0',2') + E_3^0(3',2',0') \right) 
    \right)
 \right]
 \right\} \circ \left| {\cal A}_3^{(0)} \right|^2 d\phi_5.
 \nonumber 
\eq
These subtraction terms are integrated over an one-parton phase space. We obtain
\bq
\lefteqn{
 \int\limits_1 \left( d\alpha_{3,\bar{q}gggq}^{almost} + d\alpha_{3,\bar{q}\bar{q}'gq'q \rightarrow \bar{q}ggq \rightarrow \bar{q}gq}^{almost} \right)
 = 
 \frac{1}{2}
 } & & \\
 & &
 \left\{
 \frac{N_c^2}{4} \frac{1}{2}
 \left[
               \left( 
                      D_3^0(0,1,2) - D_3^0(3,1,2)
               \right)
             \left( 
                      {\cal D}_3^0(s_{23}) + N_f {\cal E}_3^0(s_{23})
                    - {\cal D}_3^0(s_{02}) - N_f {\cal E}_3^0(s_{02})  
             \right)
     \right. \right. \nonumber \\ & & \left. \left.
     + 
               \left(
                      D_3^0(3,2,1)- D_3^0(0,2,1)
               \right)
             \left( 
                      {\cal D}_3^0(s_{01}) + N_f {\cal E}_3^0(s_{01}) 
                    - {\cal D}_3^0(s_{13}) - N_f {\cal E}_3^0(s_{13}) 
             \right)
     \right. \right. \nonumber \\ & & \left. \left.
     -        
               \left(
                      D_3^0(0,1,2) + D_3^0(3,2,1)
                    + D_3^0(0,2,1) + D_3^0(3,1,2)
               \right)
                      {\cal A}_3^0(s_{03})
     \right. \right. \nonumber \\ & & \left. \left.
     +   
               A_3^0(0,1,3)
             \left( 
                      {\cal D}_3^0(s_{02}) + N_f {\cal E}_3^0(s_{02})  
                    + {\cal D}_3^0(s_{23}) + N_f {\cal E}_3^0(s_{23})
             \right)
     \right. \right. \nonumber \\ & & \left. \left.
     + 
               A_3^0(0,2,3)
             \left( 
                      {\cal D}_3^0(s_{01}) + N_f {\cal E}_3^0(s_{01})  
                    + {\cal D}_3^0(s_{13}) + N_f {\cal E}_3^0(s_{13})
             \right)
     \right. \right. \nonumber \\ & & \left. \left.
     +   
               \left(
                      A_3^0(0,1,3) + A_3^0(0,2,3)
               \right)
                    {\cal A}_3^0(s_{03})
 \right]
%
             \right. \nonumber \\
             & & \left.
 + \frac{1}{4}
 \left[
               \left(
                      A_3^0(0,1,3) + A_3^0(0,2,3)
                    - D_3^0(0,1,2) - D_3^0(3,2,1) 
                    - D_3^0(0,2,1) - D_3^0(3,1,2)
               \right)
     \right. \right. \nonumber \\ & & \left. \left.
                      {\cal A}_3^0(s_{03})
 \right]
 \right\} \circ \left| {\cal A}_3^{(0)} \right|^2 d\phi_4,
\nonumber \\
\lefteqn{
 \int\limits_1 d\alpha_{3,\bar{q}\bar{q}'gq'q \rightarrow \bar{q}\bar{q}'q'q  \rightarrow \bar{q}gq}^{almost}
 = 
 \left\{
 \frac{N_c^2}{4}
 \left[
 \left( \frac{1}{4} + \frac{N_f-1}{2} \right)
  \left(
         E_3^0(0,1,2) \left( {\cal A}_3^0(s_{13}) - \frac{1}{2} {\cal A}_3^0(s_{02}) - \frac{1}{2} {\cal A}_3^0(s_{23}) 
         \right. \right. \right. \right.
 } & & \nonumber \\
 & &
\left. \left. \left. \left.
- \frac{1}{2} {\cal A}_3^0(s_{03}) \right)   
       + E_3^0(2,3,0) \left( {\cal A}_3^0(s_{13}) - \frac{1}{2} {\cal A}_3^0(s_{02}) - \frac{1}{2} {\cal A}_3^0(s_{01}) - \frac{1}{2} {\cal A}_3^0(s_{12}) \right)   
         \right. \right. \right. \nonumber \\ & & \left. \left. \left.
       + E_3^0(1,0,3) \left( {\cal A}_3^0(s_{02}) - \frac{1}{2} {\cal A}_3^0(s_{13}) - \frac{1}{2} {\cal A}_3^0(s_{23}) - \frac{1}{2} {\cal A}_3^0(s_{12}) \right)  
         \right. \right. \right. \nonumber \\ & & \left. \left. \left.
       + E_3^0(3,2,1) \left( {\cal A}_3^0(s_{02}) - \frac{1}{2} {\cal A}_3^0(s_{13}) - \frac{1}{2} {\cal A}_3^0(s_{01}) - \frac{1}{2} {\cal A}_3^0(s_{03}) \right)  
  \right)
 \right. \right. \nonumber \\ & & \left. \left.
 + \frac{1}{4}
  \left( 
         E_3^0(0,1,3) \left( {\cal A}_3^0(s_{12}) - \frac{1}{2} {\cal A}_3^0(s_{03}) - \frac{1}{2} {\cal A}_3^0(s_{23}) - \frac{1}{2} {\cal A}_3^0(s_{02}) \right)  
         \right. \right. \right. \nonumber \\ & & \left. \left. \left.
       + E_3^0(3,2,0) \left( {\cal A}_3^0(s_{12}) - \frac{1}{2} {\cal A}_3^0(s_{03}) - \frac{1}{2} {\cal A}_3^0(s_{01}) - \frac{1}{2} {\cal A}_3^0(s_{13}) \right)  
         \right. \right. \right. \nonumber \\ & & \left. \left. \left.
       + E_3^0(1,0,2) \left( {\cal A}_3^0(s_{03}) - \frac{1}{2} {\cal A}_3^0(s_{12}) - \frac{1}{2} {\cal A}_3^0(s_{23}) - \frac{1}{2} {\cal A}_3^0(s_{13}) \right)  
         \right. \right. \right. \nonumber \\ & & \left. \left. \left.
       + E_3^0(2,3,1) \left( {\cal A}_3^0(s_{03}) - \frac{1}{2} {\cal A}_3^0(s_{12}) - \frac{1}{2} {\cal A}_3^0(s_{01}) - \frac{1}{2} {\cal A}_3^0(s_{02}) \right)  
  \right)
 \right]
 \right. \nonumber \\
 & & \left. 
 + \frac{1}{4} 
 \left[
 \left( \frac{1}{4} + \frac{N_f-1}{2} \right)
 \left( 
    2
   \left( E_3^0(0,1,2) + E_3^0(1,0,3) 
        + E_3^0(2,3,0) + E_3^0(3,2,1) \right)
 \right. \right. \right. \nonumber \\
 & & \left. \left. \left. 
\left( {\cal A}_3^0(s_{01}) - {\cal A}_3^0(s_{02}) + {\cal A}_3^0(s_{23}) - {\cal A}_3^0(s_{13}) \right)
 \right. \right. \right. \nonumber \\
 & & \left. \left. \left.
      - \left( E_3^0(0,1,2) + E_3^0(3,2,1) \right) {\cal A}_3^0(s_{03})  
      - \left( E_3^0(1,0,3) + E_3^0(2,3,0) \right) {\cal A}_3^0(s_{12})  
 \right) 
 \right. \right. \nonumber \\
 & & \left. \left.
 + 
    \frac{1}{4}
   \left( 
      2 
   \left( E_3^0(0,1,3) + E_3^0(1,0,2) 
        + E_3^0(3,2,0) + E_3^0(2,3,1) \right)
 \right. \right. \right. \nonumber \\
 & & \left. \left. \left. 
 \left( {\cal A}_3^0(s_{01}) - {\cal A}_3^0(s_{03}) + {\cal A}_3^0(s_{23}) - {\cal A}_3^0(s_{12}) \right)
 \right. \right. \right. \nonumber \\
 & & \left. \left. \left.
      - \left( E_3^0(0,1,3) + E_3^0(2,3,1) \right) {\cal A}_3^0(s_{02})    
      - \left( E_3^0(1,0,2) + E_3^0(3,2,0) \right) {\cal A}_3^0(s_{13})  
    \right)
 \right]
 \right\} \circ \left| {\cal A}_3^{(0)} \right|^2 d\phi_4.
 \nonumber 
\eq

\subsection{The iterated double unresolved subtraction terms}
\label{sect:subtraction_terms_iter}

For the subtraction term $d\alpha_3^{iterated}$ we have two contributions corresponding to the sub-processes
$e^+ e^- \rightarrow \bar{q} g g g q$ and $e^+ e^- \rightarrow \bar{q} q \bar{q}' q' g$.
For the sub-process $e^+ e^- \rightarrow \bar{q} g g g q$ we have
\bq
\lefteqn{
 d\alpha_{3,\bar{q}gggq}^{iterated} = 
 \frac{1}{6}
} & & \\
 & &
\left\{
 \frac{N_c^2}{4}
 \left[
   \left( D_3^0(0,1,2) + D_3^0(0,2,1) + D_3^0(0,3,1) 
        + D_3^0(4,3,2) + D_3^0(4,1,3) + D_3^0(4,2,3) 
          \right. \right. \right. \nonumber \\
          & & \left. \left. \left.
        + F_3^0(1,2,3) + F_3^0(1,3,2) + F_3^0(2,1,3) 
   \right) 
   \circ
   \left( D_3^0(0',1',2') + D_3^0(3',2',1') \right)
          \right. \right. \nonumber \\
          & & \left. \left.
  +
   \left( D_3^0(0,2,3) + D_3^0(0,3,2) + D_3^0(0,1,3) 
        + D_3^0(4,2,1) + D_3^0(4,3,1) + D_3^0(4,1,2) 
          \right. \right. \right. \nonumber \\
          & & \left. \left. \left.
        + F_3^0(1,2,3) + F_3^0(1,3,2) + F_3^0(2,1,3) 
   \right)
   \circ
   \left( D_3^0(0',2',1') + D_3^0(3',1',2') \right)
 \right]
          \right. \nonumber \\
          & & \left.
 + \frac{1}{4}
 \left[
   \left( 1 + \frac{1}{N_c^2} \right)
   \left( A_3^0(0,1,4) + A_3^0(0,2,4) + A_3^0(0,3,4) \right)
   \circ
   \left( A_3^0(0',1',3') + A_3^0(0',2',3') \right)
          \right. \right. \nonumber \\
          & & \left. \left.
        - \left( A_3^0(0,1,4) + A_3^0(0,2,4) + A_3^0(0,3,4) \right)
          \right. \right. \nonumber \\
          & & \left. \left.
          \circ
          \left( D_3^0(0',1',2') + D_3^0(3',2',1') 
               + D_3^0(0',2',1') + D_3^0(3',1',2') \right)
          \right. \right. \nonumber \\
          & & \left. \left.
        - 
               \left( D_3^0(0,1,2) + D_3^0(0,2,1) + D_3^0(0,3,1) 
                    + D_3^0(0,2,3) + D_3^0(0,3,2) + D_3^0(0,1,3) 
          \right. \right. \right. \nonumber \\
          & & \left. \left. \left.
                    + D_3^0(4,2,1) + D_3^0(4,3,1) + D_3^0(4,1,2) 
                    + D_3^0(4,3,2) + D_3^0(4,1,3) + D_3^0(4,2,3) 
              \right) 
          \right. \right. \nonumber \\
          & & \left. \left.
              \circ
              \left( A_3^0(0',1',3') + A_3^0(0',2',3') \right)
 \right]
 \right\} \circ \left| {\cal A}_3^{(0)} \right|^2 d\phi_5.
 \nonumber
\eq
For the sub-process $e^+ e^- \rightarrow \bar{q} q \bar{q}' q' g$ we have two subtractions terms,
corresponding to an intermediate $\bar{q} g g q$ or $\bar{q} q \bar{q}' q'$ state:
\bq
 d\alpha_{3,\bar{q}\bar{q}'gq'q}^{iterated} & = & 
  d\alpha_{3,\bar{q}\bar{q}'gq'q \rightarrow \bar{q}ggq \rightarrow \bar{q}gq}^{iterated} 
 + d\alpha_{3,\bar{q}\bar{q}'gq'q \rightarrow \bar{q}\bar{q}'q'q \rightarrow \bar{q}gq}^{iterated}.
\eq
We have
\bq
\lefteqn{
 d\alpha_{3,\bar{q}\bar{q}'gq'q \rightarrow \bar{q}ggq \rightarrow \bar{q}gq}^{iterated} = 
 \left\{
    \left[ 
        \left( \frac{1}{4} + \frac{N_f-1}{2} \right)
        \left( E_3^0(0,1,3) + E_3^0(1,0,4) + E_3^0(3,4,0) + E_3^0(4,3,1) \right)
       \right. \right. 
 } & & \nonumber \\
 & & \left. \left.
       +
        \frac{1}{4}
        \left( E_3^0(0,1,4) + E_3^0(1,0,3) + E_3^0(4,3,0) + E_3^0(3,4,1) \right)       
    \right]
 \right. \nonumber \\ & & \left.
  \circ
  \left[
   \frac{N_c^2}{4} \left( D_3^0(0',2',1') + D_3^0(3',1',2') \right)
   - \frac{1}{4}   \left( A_3^0(0',1',3') + A_3^0(0',2',3') \right)
   \right]
 \right. \nonumber \\ & & \left.
+
 \frac{N_c^2}{4}
   \left[ 
        \left( \frac{1}{4} + \frac{N_f-1}{2} \right)
        \left( G_3^0(2,1,3) + G_3^0(2,0,4) \right)
       +
        \frac{1}{4}
        \left( G_3^0(2,0,3) + G_3^0(2,1,4) \right)       
   \right]
 \right.
 \hspace*{15mm} 
 \nonumber \\ & & \left.
  \circ
  \left( D_3^0(0',1',2') + D_3^0(3',2',1') 
       + D_3^0(0',2',1') + D_3^0(3',1',2') \right)
 \right\} \circ \left| {\cal A}_3^{(0)} \right|^2 d\phi_5.
\eq
\bq
\lefteqn{
 d\alpha_{3,\bar{q}\bar{q}'gq'q \rightarrow \bar{q}\bar{q}'q'q \rightarrow \bar{q}gq}^{iterated} 
 = 
 \left\{
 \left( \frac{1}{4} + \frac{N_f-1}{2} \right)
 \left[ \frac{N_c^2}{4}  \left( A_3^0(0,2,3) + A_3^0(1,2,4) \right)
 \right. \right.
} & & \nonumber \\
 & &
 \left. \left.
       + \frac{1}{4} \left(
                           -   A_3^0(0,2,4) -   A_3^0(1,2,3) 
                           - 2 A_3^0(0,2,3) - 2 A_3^0(1,2,4) 
                           + 2 A_3^0(0,2,1) + 2 A_3^0(3,2,4)
                     \right)
 \right]
 \right. \nonumber \\
 & & \left.
 \circ
 \left[ E_3^0(0',1',2') + E_3^0(1',0',3') 
      + E_3^0(2',3',0') + E_3^0(3',2',1') \right] 
 \right. \nonumber \\
 & & \left.
 + \frac{1}{4}
 \left[ \frac{N_c^2}{4}  \left( A_3^0(0,2,4) + A_3^0(1,2,3) \right)
 \right. \right. \nonumber \\
 & & \left. \left.
      + \frac{1}{4} \left(
                           -   A_3^0(0,2,3) -   A_3^0(1,2,4) 
                           - 2 A_3^0(0,2,4) - 2 A_3^0(1,2,3) 
                           + 2 A_3^0(0,2,1) + 2 A_3^0(3,2,4)
                    \right)
 \right]
 \right. \nonumber \\
 & & \left.
 \circ
   \left[ E_3^0(0',1',3') + E_3^0(1',0',2') 
        + E_3^0(3',2',0') + E_3^0(2',3',1') \right]
 \right\} \circ \left| {\cal A}_3^{(0)} \right|^2 d\phi_5.
\eq
These subtraction terms are integrated over an one-parton phase space. We obtain
\bq
\lefteqn{
 \int\limits_1 \left( d\alpha_{3,\bar{q}gggq}^{iterated} + d\alpha_{3,\bar{q}\bar{q}'gq'q \rightarrow \bar{q}ggq \rightarrow \bar{q}gq}^{iterated} \right)
 = 
 \frac{1}{2}
 \left\{
 \frac{N_c^2}{4}
 \left[
   \left( D_3^0(0,1,2) + D_3^0(3,2,1) \right)
 \right. \right.
 } & & \\
 & &
 \left. \left.
   \left( {\cal D}_3^0(s_{01}) + N_f {\cal E}_3^0(s_{01})
        + {\cal D}_3^0(s_{23}) + N_f {\cal E}_3^0(s_{23})
        + {\cal F}_3^0(s_{12}) + 2 N_f {\cal G}_3^0(s_{12})
   \right) 
 \right. \right. \nonumber \\ & & \left. \left.
  +
   \left( D_3^0(0,2,1) + D_3^0(3,1,2) \right)
   \right. \right. \nonumber \\ & & \left. \left.
   \left( {\cal D}_3^0(s_{02}) + N_f {\cal E}_3^0(s_{02}) 
        + {\cal D}_3^0(s_{13}) + N_f {\cal E}_3^0(s_{13})
        + {\cal F}_3^0(s_{12}) + 2 N_f {\cal G}_3^0(s_{12})
   \right)
 \right]
 \right. \nonumber \\ & & \left.
 + \frac{1}{4}
 \left[
        - 
          \left( D_3^0(0,1,2) + D_3^0(3,2,1) 
               + D_3^0(0,2,1) + D_3^0(3,1,2) \right)
          {\cal A}_3^0(s_{03})
 \right. \right. \nonumber \\ & & \left. \left.
        - 
              \left( A_3^0(0,1,3) + A_3^0(0,2,3) \right)
               \left( {\cal D}_3^0(s_{01}) + N_f {\cal E}_3^0(s_{01})  
                    + {\cal D}_3^0(s_{02}) + N_f {\cal E}_3^0(s_{02})
   \right. \right. \right. \nonumber \\ & & \left. \left. \left.
                    + {\cal D}_3^0(s_{13}) + N_f {\cal E}_3^0(s_{13})
                    + {\cal D}_3^0(s_{23}) + N_f {\cal E}_3^0(s_{23})
              \right) 
 \right. \right. \nonumber \\ & & \left. \left.
 +
   \left( 1 + \frac{1}{N_c^2} \right)
   \left( A_3^0(0,1,3) + A_3^0(0,2,3) \right)
   {\cal A}_3^0(s_{03})
 \right]
 \right\} \circ \left| {\cal A}_3^{(0)} \right|^2 d\phi_4,
\nonumber \\
\lefteqn{
 \int\limits_1 d\alpha_{3,\bar{q}\bar{q}'gq'q \rightarrow \bar{q}\bar{q}'q'q  \rightarrow \bar{q}gq}^{iterated}
 = 
 \left\{
 \left( \frac{1}{4} + \frac{N_f-1}{2} \right)
 \left[ E_3^0(0,1,2) + E_3^0(1,0,3) 
      + E_3^0(2,3,0) + E_3^0(3,2,1) \right] 
 \right. 
 } & & \nonumber \\
 & &
 \left. 
 \left[ \frac{N_c^2}{4}  \left( {\cal A}_3^0(s_{02}) + {\cal A}_3^0(s_{13}) \right)
 \right. \right. \nonumber \\ & & \left. \left.
       + \frac{1}{4} \left(
                           -   {\cal A}_3^0(s_{03}) -   {\cal A}_3^0(s_{12}) 
                           - 2 {\cal A}_3^0(s_{02}) - 2 {\cal A}_3^0(s_{13}) 
                           + 2 {\cal A}_3^0(s_{01}) + 2 {\cal A}_3^0(s_{23})
                     \right)
 \right]
 \right. \nonumber \\ & & \left.
 + \frac{1}{4}
   \left[ E_3^0(0,1,3) + E_3^0(1,0,2) 
        + E_3^0(3,2,0) + E_3^0(2,3,1) \right]
 \right. \nonumber \\ & & \left.
 \left[ \frac{N_c^2}{4}  \left( {\cal A}_3^0(s_{03}) + {\cal A}_3^0(s_{12}) \right)
 \right. \right. \nonumber \\ & & \left. \left.
      + \frac{1}{4} \left(
                           -   {\cal A}_3^0(s_{02}) -   {\cal A}_3^0(s_{13}) 
                           - 2 {\cal A}_3^0(s_{03}) - 2 {\cal A}_3^0(s_{12}) 
                           + 2 {\cal A}_3^0(s_{01}) + 2 {\cal A}_3^0(s_{23})
                    \right)
 \right]
 \right\} \circ \left| {\cal A}_3^{(0)} \right|^2 d\phi_4.
 \nonumber 
\eq

\subsection{The soft subtraction terms}
\label{sect:subtraction_terms_soft}

In the four-parton channel the poles from $d\sigma_4^{(1)}$ and $d\alpha_4^{single}$ cancel each other.
The other terms which we defined up to now in the four-parton channel are
\bq
 d\alpha^{loop}_3 + d\alpha^{product}_3 - \int\limits_1 d\alpha^{almost}_3 + \int\limits_1 d\alpha^{iterated}_3.
\eq
We now study the infrared pole structure of the combination of these terms.
In the four-parton channel there are two sub-channels, the $\bar{q}ggq$ sub-channel and the
$\bar{q} \bar{q}' q' q$ sub-channel.
We first consider the $\bar{q}ggq$ sub-channel. For the infrared poles proportional to the 
unintegrated three-parton antenna function $D_3^0(0,1,2)$ we find
\bq
\label{poles_lc}
\lefteqn{
 \left[ d\alpha^{loop}_3 + d\alpha^{product}_3 
 - \int\limits_1 d\alpha^{almost}_3 + \int\limits_1 d\alpha^{iterated}_3
 \right]_{Poles, D_3^0(0,1,2)}
 = 
 \frac{1}{2} 
 \frac{N_c^2}{4}
 \frac{1}{2}
 D_3^0(0,1,2) 
}
\\
 & &
 \left.
 \left[ 
        {\cal D}_3^0(s_{\tilde{0}\tilde{2}}) - {\cal D}_3^0(s_{02})
      - {\cal A}_3^0(s_{\tilde{0}3}) + {\cal A}_3^0(s_{03})
      - {\cal D}_3^0(s_{\tilde{2}3}) + {\cal D}_3^0(s_{23})
 \right]
 \circ \left| {\cal A}_3^{(0)} \right|^2 d\phi_4
 \right|_{Poles}.
\nonumber
\eq
There are three similar terms with poles proportional to $D_3^0(0,2,1)$, $D_3^0(3,2,1)$ and $D_3^0(3,1,2)$.
For the infrared poles proportional to the unintegrated three-parton antenna function $A_3^0(0,1,3)$ we find
\bq
\label{poles_sc}
\lefteqn{
 \left[ d\alpha^{loop}_3 + d\alpha^{product}_3 
 - \int\limits_1 d\alpha^{almost}_3 + \int\limits_1 d\alpha^{iterated}_3
 \right]_{Poles, A_3^0(0,1,3)}
 = 
 \frac{1}{2} 
 \left( \frac{N_c^2}{4} \frac{1}{2} + \frac{1}{4} \right)
 A_3^0(0,1,3)
}
\nonumber \\
 & &
 \left.
 \left[ 
      - {\cal A}_3^0(s_{\tilde{0}\tilde{3}}) + {\cal A}_3^0(s_{03})
      + {\cal D}_3^0(s_{\tilde{0}2}) - {\cal D}_3^0(s_{02})
      + {\cal D}_3^0(s_{\tilde{2}3}) - {\cal D}_3^0(s_{23})
 \right\}
 \circ \left| {\cal A}_3^{(0)} \right|^2 d\phi_4
 \right|_{Poles}.
\nonumber \\
\eq
There is a similar term proportional to $A_3^0(0,2,3)$.
Let us now consider the sub-channel $\bar{q} \bar{q}' q' q$.
For the infrared poles proportional to the unintegrated three-parton antenna function $E_3^0(0,1,2)$ we find
\bq
\label{poles_nf}
\lefteqn{
 \left[ d\alpha^{loop}_3 + d\alpha^{product}_3 
 - \int\limits_1 d\alpha^{almost}_3 + \int\limits_1 d\alpha^{iterated}_3
 \right]_{Poles, E_3^0(0,1,2)}
 = 
 \left( \frac{1}{4} + \frac{N_f-1}{2} \right)
 \frac{N_c^2}{4}
 \frac{1}{2}
 E_3^0(0,1,2) 
}
\nonumber \\
 & &
 \left.
 \left[ 
        {\cal D}_3^0(s_{\tilde{0}\tilde{2}}) - 2 {\cal A}_3^0(s_{01})
      - {\cal A}_3^0(s_{\tilde{0}3}) + {\cal A}_3^0(s_{02}) + {\cal A}_3^0(s_{03})
      - {\cal D}_3^0(s_{\tilde{2}3}) + {\cal A}_3^0(s_{23})
 \right]
 \right.
 \nonumber \\
 & &
 \left.
 \circ \left| {\cal A}_3^{(0)} \right|^2 d\phi_4
 \right|_{Poles}.
\hspace*{120mm}
\eq
There are three similar terms with poles proportional to $E_3^0(1,0,3)$, $E_3^0(2,3,0)$ and $E_3^0(3,2,1)$.
In addition there are four terms with proportional to $E_3^0(0,1,3)$,
$E_3^0(1,0,2)$, $E_3^0(3,2,0)$ and $E_3^0(2,3,1)$ for which one replaces
$1/4+(N_f-1)/2$ by $1/4$.
We can re-write the terms in eq.~(\ref{poles_nf}) as
\bq
\label{poles_nf_rewritten}
\lefteqn{
        {\cal D}_3^0(s_{012}) - 2 {\cal A}_3^0(s_{01})
      - {\cal A}_3^0(s_{\tilde{0}3}) + {\cal A}_3^0(s_{02}) + {\cal A}_3^0(s_{03})
      - {\cal D}_3^0(s_{\tilde{2}3}) + {\cal A}_3^0(s_{23})
= } & & 
\nonumber \\
 & &
 \left[
        {\cal D}_3^0(s_{012}) - {\cal A}_3^0(s_{02}) 
      - {\cal A}_3^0(s_{\tilde{0}3}) + {\cal A}_3^0(s_{03})
      - {\cal D}_3^0(s_{\tilde{2}3}) + {\cal A}_3^0(s_{23})
 \right]
 \nonumber \\
 & &
 + 2 \left[
 {\cal A}_3^0(s_{02}) - {\cal A}_3^0(s_{01})
 \right].
\eq
Terms with
\bq
\label{Nf_asym_poles}
 \int\limits_1
 E_3^0(0,1,2) 
 \left[
 {\cal A}_3^0(s_{02}) - {\cal A}_3^0(s_{01})
 \right]
 \circ \left| {\cal A}_3^{(0)} \right|^2 d\phi_4
\eq
vanish after integration over the unresolved phase space for a momentum mapping symmetric
in $(p_1,p_2)$. On the other hand, the first term on the right hand side of eq.~(\ref{poles_nf_rewritten})
has the same structure as the terms in eq.~(\ref{poles_lc}) and eq.~(\ref{poles_sc}).
The integrated antenna functions ${\cal A}_3^0(s)$ and ${\cal D}_3^0(s)$
have the form
\bq
\label{basic_integrated_antenna_fct}
 \frac{\alpha_s}{\pi} \left( \frac{s}{\mu^2} \right)^{-\eps} 
 \left[ \frac{1}{\eps^2} + \frac{c}{\eps} \right] + {\cal O}(\eps).
\eq
Due to the structure of alternating signs in eq.(\ref{poles_lc}), (\ref{poles_sc}) 
and (\ref{poles_nf_rewritten}) the coefficient $c$ of the single pole in 
eq.~(\ref{basic_integrated_antenna_fct})
will drop out and only the double pole contributes.
The double pole is related to soft gluons.
Generically we have to consider terms of the form
\bq
\label{generic_eps_poles}
 X_3^0(a,i,b)
 \left[ 
  {\cal S}_3^0(s_{\tilde{a}\tilde{b}}) - {\cal S}_3^0(s_{ab})
  - {\cal S}_3^0(s_{\tilde{a}j}) + {\cal S}_3^0(s_{aj})
  - {\cal S}_3^0(s_{\tilde{b}j}) + {\cal S}_3^0(s_{bj})
 \right]
 \circ \left| {\cal A}_3^{(0)} \right|^2 d\phi_4
\eq
with
\bq
\label{integrated_soft_antenna}
 {\cal S}_3^0(s)
 & = &
 \frac{\alpha_s}{\pi} \left( \frac{s}{\mu^2} \right)^{-\eps} 
  \left[ \frac{1}{\eps^2} + \frac{2}{\eps} + 6 - \frac{7}{12} \pi^2 \right] + {\cal O}(\eps).
\eq
Neglecting finite terms this is equivalent to
\bq
\label{eq_eps_poles}
\frac{\alpha_s}{\pi}
 X_3^0(a,i,b)
\frac{1}{\eps}
\left[ \ln \frac{s_{\tilde{a}j} s_{j\tilde{b}}}{s_{\tilde{a}\tilde{b}}} - \ln \frac{s_{aj} s_{jb}}{s_{ab}} \right]
 \circ \left| {\cal A}_3^{(0)} \right|^2 d\phi_4.
\eq
The three-parton amplitude ${\cal A}_3^{(0)}$ is evaluated with the momenta
$\tilde{p}_a$, $\tilde{p}_b$ and $p_j$.
The three-parton amplitude and the observable are independent of particle $i$.
We now investigate the behaviour of eq.~(\ref{eq_eps_poles}) if we integrate over the
phase space of particle $i$.
It is convenient to choose a specific frame, which we take as the
centre-of-mass frame of $\tilde{p}_{a}+\tilde{p}_{b}$ with $\tilde{p}_{a}$ and $p_a$ along the positive
$z$-axis and $p_j$ in the $x-z$ plane.
In this frame we have
\bq
\label{def_frame}
 \tilde{p}_a & = & \frac{1}{2} \sqrt{s_{\tilde{a}\tilde{b}}} (1,0,0,1),
 \nonumber \\
 \tilde{p}_b & = & \frac{1}{2} \sqrt{s_{\tilde{a}\tilde{b}}} (1,0,0,-1),
 \nonumber \\
 p_a & = & E_a (1,0,0,1),
 \nonumber \\
 p_b & = & E_b (1,\sin\theta_b\cos\phi,\sin\theta_b\sin\phi,\cos\theta_b),
 \nonumber \\
 p_j & = & E_j (1,\sin\theta_j,0,\cos\theta_j).
\eq
One finds
\bq
\label{phi_integrand}
 \ln \frac{s_{\tilde{a}j} s_{j\tilde{b}}}{s_{\tilde{a}\tilde{b}}} - \ln \frac{s_{aj} s_{jb}}{s_{ab}}
 & = &
 \ln \left( \frac{(1+\cos\theta_j)(1-\cos\theta_b)}{2 (1-\cos\theta_b\cos\theta_j -\sin\theta_b\sin\theta_j\cos \phi)} \right).
\eq
We focus on the integral over the azimuthal angle $\phi$ of particle $i$ in this frame.
In eq.~(\ref{eq_eps_poles}) only the square bracket depends on the azimuthal angle $\phi$.
With the help of eq.~(\ref{phi_integrand}) the relevant integral is given by
\bq
 I & = & 
\frac{1}{2\pi} \int\limits_0^{2\pi} d\phi
 \ln \left( \frac{(1+\cos\theta_j)(1-\cos\theta_b)}{2 (1-\cos\theta_b\cos\theta_j -\sin\theta_b\sin\theta_j\cos \phi)} \right),
\eq
The integral equals
\bq
 I & = &
 \ln \left( \frac{1-\cos\theta_b\cos\theta_j+(\cos\theta_j-\cos\theta_b)}{1-\cos\theta_b\cos\theta_j+|\cos\theta_j-\cos\theta_b|} \right).
\eq
The integral is zero for $\theta_j < \theta_b$ but non-zero for $\theta_j > \theta_b$.
Therefore the $1/\eps$-poles of eq.~(\ref{generic_eps_poles}) vanish after integration over the azimuthal angle
for $\theta_j < \theta_b$.
However, they do not vanish for $\theta_j > \theta_b$ and an additional subtraction term $d\alpha_3^{soft}$
is needed to cancel the explicit poles of
\bq
 d\alpha^{loop}_{3} + d\alpha^{product}_{3} - \int\limits_1 d\alpha^{almost}_{3} + \int\limits_1 d\alpha^{iterated}_{3}.
\eq
We introduce the notation
\bq 
 S_3^0(i,j,k;a,b) & = & 8 \pi \alpha_s \frac{2s_{ab}}{s_{aj}s_{jb}},
\eq
together with the convention that the phase space mapping $(i,j,k) \rightarrow (\tilde{i},\tilde{k})$
should be used.
In the $\bar{q}gggq$-channel the soft subtraction term can be taken as
\bq
\lefteqn{
 d\alpha^{soft}_{3,\bar{q}gggq} =  
 \frac{1}{6}
} & & \\
 & &
 \left\{
 \frac{N_c^2}{4} \frac{1}{2} 
 \left[ 
  \sum\limits_{(i,j)\in\{(1,3),(2,3),(3,2)\}}
  \left(
         S_3^0(0,i,j;0'',1'') - S_3^0(0,i,j;0'',2'') - S_3^0(0,i,j;1'',2'')
 \right. \right. \right. \nonumber \\
 & & \left. \left. \left.
       - S_3^0(0,i,j;0',2')   + S_3^0(0,i,j;0',3')   + S_3^0(0,i,j;2',3') 
  \right) \circ D_3^0(0',1',2')
 \right. \right. \nonumber \\
 & & \left. \left.
 +
  \sum\limits_{(i,j)\in\{(1,2),(2,1),(3,1)\}}
  \left(
         S_3^0(0,i,j;0'',1'') - S_3^0(0,i,j;0'',2'') - S_3^0(0,i,j;1'',2'')
 \right. \right. \right. \nonumber \\
 & & \left. \left. \left.
       - S_3^0(0,i,j;0',1')   + S_3^0(0,i,j;0',3')   + S_3^0(0,i,j;1',3') 
  \right) \circ D_3^0(0',2',1')
 \right. \right. \nonumber \\
 & & \left. \left.
 +
  \sum\limits_{(i,j)\in\{(1,2),(2,1),(1,3)\}}
  \left(
         S_3^0(i,j,4;1'',2'') - S_3^0(i,j,4;0'',1'') - S_3^0(i,j,4;0'',2'')
 \right. \right. \right. \nonumber \\
 & & \left. \left. \left.
       - S_3^0(i,j,4;1',3')   + S_3^0(i,j,4;0',1')   + S_3^0(i,j,4;0',3') 
  \right) \circ D_3^0(3',2',1')
 \right. \right. \nonumber \\
 & & \left. \left.
 +
  \sum\limits_{(i,j)\in\{(2,3),(3,1),(3,2)\}}
  \left(
         S_3^0(i,j,4;1'',2'') - S_3^0(i,j,4;0'',1'') - S_3^0(i,j,4;0'',2'')
 \right. \right. \right. \nonumber \\
 & & \left. \left. \left.
       - S_3^0(i,j,4;2',3')   + S_3^0(i,j,4;0',2')   + S_3^0(i,j,4;0',3') 
  \right) \circ D_3^0(3',1',2')
 \right] 
 \right. \nonumber \\
 & & \left.
 -
 \left( \frac{N_c^2}{4} \frac{1}{2} + \frac{1}{4} \right)
 \left[ 
  \sum\limits_{i\in\{1,2,3\}}
  \left(
         S_3^0(0,i,4;0'',2'') - S_3^0(0,i,4;0'',1'') - S_3^0(0,i,4;1'',2'')
 \right. \right. \right. \nonumber \\
 & & \left. \left. \left.
       - S_3^0(0,i,4;0',3')   + S_3^0(0,i,4;0',2')   + S_3^0(0,i,4;2',3') 
  \right) \circ A_3^0(0',1',3')
 \right. \right. \nonumber \\
 & & \left. \left.
+
  \sum\limits_{i\in\{1,2,3\}}
  \left(
         S_3^0(0,i,4;0'',2'') - S_3^0(0,i,4;0'',1'') - S_3^0(0,i,4;1'',2'')
 \right. \right. \right. \nonumber \\
 & & \left. \left. \left.
       - S_3^0(0,i,4;0',3')   + S_3^0(0,i,4;0',1')   + S_3^0(0,i,4;1',3') 
  \right) \circ A_3^0(0',2',3')
 \right] 
 \right\} \circ \left| {\cal A}_3^{(0)} \right|^2 d\phi_5.
\nonumber
\eq
In the $\bar{q}\bar{q}'gq'q$-channel we have
\bq
\lefteqn{
 d\alpha_{3,\bar{q}\bar{q}'gq'q}^{soft} 
 = 
 \frac{N_c^2}{4} \frac{1}{2}
} & & \\
& & 
 \left\{
 \left( \frac{1}{4} + \frac{N_f-1}{2} \right)
 \left[
  \left(
         S_3^0(0,2,3;0'',1'') - S_3^0(0,2,3;0'',2'') - S_3^0(0,2,3;1'',2'')
 \right. \right. \right. \nonumber \\
 & & \left. \left. \left.
       - S_3^0(0,2,3;0',2')   + S_3^0(0,2,3;0',3')   + S_3^0(0,2,3;2',3') 
  \right) \circ E_3^0(0',1',2')
 \right. \right. \nonumber \\
 & & \left. \left.
 +
  \left(
         S_3^0(1,2,4;0'',1'') - S_3^0(1,2,4;0'',2'') - S_3^0(1,2,4;1'',2'')
 \right. \right. \right. \nonumber \\
 & & \left. \left. \left.
       - S_3^0(1,2,4;1',3')   + S_3^0(1,2,4;1',2')   + S_3^0(1,2,4;2',3') 
  \right) \circ E_3^0(1',0',3')
 \right. \right. \nonumber \\
 & & \left. \left.
 +
  \left(
         S_3^0(0,2,3;1'',2'') - S_3^0(0,2,3;0'',1'') - S_3^0(0,2,3;0'',2'')
 \right. \right. \right. \nonumber \\
 & & \left. \left. \left.
       - S_3^0(0,2,3;0',2')   + S_3^0(0,2,3;0',1')   + S_3^0(0,2,3;1',2') 
  \right) \circ E_3^0(2',3',0')
 \right. \right. \nonumber \\
 & & \left. \left.
 +
  \left(
         S_3^0(1,2,4;1'',2'') - S_3^0(1,2,4;0'',1'') - S_3^0(1,2,4;0'',2'')
 \right. \right. \right. \nonumber \\
 & & \left. \left. \left.
       - S_3^0(1,2,4;1',3')   + S_3^0(1,2,4;0',1')   + S_3^0(1,2,4;0',3') 
  \right) \circ E_3^0(3',2',1')
 \right]
 \right. \nonumber \\
 & & \left.
 + \frac{1}{4}
 \left[
  \left(
         S_3^0(0,2,4;0'',1'') - S_3^0(0,2,4;0'',2'') - S_3^0(0,2,4;1'',2'')
 \right. \right. \right. \nonumber \\
 & & \left. \left. \left.
       - S_3^0(0,2,4;0',3')   + S_3^0(0,2,4;0',2')   + S_3^0(0,2,4;2',3') 
  \right) \circ E_3^0(0',1',3')
 \right. \right. \nonumber \\
 & & \left. \left.
 +
  \left(
         S_3^0(1,2,3;0'',1'') - S_3^0(1,2,3;0'',2'') - S_3^0(1,2,3;1'',2'')
 \right. \right. \right. \nonumber \\
 & & \left. \left. \left.
       - S_3^0(1,2,3;1',2')   + S_3^0(1,2,3;1',3')   + S_3^0(1,2,3;2',3') 
  \right) \circ E_3^0(1',0',2')
 \right. \right. \nonumber \\
 & & \left. \left.
 +
  \left(
         S_3^0(0,2,4;1'',2'') - S_3^0(0,2,4;0'',1'') - S_3^0(0,2,4;0'',2'')
 \right. \right. \right. \nonumber \\
 & & \left. \left. \left.
       - S_3^0(0,2,4;0',3')   + S_3^0(0,2,4;0',1')   + S_3^0(0,2,4;1',3') 
  \right) \circ E_3^0(3',2',0')
 \right. \right. \nonumber \\
 & & \left. \left.
 +
  \left(
         S_3^0(1,2,3;1'',2'') - S_3^0(1,2,3;0'',1'') - S_3^0(1,2,3;0'',2'')
 \right. \right. \right. \nonumber \\
 & & \left. \left. \left.
       - S_3^0(1,2,3;1',2')   + S_3^0(1,2,3;0',1')   + S_3^0(1,2,3;0',2') 
  \right) \circ E_3^0(2',3',1')
 \right]
 \right\} \circ \left| {\cal A}_3^{(0)} \right|^2 d\phi_5.
 \nonumber
\eq
In these subtraction terms the soft eikonal factors contain apart from the soft gluon momentum 
only momenta, which are not affected
by the phase space mapping.
The soft subtraction term $d\alpha_3^{soft}$ is only singular in the single soft limit, in all other limits it is integrable.
For the integrated version we need therefore the integral over the soft eikonal factor for the case
where the hard partons in the eikonal
factor are independent of the phase space mapping. This integral is given by \cite{GehrmannDeRidder:2007jk}
\bq
 {\cal S}_3^0(\tilde{p}_i,\tilde{p}_k,p_a,p_b) 
 & = & S_\eps^{-1} \mu^{2\eps} \int d\phi_{unres,ijk} \; S_3^0(i,j,k;a,b)
 \\
 & = &
 \frac{\alpha_s}{\pi} 
 \left( \frac{s_{\tilde{i}\tilde{k}}}{\mu^2}\right)^{-\eps}
 \left[ \frac{1}{\eps^2} - \frac{1}{\eps} \ln(x) - \mbox{Li}_2\left(-\frac{1-x}{x}\right) - \frac{7}{12} \pi^2 \right]
 + {\cal O}(\eps),
 \nonumber
\eq
where
\bq
 x & = & \frac{s_{ab}s_{\tilde{i}\tilde{k}}}
              {\left(s_{a\tilde{i}}+s_{a\tilde{k}}\right)\left(s_{b\tilde{i}}+s_{b\tilde{k}}\right)}.
\eq
Integrated over a one-particle unresolved phase space one obtains
\bq
\lefteqn{
 \int\limits_1 d\alpha^{soft}_{3,\bar{q}gggq} =  
 \frac{1}{2}
} & & \\
 & &
 \left\{
 \frac{N_c^2}{4} \frac{1}{2} 
 \left[ 
  D_3^0(0,1,2)
  \left(
         {\cal S}_3^0(0,2;0',1') - {\cal S}_3^0(0,2;0',2') - {\cal S}_3^0(0,2;1',2')
 \right. \right. \right. \nonumber \\
 & & \left. \left. \left.
       - {\cal S}_3^0(0,2;0,2)   + {\cal S}_3^0(0,2;0,3)   + {\cal S}_3^0(0,2;2,3) 
  \right) 
 \right. \right. \nonumber \\
 & & \left. \left.
 +
  D_3^0(0,2,1)
  \left(
         {\cal S}_3^0(0,1;0',1') - {\cal S}_3^0(0,1;0',2') - {\cal S}_3^0(0,1;1',2')
 \right. \right. \right. \nonumber \\
 & & \left. \left. \left.
       - {\cal S}_3^0(0,1;0,1)   + {\cal S}_3^0(0,1;0,3)   + {\cal S}_3^0(0,1;1,3) 
  \right) 
 \right. \right. \nonumber \\
 & & \left. \left.
 +
  D_3^0(3,2,1)
  \left(
         {\cal S}_3^0(1,3;1',2') - {\cal S}_3^0(1,3;0',1') - {\cal S}_3^0(1,3;0',2')
 \right. \right. \right. \nonumber \\
 & & \left. \left. \left.
       - {\cal S}_3^0(1,3;1,3)   + {\cal S}_3^0(1,3;0,1)   + {\cal S}_3^0(1,3;0,3) 
  \right) 
 \right. \right. \nonumber \\
 & & \left. \left.
 +
  D_3^0(3,1,2)
  \left(
         {\cal S}_3^0(2,3;1',2') - {\cal S}_3^0(2,3;0',1') - {\cal S}_3^0(2,3;0',2')
 \right. \right. \right. \nonumber \\
 & & \left. \left. \left.
       - {\cal S}_3^0(2,3;2,3)   + {\cal S}_3^0(2,3;0,2)   + {\cal S}_3^0(2,3;0,3) 
  \right) 
 \right] 
 \right. \nonumber \\
 & & \left.
 -
 \left( \frac{N_c^2}{4} \frac{1}{2} + \frac{1}{4} \right)
 \left[ 
  A_3^0(0,1,3)
  \left(
         {\cal S}_3^0(0,3;0',2') - {\cal S}_3^0(0,3;0',1') - {\cal S}_3^0(0,3;1',2')
 \right. \right. \right. \nonumber \\
 & & \left. \left. \left.
       - {\cal S}_3^0(0,3;0,3)   + {\cal S}_3^0(0,3;0,2)   + {\cal S}_3^0(0,3;2,3) 
  \right) 
 \right. \right. \nonumber \\
 & & \left. \left.
+
  A_3^0(0,2,3)
  \left(
         {\cal S}_3^0(0,3;0',2') - {\cal S}_3^0(0,3;0',1') - {\cal S}_3^0(0,3;1',2')
 \right. \right. \right. \nonumber \\
 & & \left. \left. \left.
       - {\cal S}_3^0(0,3;0,3)   + {\cal S}_3^0(0,3;0,1)   + {\cal S}_3^0(0,3;1,3) 
  \right) 
 \right] 
 \right\} \circ \left| {\cal A}_3^{(0)} \right|^2 d\phi_4.
\nonumber
\eq
\bq
\lefteqn{
 d\alpha_{3,\bar{q}\bar{q}'gq'q}^{soft} 
 = 
 \frac{N_c^2}{4} \frac{1}{2}
} & & \\
& & 
 \left\{
 \left( \frac{1}{4} + \frac{N_f-1}{2} \right)
 \left[
  E_3^0(0,1,2)
  \left(
         {\cal S}_3^0(0,2;0',1') - {\cal S}_3^0(0,2;0',2') - {\cal S}_3^0(0,2;1',2')
 \right. \right. \right. \nonumber \\
 & & \left. \left. \left.
       - {\cal S}_3^0(0,2;0,2)   + {\cal S}_3^0(0,2;0,3)   + {\cal S}_3^0(0,2;2,3) 
  \right) 
 \right. \right. \nonumber \\
 & & \left. \left.
 +
  E_3^0(1,0,3)
  \left(
         {\cal S}_3^0(1,3;0',1') - {\cal S}_3^0(1,3;0',2') - {\cal S}_3^0(1,3;1',2')
 \right. \right. \right. \nonumber \\
 & & \left. \left. \left.
       - {\cal S}_3^0(1,3;1,3)   + {\cal S}_3^0(1,3;1,2)   + {\cal S}_3^0(1,3;2,3) 
  \right) 
 \right. \right. \nonumber \\
 & & \left. \left.
 +
  E_3^0(2,3,0)
  \left(
         {\cal S}_3^0(0,2;1',2') - {\cal S}_3^0(0,2;0',1') - {\cal S}_3^0(0,2;0',2')
 \right. \right. \right. \nonumber \\
 & & \left. \left. \left.
       - {\cal S}_3^0(0,2;0,2)   + {\cal S}_3^0(0,2;0,1)   + {\cal S}_3^0(0,2;1,2) 
  \right) 
 \right. \right. \nonumber \\
 & & \left. \left.
 +
  E_3^0(3,2,1)
  \left(
         {\cal S}_3^0(1,3;1',2') - {\cal S}_3^0(1,3;0',1') - {\cal S}_3^0(1,3;0',2')
 \right. \right. \right. \nonumber \\
 & & \left. \left. \left.
       - {\cal S}_3^0(1,3;1,3)   + {\cal S}_3^0(1,3;0,1)   + {\cal S}_3^0(1,3;0,3) 
  \right) 
 \right]
 \right. \nonumber \\
 & & \left.
 + \frac{1}{4}
 \left[
  E_3^0(0,1,3)
  \left(
         {\cal S}_3^0(0,3;0',1') - {\cal S}_3^0(0,3;0',2') - {\cal S}_3^0(0,3;1',2')
 \right. \right. \right. \nonumber \\
 & & \left. \left. \left.
       - {\cal S}_3^0(0,3;0,3)   + {\cal S}_3^0(0,3;0,2)   + {\cal S}_3^0(0,3;2,3) 
  \right) 
 \right. \right. \nonumber \\
 & & \left. \left.
 +
  E_3^0(1,0,2)
  \left(
         {\cal S}_3^0(1,2;0',1') - {\cal S}_3^0(1,2;0',2') - {\cal S}_3^0(1,2;1',2')
 \right. \right. \right. \nonumber \\
 & & \left. \left. \left.
       - {\cal S}_3^0(1,2;1,2)   + {\cal S}_3^0(1,2;1,3)   + {\cal S}_3^0(1,2;2,3) 
  \right) 
 \right. \right. \nonumber \\
 & & \left. \left.
 +
  E_3^0(3,2,0)
  \left(
         {\cal S}_3^0(0,3;1',2') - {\cal S}_3^0(0,3;0',1') - {\cal S}_3^0(0,3;0',2')
 \right. \right. \right. \nonumber \\
 & & \left. \left. \left.
       - {\cal S}_3^0(0,3;0,3)   + {\cal S}_3^0(0,3;0,1)   + {\cal S}_3^0(0,3;1,3) 
  \right) 
 \right. \right. \nonumber \\
 & & \left. \left.
 +
  E_3^0(2,3,1)
  \left(
         {\cal S}_3^0(1,2;1',2') - {\cal S}_3^0(1,2;0',1') - {\cal S}_3^0(1,2;0',2')
 \right. \right. \right. \nonumber \\
 & & \left. \left. \left.
       - {\cal S}_3^0(1,2;1,2)   + {\cal S}_3^0(1,2;0,1)   + {\cal S}_3^0(1,2;0,2) 
  \right) 
 \right]
 \right\} \circ \left| {\cal A}_3^{(0)} \right|^2 d\phi_4.
 \nonumber
\eq
These integrated contributions cancel all explicit poles of the form as in eq.~(\ref{generic_eps_poles})
point by point in the four parton phase space.
Only in the colour factor $N_f N_c$ remain some explicit poles, which are of the form as in eq.~(\ref{Nf_asym_poles}).
The contribution in eq.~(\ref{Nf_asym_poles}) vanishes to all order in $\eps$ after the integration over the
unresolved phase space in $D$ dimensions.


\section{Infrared finiteness}
\label{sect_finiteness}

In this section I show that the three-parton, four-parton and five-parton contributions to the NNLO correction are 
individually infrared finite.
There are two types of potential divergences which need to be checked: 
\\
(i) Explicit poles in the dimensional
regularisation parameter $\eps=(4-D)/2$, which occur in individual stages after one or two integrations
have been performed analytically.
\\
(ii) Singularities of the phase-space integration.
\\
\\
We begin with the three-parton contribution. For this contribution only the cancellation of the explicit poles in $\eps$ needs
to be checked.
It is easily verified that the combination
\bq
 \int {\cal O}^{(3)}_{3} \circ \left( d\sigma_3^{(2)} 
               + d\alpha^{double}_{3} 
               + d\alpha^{loop}_{3} + d\alpha^{product}_{3}
        \right)
\eq
is free of explicit poles in $\eps$.
\\
\\
For the four-parton contribution we have to check the explicit poles and the phase-space singularities.
The four-parton contribution is given by
\bq
 \int \left[ {\cal O}^{(3)}_{4} \left( d\sigma_{4}^{(1)} + d\alpha^{single}_{4} \right)
    - {\cal O}^{(3)}_{3} \circ \left( d\alpha^{loop}_{3} + d\alpha^{product}_{3}
                                    - d\alpha^{almost}_{3} - d\alpha^{soft}_{3} + d\alpha^{iterated}_{3}
\right) \right].
\nonumber 
\eq
The combination 
\bq
 \int {\cal O}^{(3)}_{4} \left( d\sigma_{4}^{(1)} + d\alpha^{single}_{4} \right)
\eq
is free of explicit poles in $\eps$. This is just the cancellation of the explicit poles in $\eps$ of NLO calculations.
On the other hand as discussed in detail in section~\ref{sect:subtraction_terms_soft}  the explicit poles of the combination
\bq
 \int 
    {\cal O}^{(3)}_{3} \circ \left( d\alpha^{loop}_{3} + d\alpha^{product}_{3}
                                    - d\alpha^{almost}_{3} - d\alpha^{soft}_{3} + d\alpha^{iterated}_{3}
\right)
\eq
cancel point by point of the four-particle phase space for the colour
factors $N_c^2$, $N_c^0$, $N_c^{-2}$, $N_f/N_c$ and $N_f^2$.
In the colour factor $N_f N_c$ the explicit poles in $\eps$ cancel after integration over a one-particle
unresolved phase space in $D$ dimensions. This is related to eq.~(\ref{Nf_asym_poles}).
Alternatively they cancel after symmetrisation in the quark momenta.
Due to this anti-symmetry finite terms obtained from expressing the $D$-dimensional unresolved phase space measure
as an effective four-dimensional unresolved phase space measure integrate to zero and can be dropped in the numerical program.
It is possible to obtain also for the colour factor $N_f N_c$ 
a cancellation of the poles point by point in the four-particle phase space
at the expense of introducing additional subtraction terms.
However a cancellation of the poles after integration of the one-particle unresolved phase space is sufficient.
\\
We now consider the phase-space singularities of the four-parton contribution.
The following combinations are individually integrable over the $D$-dimensional four-parton phase space:
\bq
 & &
 \int \left( {\cal O}^{(3)}_{4} d\sigma_{4}^{(1)} - {\cal O}^{(3)}_{3} \circ d\alpha^{loop}_{3} \right),
 \nonumber \\
 & &
 \int \left( {\cal O}^{(3)}_{4} d\alpha^{single}_{4} - {\cal O}^{(3)}_{3} \circ d\alpha^{iterated}_{3} \right),
 \nonumber \\
 & &
 \int {\cal O}^{(3)}_{3} \circ \left( d\alpha^{product}_{3} - d\alpha^{almost}_{3} \right),
 \nonumber \\
 & &
 \int {\cal O}^{(3)}_{3} \circ d\alpha^{soft}_{3}.
\eq
We now consider the five-parton contribution. Here only phase-space singularities have to be considered.
The five-parton contribution is given by
\bq
 \int \left[ {\cal O}^{(3)}_{5} \; d\sigma_{5}^{(0)} 
             - {\cal O}^{(3)}_{4} \circ d\alpha^{single}_{4}
             - {\cal O}^{(3)}_{3} \circ 
\left( d\alpha^{double}_{3} + d\alpha^{almost}_{3} + d\alpha^{soft}_{3} 
                           - d\alpha^{iterated}_{3} \right) \right].
\eq
The potential double-unresolved singularities are: 
double soft, soft-collinear, triple collinear and two pairs of collinear particles.
The single-unresolved cases, which need to be checked are one soft particle and one pair of collinear particles.
The five-parton contribution consists of two channels, the $\bar{q} g g g q$-channel and the $\bar{q} \bar{q}' g q' q$-channel.
It is sufficient to check one representative case for each type of singularity.
We have the following cases:
\begin{itemize}
\item Double unresolved, $0_{\bar{q}} 1_g 2_g 3_g 4_q$-channel:
\begin{itemize}
 \item Double soft: $1_g$, $2_g$ soft.
 \item Soft-collinear: $1_g$ soft, $2_g || 3_g$ collinear.
 \item Soft-collinear: $1_g$ soft, $3_g || 4_q$ collinear.
 \item Triple collinear: $0_{\bar{q}} || 1_g || 2_g$.
 \item Triple collinear: $1_g || 2_g || 3_g$.
 \item Two collinear pairs: $0_{\bar{q}} || 1_g$ and $2_g || 3_g$.
 \item Two collinear pairs: $0_{\bar{q}} || 1_g$ and $3_g || 4_q$.
\end{itemize}
\item Single unresolved, $0_{\bar{q}} 1_g 2_g 3_g 4_q$-channel:
\begin{itemize}
 \item Soft: $1_g$ soft.
 \item Collinear: $0_{\bar{q}} || 1_g$.
 \item Collinear: $1_g || 2_g$.
\end{itemize}
\item Double unresolved, $0_{\bar{q}} 1_{\bar{q}'} 2_g 3_{q'} 4_q$-channel:
\begin{itemize}
 \item Double soft: $1_{\bar{q}'}$, $3_{q'}$ soft.
 \item Soft-collinear: $2_g$ soft, $1_{\bar{q}'} || 3_{q'}$ collinear.
 \item Triple collinear: $1_{\bar{q}'} || 2_g || 3_{q'}$.
 \item Triple collinear: $0_{\bar{q}} || 1_{\bar{q}'} || 3_{q'}$.
 \item Two collinear pairs: $0_{\bar{q}} || 2_g$ and $1_{\bar{q}'} || 3_{q'}$.
\end{itemize}
\item Single unresolved, $0_{\bar{q}} 1_{\bar{q}'} 2_g 3_{q'} 4_q$-channel:
\begin{itemize}
 \item Soft: $2_g$ soft.
 \item Collinear: $1_{\bar{q}'} || 3_{q'}$.
 \item Collinear: $1_{\bar{q}'} || 2_g$.
\end{itemize}
\end{itemize}
In the double unresolved limits we have
\bq
& &
\lim\limits_{\mbox{\scriptsize double unresolved}}
 {\cal O}^{(3)}_{5} \; d\sigma_{5}^{(0)} 
 - {\cal O}^{(3)}_{3} \circ d\alpha^{double}_{3}
 - {\cal O}^{(3)}_{3} \circ d\alpha^{almost}_{3}
= \mbox{integrable},
 \nonumber \\
& &
\lim\limits_{\mbox{\scriptsize double unresolved}}
 {\cal O}^{(3)}_{4} \circ d\alpha^{single}_{4}
 - {\cal O}^{(3)}_{3} \circ d\alpha^{iterated}_{3} 
 = \mbox{integrable},
 \nonumber \\
& &
\lim\limits_{\mbox{\scriptsize double unresolved}}
 {\cal O}^{(3)}_{3} \circ d\alpha^{soft}_{3} 
= \mbox{integrable}.
\eq
For the single unresolved it is convenient to split the NLO subtraction term
$d\alpha^{single}_{4}$ into
\bq
 d\alpha^{single}_{4} & = & d\alpha^{single,relevant}_{4} + d\alpha^{single,remainder}_{4},
\eq
where $d\alpha^{single,relevant}_{4}$ contains the antenna subtraction terms singular in the single
unresolved limit under consideration and $d\alpha^{single,remainder}_{4}$ contains the remaining terms.
Although the antenna terms of $d\alpha^{single,remainder}_{4}$ are finite in this particular limit, 
the matrix element multiplying these antenna functions can become singular in an NNLO calculation.
(In an NLO calculation the infrared-safe observable ensures that there is no contribution from these regions
to the NLO calculation.)
In a similar way we split $d\alpha^{iterated}_{3}$
\bq
 d\alpha^{iterated}_{3} & = & d\alpha^{iterated,relevant}_{3} + d\alpha^{iterated,remainder}_{3},
\eq
such that $d\alpha^{iterated,relevant}_{3}$ is an approximation to $d\alpha^{single,relevant}_{4}$ and
$d\alpha^{iterated,remainder}_{3}$ is an approximation to $d\alpha^{single,remainder}_{4}$.
We then have in single unresolved limits
\bq
& &
\lim\limits_{\mbox{\scriptsize single unresolved}}
 {\cal O}^{(3)}_{5} \; d\sigma_{5}^{(0)} 
 - {\cal O}^{(3)}_{4} \circ d\alpha^{single,relevant}_{4}
= \mbox{integrable},
 \\
& &
\lim\limits_{\mbox{\scriptsize single unresolved}}
 {\cal O}^{(3)}_{4} \circ d\alpha^{single,remainder}_{4}
 - {\cal O}^{(3)}_{3} \circ d\alpha^{iterated,remainder}_{3}
= \mbox{integrable},
 \nonumber \\
& &
\lim\limits_{\mbox{\scriptsize single unresolved}}
 {\cal O}^{(3)}_{3} \circ \left( d\alpha^{iterated,relevant}_{3}
 - d\alpha^{double}_{3}
 - d\alpha^{almost}_{3}
 - d\alpha^{soft}_{3} \right)
= \mbox{integrable}.
 \nonumber
\eq
The explicit results for the non-trivial singular limits are listed in the appendix~\ref{sect:limits}.


\section{Conclusions}
\label{sect_conclusions}

In this paper I gave detailed information on the structure of the infrared singularities for the process
$e^+ e^- \rightarrow \mbox{3 jets}$ at next-to-next-to-leading order in perturbation theory.
Particular emphasis was put on singularities associated to soft gluons.
The knowledge how to disentangle singularities at NNLO for this process will also be useful for other
processes with three or more hard coloured partons, if calculated at NNLO.
In this paper a complete list of the subtraction terms used in the numerical program ``Mercutio2''
was given.


\begin{appendix}

\section{Alternative soft subtraction term}
\label{sect:alternative_soft_term}

In this appendix I present an alternative form for the soft subtraction terms.
The starting point is eq.~(\ref{generic_eps_poles})
\bq
\label{starting_point}
I & = & 
 X_3^0(a,i,b)
 \left[ 
  {\cal S}_3^0(s_{\tilde{a}\tilde{b}}) - {\cal S}_3^0(s_{ab})
  - {\cal S}_3^0(s_{\tilde{a}j}) + {\cal S}_3^0(s_{aj})
  - {\cal S}_3^0(s_{\tilde{b}j}) + {\cal S}_3^0(s_{bj})
 \right]
 \circ \left| {\cal A}_3^{(0)} \right|^2 d\phi_4,
\nonumber \\
\eq
which in the frame defined by eq.~(\ref{def_frame})
\bq
\label{def_frame_2}
 \tilde{p}_a & = & \frac{1}{2} \sqrt{s_{\tilde{a}\tilde{b}}} (1,0,0,1),
 \nonumber \\
 \tilde{p}_b & = & \frac{1}{2} \sqrt{s_{\tilde{a}\tilde{b}}} (1,0,0,-1),
 \nonumber \\
 p_a & = & E_a (1,0,0,1),
 \nonumber \\
 p_b & = & E_b (1,\sin\theta_b\cos\phi,\sin\theta_b\sin\phi,\cos\theta_b),
 \nonumber \\
 p_j & = & E_j (1,\sin\theta_j,0,\cos\theta_j),
\eq
leaves uncancelled explicit poles in $1/\eps$ for $\theta_j>\theta_b$.
The condition $\theta_j>\theta_b$ is equivalent to $\cos \theta_j< \cos \theta_b$ or,
expressed in invariants,
\bq
 \frac{2p_ap_j}{2p_{aib}p_j} & > & \frac{2p_ap_b}{2p_{aib}p_b}.
\eq
Let us define two new momenta $\hat{p}_a$ and $\hat{p}_b$ which are related to $p_a$, $p_i$ and $p_b$
by the Catani-Seymour mapping with particle $a$ being the spectator and particle $b$ being the emitter:
\bq
 \hat{p}_a = \frac{s_{aib}}{s_{ai}+s_{ab}} p_a,
 & &
 \hat{p}_b = p_b + p_i - \frac{s_{ib}}{s_{ai}+s_{ab}} p_a.
\eq
As subtraction term to eq.~(\ref{starting_point}) we can use
\bq
\label{angular_subtraction_term}
A & = &
 X_3^0(a,i,b)
 \Theta\left(\theta_j>\theta_b\right)
 \left[ 
  {\cal S}_3^0(s_{b\hat{b}}) - {\cal S}_3^0(s_{ab})
  - {\cal S}_3^0(s_{j\hat{b}}) + {\cal S}_3^0(s_{aj})
 \right]
 \circ \left| {\cal A}_3^{(0)} \right|^2 d\phi_4.
\eq
We then have
\bq
\lefteqn{
I-A = 
 \frac{\alpha_s}{\pi} 
 X_3^0(a,i,b)
 \frac{1}{\eps}
} & & \\
 & &
 \left[
 \Theta\left(\theta_j<\theta_b\right)
 \ln \left( \frac{(1+\cos\theta_j)(1-\cos\theta_b)}{2 (1-\cos\theta_b\cos\theta_j -\sin\theta_b\sin\theta_j\cos \phi)} \right)
\right. \nonumber \\
 & & \left.
+
 \Theta\left(\theta_j>\theta_b\right)
 \ln \left( \frac{(1-\cos\theta_j)(1+\cos\theta_b)}{2 (1-\cos\theta_b\cos\theta_j -\sin\theta_b\sin\theta_j\cos \phi)} \right)
 \right]
 \circ \left| {\cal A}_3^{(0)} \right|^2 d\phi_4
 + {\cal O}(\eps).
 \nonumber
\eq
Integration over the azimuthal angle now leads to zero for the pole terms proportional to $1/\eps$
in both cases $\theta_j<\theta_b$ and $\theta_j>\theta_b$.
We then have to lift eq.~(\ref{angular_subtraction_term}) to the five-parton phase space.
The terms with ${\cal S}_3^0(s_{ab})$ and ${\cal S}_3^0(s_{aj})$ pose no problem.
The terms with ${\cal S}_3^0(s_{b\hat{b}})$ and ${\cal S}_3^0(s_{j\hat{b}})$ are more subtle.
Let us first consider the term ${\cal S}_3^0(s_{j\hat{b}})$. Lifting ${\cal S}_3^0$ gives an eikonal factor
$S_3^0$ which depends apart from the momentum of the soft gluon on two hard momenta. We would like these two
hard momenta to be such that under the momentum mapping they are mapped onto $p_j$ and $p_{\hat{b}}$.
We can do this by a momentum mapping, which maps five momenta to four momenta.
Let us denote the five momenta by $p_i,p_j,p_k,p_l,p_m$, with $p_j$ being the soft particle.
The eikonal factor is given by
\bq 
 S_3^0(i,j,n) & = & 8 \pi \alpha_s \frac{2s_{in}}{s_{ij}s_{jn}},
\eq
where $p_n$ is a linear combination of $p_k$, $p_l$ and $p_m$:
\bq
 p_n & = & p_k + p_l - \frac{y_2}{1-y_2} p_m,
\;\;\;
y_2 = \frac{s_{kl}}{s_{klm}}.
\eq
The $5\rightarrow 4$-mapping is given by
\bq
 \tilde{p}_i = p_i + p_j - \frac{y_1}{1-y_1} p_n,
 & &
 \tilde{p}_k = \frac{1}{1-y_1} p_k,
 \nonumber \\
 \tilde{p}_l = \frac{1}{1-y_1} p_l,
 & &
 \tilde{p}_m = \frac{1-y_1-y_2}{(1-y_1)(1-y_2)} p_k,
\eq
with
\bq
 y_1 = \frac{s_{ij}}{s_{ijm}}.
\eq
$p_n$ is transformed to
\bq
 \tilde{p}_n & = & \frac{1}{1-y_1} p_n
 = \tilde{p}_k + \tilde{p_l} - \frac{\tilde{y}_2}{1-\tilde{y}_2} \tilde{p}_m,
\;\;\;
\tilde{y}_2 = \frac{s_{\tilde{k}\tilde{l}}}{s_{\tilde{k}\tilde{l}\tilde{m}}} = \frac{y_2}{1-y_1}.
\eq
Integration over the unresolved phase space yields
\bq
S_\eps^{-1} \mu^{2\eps} \int d\phi_{unres} \; S_3^0(i,j,n)
 & = & 
 {\cal S}_3^0(s_{\tilde{i}\tilde{n}}),
\eq
with ${\cal S}_3^0(s)$ given by eq.~(\ref{integrated_soft_antenna}).
\\
To lift the term ${\cal S}_3^0(s_{b\hat{b}})$ we consider four momenta $p_i,p_j,p_k,p_l$ and map them to three
momenta $\tilde{p}_j$, $\tilde{p}_k$ and $\tilde{p}_l$. In this case we denote by $p_i$ the soft particle.
The $4\rightarrow 3$-mapping is given by
\bq
 \tilde{p}_j & = & p_i + p_j - y_1 \tilde{p}_n,
 \nonumber \\
 \tilde{p}_k & = & p_k + y_1 \tilde{p}_n - \frac{a}{1-a} p_l,
 \nonumber \\
 \tilde{p}_l & = & \frac{1}{1-a} p_l,
\eq
with 
\bq
 \tilde{p}_n = p_i + p_j + p_k - \frac{s_{ijk}}{2p_lp_{ijk}} p_l,
 \;\;\;
 y_1 = \frac{s_{ij}}{s_{i\tilde{n}}s_{j\tilde{n}}}.
 \;\;\;
 a = \frac{y_1 s_{k\tilde{n}}}{y_1 s_{k\tilde{n}}+y_1 s_{l\tilde{n}}+s_{kl}}.
\eq
Note that 
\bq
 \tilde{p}_n & = & \tilde{p}_j + \tilde{p}_k - \frac{s_{\tilde{j}\tilde{k}}}{2p_{\tilde{l}}p_{\tilde{j}\tilde{k}}} \tilde{p}_l.
\eq
With 
\bq
 p_n & = & (1-y_1) \tilde{p}_n,
\eq
the eikonal factor is given by
\bq 
 S_3^0(j,i,n) & = & 8 \pi \alpha_s \frac{2s_{jn}}{s_{ji}s_{in}}.
\eq
Integration over the unresolved phase space yields
\bq
S_\eps^{-1} \mu^{2\eps} \int d\phi_{unres} \; S_3^0(j,i,n)
 & = & 
 {\cal S}_3^0(s_{\tilde{j}\tilde{n}}).
\eq
We remark that the subtraction term of eq.~(\ref{angular_subtraction_term}) 
treats the partons $a$ and $b$ asymmetric for the construction of the hatted momenta, 
with $a$ being the spectator and $b$ being the emitter.
A more symmetric version is given by
\bq
\lefteqn{
A = 
 X_3^0(a,i,b)
 \left\{
 \frac{s_{ai}}{s_{ai}+s_{ib}}
 \Theta\left( \frac{2p_ap_j}{2p_{aib}p_j} - \frac{2p_ap_b}{2p_{aib}p_b}\right)
 \circ
 \left[ 
  {\cal S}_3^0(s_{b\hat{b}}) - {\cal S}_3^0(s_{ab})
  - {\cal S}_3^0(s_{j\hat{b}}) + {\cal S}_3^0(s_{aj})
 \right]
 \right. }
 \nonumber \\
 & & \left.
 +
 \frac{s_{ib}}{s_{ai}+s_{ib}}
 \Theta\left( \frac{2p_bp_j}{2p_{aib}p_j} - \frac{2p_ap_b}{2p_{aib}p_a}\right)
 \circ
 \left[ 
  {\cal S}_3^0(s_{a\hat{a}}) - {\cal S}_3^0(s_{ab})
  - {\cal S}_3^0(s_{\hat{a}j}) + {\cal S}_3^0(s_{jb})
 \right]
\right\}
 \circ \left| {\cal A}_3^{(0)} \right|^2 d\phi_4.
\nonumber
\eq
Finally we remark that the expressions of eq.~(\ref{starting_point}) and eq.~(\ref{angular_subtraction_term})
give rise to finite terms, which have to be taken into account.
The first source of these terms is straightforward and arises from the expansion of the functions
${\cal S}_3^0$ to order $\eps^0$, which occur in eq.~(\ref{starting_point}) and eq.~(\ref{angular_subtraction_term}).
The $\eps^1$ terms of the function $X_3^0(a,i,b)$ give no contribution to the finite terms, they vanish for the same
reasons as the pole terms.
However the $D$-dimensional phase space measure $d\phi_{unres}$ gives an additional contribution at order $\eps^0$.
Let us consider again the frame defined in eq.~(\ref{def_frame}) and in eq.~(\ref{def_frame_2}).
In four dimensions we parametierised $p_b$ by
\bq
 p_b & = & E_b (1,\sin\theta_b\cos\phi,\sin\theta_b\sin\phi,\cos\theta_b).
\eq
$\theta_b$ is the polar angle and $\phi$ the azimuthal angle.
In $D$-dimensions we can parameterise $p_b$ by
\bq
 p_b & = & E_b (1,\sin\theta_1\cos\theta_2,\sin\theta_1\sin\theta_2\cos\theta_3,\cos\theta_1,...).
\eq
The dots indicate the components in the additional dimensions. $\theta_1$, $\theta_2$ and $\theta_3$ are polar angles.
As in the chosen frame all other particles lie in the $x-z$ plane, nothing depends on $\theta_3$ and we can integrate
over $\theta_3$ and all additional angles $\theta_4$, ..., $\theta_{D-2}$.
The unresolved phase space in $D$-dimensions is then given by
\bq
 \int d\phi_{unres,aib} & = & 
 \frac{s_{aib}}{32\pi^3} 
 (4\pi)^\eps \left(x_1x_2x_3s_{aib}\right)^{-\eps} 
 \int d^3x \; \delta\left(1-x_1-x_2-x_3\right) 
 \nonumber \\
 & &
 2 \frac{\Gamma\left(\frac{1}{2}\right)}{\Gamma\left(\frac{1}{2}-\eps\right)}
 \int\limits_0^\pi d\theta_2 \left( \sin \theta_2 \right)^{-2\eps},
\eq
where 
\bq
 x_1 = \frac{s_{ai}}{s_{aib}},
 \;\;\;
 x_2 = \frac{s_{ib}}{s_{aib}},
 \;\;\;
 x_3 = \frac{s_{ab}}{s_{aib}}.
\eq
We would like to write this as an effective four-dimensional phase space. For an integrand which depends on the polar angle
$\theta_2$ only through $\cos\theta_2$ we may write
\bq
 2 \int\limits_0^\pi d\theta_2 \left( \sin \theta_2 \right)^{-2\eps} f\left(\cos\theta_2\right)
 & = & 
 \int\limits_0^{2\pi} d\phi \left| \sin \phi \right|^{-2\eps} f\left(\cos\phi\right).
\eq
This allows us to treat the polar angle $\theta_2$ of the $D$-dimensional parameterisation 
as the azimuthal angle $\phi$ of the four-dimensional parameterisation.
We can therefore write the unresolved phase space in $D$-dimensions effectively as a phase space in four dimensions with
additional $\eps$-dependent factors:
\bq
 \int d\phi_{unres,aib} & = & 
 \frac{s_{aib}}{32\pi^3} 
 (4\pi)^\eps \left(x_1x_2x_3s_{aib}\right)^{-\eps} 
 \int d^3x \; \delta\left(1-x_1-x_2-x_3\right) 
 \nonumber \\
 & &
 \frac{\Gamma\left(\frac{1}{2}\right)}{\Gamma\left(\frac{1}{2}-\eps\right)}
 \int\limits_0^{2\pi} d\phi \left| \sin \phi \right|^{-2\eps}.
\eq
Expansion of $| \sin \phi |^{-2\eps}$ to order $\eps^1$ gives a finite contribution when combined with the $1/\eps$ poles of
${\cal S}_3^0$. The $\eps^1$-terms of the factors $(4\pi)^\eps (x_1x_2x_3s_{aib})^{-\eps}$
and $\Gamma(1/2)/\Gamma(1/2-\eps)$ can be dropped, as they vanish after integration over the unresolved phase space for the same
reason as the pole terms.

\section{Singular limits of the five parton contribution}
\label{sect:limits}

In this appendix limit I list the explicit forms of the non-trivial singular limits of the five parton contributions.
The double unresolved limits of $d\sigma_5^{(0)}$ have been considered in 
\cite{Berends:1989zn,Gehrmann-DeRidder:1998gf,Campbell:1998hg,Catani:1998nv,Catani:1999ss,DelDuca:1999ha,Kosower:2002su},
the singular limits of the four-parton tree level antenna functions $X_4^0$ have been taken from \cite{Gehrmann-DeRidder:2005cm}.
The singular functions are denoted as follows:
The single soft factor is denoted by $S_{ijk}$, where $j$ is a soft particle between particles $i$ and $k$.
The double soft factor is denoted by $S_{ijkl}$, in this case particles $j$ and $k$ are soft.
The singular function for two collinear particles $i$ and $j$ is denoted by $P_{ij}$. No distinction is made in the notation between
$g\rightarrow g g$-, $g \rightarrow q \bar{q}$- and $q \rightarrow q g$-splittings. This can be deduced from the involved particles.
If particles $i$ and $j$ become collinear the resulting momentum is denoted by $(ij)$.
The singular function for the triple collinear limit is denoted by $P_{ijk}$.
Finally the soft-collinear function is denoted by $SC_{ij(kl)}$ with $j$ the soft particle and $k$ and $l$ the collinear pair.
We note that in the soft-collinear limit the following combinations
\bq
 & &
 \left( SC_{ij(kl)} - S_{ijk} \right) P_{kl},
 \nonumber \\
 & &
 \left( SC_{ij(kl)} - SC_{ij(lk)} \right) P_{kl},
 \nonumber \\
 & &
 \left( S_{ij(kl)} - S_{ijk} \right) P_{kl}
\eq
are all integrable.
\\
\\
In the double unresolved limit we have for the $0_{\bar{q}} 1_g 2_g 3_g 4_q$-channel:
\\
\\
Double soft: $1_g$, $2_g$ soft.
\bq
\lefteqn{
 \lim\limits_{1,2 \rightarrow 0} d\sigma_5^{(0)} 
 =
 \frac{1}{6}  
 \left\{
  \frac{N_c^2}{4}
   \left[ 
          S_{0123} + S_{0213} + S_{013} S_{324} + S_{023} S_{314} + S_{3124} + S_{3214}
   \right]
 \right.
}
 \nonumber \\
 & & \left.
  + \frac{1}{4} 
    \left[
           - S_{0124} - S_{0214} - S_{013} S_{024} - S_{314} S_{024} - S_{014} S_{023} - S_{014} S_{324}
           + S_{014} S_{024}
    \right]
 \right. \nonumber \\
 & & \left.
  + \frac{1}{4N_c^2} S_{014} S_{024}
 \right\}
 \circ \left| {\cal A}_3^{(0)} \right|^2 d\phi_5,
 \nonumber \\
\lefteqn{
 \lim\limits_{1,2 \rightarrow 0} \left( - d\alpha^{double}_3 \right)
 =
 \frac{1}{6}  
 \left\{
  \frac{N_c^2}{4}
   \left[
          - S_{0123} - S_{0213}  
          - S_{3124} - S_{3214} 
          - S_{013} S_{023} - S_{314} S_{324}
          + S_{014} S_{024} 
   \right]
 \right. 
}
 \nonumber \\
 & &
 \left.
  + \frac{1}{4} 
    \left[
           S_{0124} + S_{0214}
           + S_{014} S_{024}
    \right]
  + \frac{1}{4N_c^2} 
    \left[
           - S_{014} S_{024}
    \right]
 \right\}
 \circ \left| {\cal A}_3^{(0)} \right|^2 d\phi_5,
 \nonumber \\
\lefteqn{
 \lim\limits_{1,2 \rightarrow 0} \left( - d\alpha^{almost}_3 \right)
 =
 \frac{1}{6}  
 \left\{
  \frac{N_c^2}{4}
   \left[ 
          - S_{013} S_{324} - S_{023} S_{314}
          + S_{013} S_{023} + S_{314} S_{324}
          - S_{014} S_{024}
   \right]
\right.
}
 \nonumber \\
 & &
\left.
  + \frac{1}{4} 
    \left[
           S_{013} S_{024} + S_{023} S_{014}
           + S_{314} S_{024} + S_{324} S_{014}
           - 2 S_{014} S_{024}
    \right]
 \right\}
 \circ \left| {\cal A}_3^{(0)} \right|^2 d\phi_5.
\hspace*{30mm}
\eq
Soft-collinear: $1_g$ soft, $2_g || 3_g$ collinear.
\bq
\lefteqn{
 \lim\limits_{1 \rightarrow 0, \; 2||3} d\sigma_5^{(0)}
 =
 \frac{1}{6} 
 \left\{
  \frac{N_c^2}{4}
   \left[
          SC_{01(23)} P_{23} + SC_{01(32)} P_{23} + SC_{41(32)} P_{23} + SC_{41(23)} P_{23}
   \right] 
 \right.
}
 \nonumber \\
 & &
 \left.
  + \frac{1}{4}
   \left[
          - 2 S_{014} P_{23} 
   \right]
 \right\}
 \circ \left| {\cal A}_3^{(0)} \right|^2 d\phi_5,
 \hspace*{100mm}
 \nonumber \\
\lefteqn{
 \lim\limits_{1 \rightarrow 0, \; 2||3} \left( - d\alpha^{double}_3 \right)
 =
 \frac{1}{6} 
 \left\{
  \frac{N_c^2}{4}
   \left[ 
          - SC_{01(23)} P_{23} - SC_{01(32)} P_{23} - SC_{41(23)} P_{23} - SC_{41(32)} P_{23} 
   \right]
 \right\}
}
 \nonumber \\
 & &
  \circ \left| {\cal A}_3^{(0)} \right|^2 d\phi_5,
 \nonumber \\
\lefteqn{
 \lim\limits_{1 \rightarrow 0, \; 2||3} \left( - d\alpha^{almost}_3 \right)
 =
 \frac{1}{6} 
 \left\{
   \frac{1}{4} 
    \left[
            2 S_{014} P_{23}
    \right]
 \right\}
 \circ \left| {\cal A}_3^{(0)} \right|^2 d\phi_5.
}
\eq
Soft-collinear: $1_g$ soft, $3_g || 4_q$ collinear.
\bq
\lefteqn{
 \lim\limits_{1 \rightarrow 0, \; 3||4} d\sigma_5^{(0)}
 =
 \frac{1}{6}  
 \left\{
  \frac{N_c^2}{4}
   \left[
          S_{012} P_{34} + SC_{21(34)} P_{34} 
   \right] 
   \right. 
}
 \nonumber \\
 & & \left.
  + \frac{1}{4}
   \left[
          - S_{012} P_{34} - SC_{01(34)} P_{34} - SC_{21(43)} P_{34} 
   \right]
  + \frac{1}{4N_c^2} SC_{01(43)} P_{34}
 \right\}
 \circ \left| {\cal A}_3^{(0)} \right|^2 d\phi_5,
 \nonumber \\
\lefteqn{
 \lim\limits_{1 \rightarrow 0, \; 3||4} \left( - d\alpha^{double}_3 \right)
 =
 \frac{1}{6} 
 \left\{
  \frac{N_c^2}{4}
   \left[ 
          - SC_{21(34)} P_{34} - SC_{21(43)} P_{34} + SC_{01(43)} P_{34}
   \right]
 \right.
}
 \nonumber \\
 & &
 \left.
  + \frac{1}{4}
   \left[
           SC_{01(34)} P_{34} + SC_{01(43)} P_{34}
   \right]
  + \frac{1}{4N_c^2} 
   \left[
           - SC_{01(43)} P_{34}
   \right]
 \right\}
 \circ \left| {\cal A}_3^{(0)} \right|^2 d\phi_5,
 \nonumber \\
\lefteqn{
 \lim\limits_{1 \rightarrow 0, \; 3||4} \left( - d\alpha^{almost}_3 \right)
 =
 \frac{1}{6}  
 \left\{
  \frac{N_c^2}{4}
   \left[ 
           - S_{012} P_{34} + S_{21(34)} P_{34} - S_{01(34)} P_{34}
   \right]
 \right.
}
 \nonumber \\
 & &
 \left.
  + \frac{1}{4} 
    \left[ 
           S_{012} P_{34} + S_{21(34)} P_{34} - S_{01(34)} P_{34}
    \right]
 \right\}
 \circ \left| {\cal A}_3^{(0)} \right|^2 d\phi_5.
\eq
Triple collinear: $0_{\bar{q}} || 1_g || 2_g$.
\bq
\lefteqn{
 \lim\limits_{0 || 1 || 2} d\sigma_5^{(0)}
 =
 \frac{1}{6}  
 \left\{
  \frac{N_c^2}{4}
   \left[
          P_{012} + P_{021} 
   \right]
  + \frac{1}{4} 
    \left[
           - P_{012} - P_{021} - P_{012}^{sc} 
    \right]
  + \frac{1}{4N_c^2} 
    \left[
          P_{012}^{sc} 
    \right]
 \right\}
 \circ \left| {\cal A}_3^{(0)} \right|^2 d\phi_5,
} & &
 \nonumber \\
\lefteqn{
 \lim\limits_{0 || 1 || 2} \left( - d\alpha^{double}_3 \right)
 =
 \frac{1}{6}  
 \left\{
  \frac{N_c^2}{4}
   \left[ 
          - P_{012} - P_{021} 
   \right]
  + \frac{1}{4} 
    \left[
           P_{012} + P_{021} + P_{012}^{sc}
    \right]
\right.
} & &
 \nonumber \\
 & &
\left.
  + \frac{1}{4N_c^2} 
    \left[
           - P_{012}^{sc}
    \right]
 \right\}
 \circ \left| {\cal A}_3^{(0)} \right|^2 d\phi_5,
 \hspace*{80mm}
 \nonumber \\
\lefteqn{
 \lim\limits_{0 || 1 || 2} \left( - d\alpha^{almost}_3 \right)
 = \mbox{integrable}.
} & &
\eq
Triple collinear: $1_g || 2_g || 3_g$.
\bq
& &
 \lim\limits_{1 || 2 || 3} d\sigma_5^{(0)}
 =
 \frac{1}{6}  
 \left\{
  \frac{N_c^2}{4}
   \left[
          P_{123} + P_{132} + P_{213} + P_{231} + P_{312} + P_{321}
   \right]
 \right\}
 \circ \left| {\cal A}_3^{(0)} \right|^2 d\phi_5,
 \nonumber \\
& &
 \lim\limits_{1 || 2 || 3} \left( - d\alpha^{double}_3 \right)
 =
 \frac{1}{6}  
 \left\{
  \frac{N_c^2}{4}
   \left[ 
          - P_{123} - P_{132} - P_{213} - P_{231} - P_{312} - P_{321}
   \right]
 \right\}
 \circ \left| {\cal A}_3^{(0)} \right|^2 d\phi_5,
 \nonumber \\
& &
 \lim\limits_{1 || 2 || 3} \left( - d\alpha^{almost}_3 \right)
 = \mbox{integrable}.
\eq
Two collinear pairs: $0_{\bar{q}} || 1_g$ and $2_g || 3_g$.
\bq
& &
 \lim\limits_{0 || 1, \; 2 || 3} d\sigma_5^{(0)}
 =
 \frac{1}{6}  
 \left\{
  \frac{N_c^2}{4}
   \left[
          P_{01} P_{23} + P_{01} P_{32} 
   \right]
 + \frac{1}{4} 
   \left[
          - P_{01} P_{23} - P_{01} P_{32}
   \right]
 \right\}
 \circ \left| {\cal A}_3^{(0)} \right|^2 d\phi_5,
 \nonumber \\
& &
 \lim\limits_{0 || 1, \; 2 || 3} \left( - d\alpha^{double}_3 \right)
 =
 \frac{1}{6}  
 \left\{
  \frac{N_c^2}{4}
   \left[ 
          - P_{01} P_{23} - P_{01} P_{32}
   \right]
 \right\}
 \circ \left| {\cal A}_3^{(0)} \right|^2 d\phi_5,
 \nonumber \\
& &
 \lim\limits_{0 || 1, \; 2 || 3} \left( - d\alpha^{almost}_3 \right)
 =
 \frac{1}{6}  
 \left\{
  \frac{1}{4} 
    \left[
           P_{01} P_{23} + P_{01} P_{32}
    \right]
 \right\}
 \circ \left| {\cal A}_3^{(0)} \right|^2 d\phi_5.
\eq
Two collinear pairs: $0_{\bar{q}} || 1_g$ and $3_g || 4_q$.
\bq
& &
 \lim\limits_{0 || 1, \; 3 || 4} d\sigma_5^{(0)}
 =
 \frac{1}{6} 
 \left\{
  \frac{N_c^2}{4}
   \left[
          P_{01} P_{34}  
   \right]
 + \frac{1}{4} 
   \left[
          - P_{01} P_{43} - P_{10} P_{34} 
   \right]
 + \frac{1}{4N_c^2} P_{01} P_{34}
 \right\}
 \circ \left| {\cal A}_3^{(0)} \right|^2 d\phi_5,
 \nonumber \\
& &
 \lim\limits_{0 || 1, \; 3 || 4} \left( - d\alpha^{double}_3 \right)
 =
 \frac{1}{6}  
 \left\{
  \frac{N_c^2}{4}
   \left[ 
          P_{01} P_{34}
   \right]
  + \frac{1}{4} 
    \left[
           2 P_{01} P_{34}
    \right]
  + \frac{1}{4N_c^2} 
    \left[
           - P_{01} P_{34}
    \right]
 \right\}
 \circ \left| {\cal A}_3^{(0)} \right|^2 d\phi_5,
 \nonumber \\
& &
 \lim\limits_{0 || 1, \; 3 || 4} \left( - d\alpha^{almost}_3 \right)
 =
 \frac{1}{6}  
 \left\{
  \frac{N_c^2}{4}
   \left[ 
          2 P_{01} P_{34}
   \right]
 \right\}
 \circ \left| {\cal A}_3^{(0)} \right|^2 d\phi_5.
\eq
In the single unresolved limit we have for the $0_{\bar{q}} 1_g 2_g 3_g 4_q$-channel:
\\
\\
Soft: $1_g$ soft.
\bq
\lefteqn{
 \lim\limits_{1 \rightarrow 0} \left( - d\alpha^{double}_3 \right)
 = 
 \frac{1}{6} 
 \left\{
  \frac{N_c^2}{4}
   \left[ 
          - \left( D_3^0(0,2,3) + D_3^0(0,3,2) \right)
            \left( S_{012} + S_{013} + S_{213} \right)
 \right. \right.
} & &
 \nonumber \\
& &
 \left. \left.
          - \left( D_3^0(4,3,2) + D_3^0(4,2,3) \right)
            \left( S_{214} + S_{314} + S_{213} \right)
          + \left( A_3^0(0,2,4) + A_3^0(0,3,4) \right) S_{014}
   \right]
 \right. \nonumber \\
 & & \left.
  + \frac{1}{4} 
    \left[
           A_3^0(0,2,4) \left( S_{012} + S_{214} \right)
         + A_3^0(0,3,4) \left( S_{013} + S_{314} \right)
         + \left( A_3^0(0,2,4) + A_3^0(0,3,4) \right) S_{014}
    \right]
 \right. \nonumber \\
 & & \left.
  + \frac{1}{4N_c^2} 
    \left[
         - \left( A_3^0(0,2,4) + A_3^0(0,3,4) \right) S_{014}
    \right]
 \right\}
 \circ \left| {\cal A}_3^{(0)} \right|^2 d\phi_5,
 \nonumber \\
\lefteqn{
 \lim\limits_{1 \rightarrow 0} \; d\alpha^{iterated,relevant}_3 
 = 
 \frac{1}{6}  
 \left\{
  \frac{N_c^2}{4}
   \left[ 
          \left( D_3^0(0,2,3) + D_3^0(4,3,2) \right)
          \left( S_{012} + S_{213} + S_{314} \right)
 \right. \right.
} & &
 \nonumber \\
 & &
 \left. \left.
        + \left( D_3^0(0,3,2) + D_3^0(4,2,3) \right)
          \left( S_{013} + S_{213} + S_{214} \right) 
   \right]
 \right. \nonumber \\
 & & \left.
  + \frac{1}{4} 
    \left[
           - \left( A_3^0(0,2,4) + A_3^0(0,3,4) \right) 
           \left( S_{012} + S_{013} + S_{214} + S_{314} \right)
 \right. \right. \nonumber \\
 & & \left. \left.
         - \left( D_3^0(0,2,3) + D_3^0(0,3,2) + D_3^0(4,3,2) + D_3^0(4,2,3) 
                - A_3^0(0,2,4) - A_3^0(0,3,4) \right)
           S_{014}
    \right]
 \right. \nonumber \\
 & & \left.
  + \frac{1}{4N_c^2} 
    \left[
           \left( A_3^0(0,2,4) + A_3^0(0,3,4) \right) S_{014}
    \right]
 \right\}
 \circ \left| {\cal A}_3^{(0)} \right|^2 d\phi_5,
 \nonumber \\
\lefteqn{
 \lim\limits_{1 \rightarrow 0} \left( - d\alpha^{almost}_3 \right)
 = 
 \frac{1}{6}  
 \left\{
  \frac{N_c^2}{4}
  \frac{1}{2}
   \left[ 
         D_3^0(0,2,3) \left( S_{013} - S_{314} + S_{014} + S_{\widetilde{02}1\widetilde{32}} - S_{\widetilde{32}14} - S_{\widetilde{02}14} \right)
 \right. \right. } & & \nonumber \\
 & & \left. \left.
       + D_3^0(0,3,2) \left( S_{012} - S_{214} + S_{014} + S_{\widetilde{03}1\widetilde{23}} - S_{\widetilde{23}14} - S_{\widetilde{03}14} \right)
 \right. \right. \nonumber \\
 & & \left. \left.
       + D_3^0(4,3,2) \left( S_{214} - S_{012} + S_{014} + S_{\widetilde{23}1\widetilde{43}} - S_{01\widetilde{23}} - S_{01\widetilde{43}} \right)
 \right. \right. \nonumber \\
 & & \left. \left.
       + D_3^0(4,2,3) \left( S_{314} - S_{013} + S_{014} + S_{\widetilde{32}1\widetilde{42}} - S_{01\widetilde{32}} - S_{01\widetilde{42}} \right)
 \right. \right. \nonumber \\
 & & \left. \left.
       + A_3^0(0,2,4) \left(-S_{013} - S_{314} - S_{014} - S_{\widetilde{02}1\widetilde{42}} + S_{\widetilde{02}13} + S_{\widetilde{24}13} \right)
 \right. \right. \nonumber \\
 & & \left. \left.
       + A_3^0(0,3,4) \left(-S_{012} - S_{214} - S_{014} - S_{\widetilde{03}1\widetilde{43}} + S_{\widetilde{03}12} + S_{\widetilde{34}12} \right)
   \right]
 \right. \nonumber \\
 & &
 \left.
  + \frac{1}{4} 
    \left[
           \left( D_0^3(0,2,3) + D_3^0(0,3,2) + D_3^0(4,2,3) + D_3^0(4,3,2) \right) S_{014}
 \right. \right. \nonumber \\
 & & \left. \left.
       + A_3^0(0,2,4) \left( - S_{014} - S_{\widetilde{02}1\widetilde{42}} + S_{\widetilde{02}13} + S_{\widetilde{24}13} \right)
 \right. \right. \nonumber \\
 & & \left. \left.
       + A_3^0(0,3,4) \left( - S_{014} - S_{\widetilde{03}1\widetilde{43}} + S_{\widetilde{03}12} + S_{\widetilde{34}12} \right)
    \right]
 \right\}
 \circ \left| {\cal A}_3^{(0)} \right|^2 d\phi_5,
\nonumber \\
\lefteqn{
 \lim\limits_{1 \rightarrow 0} \left( - d\alpha^{soft}_3 \right)
 = 
 \frac{1}{6}  
 \left\{
  \frac{N_c^2}{4}
  \frac{1}{2}
   \left[ 
         D_3^0(0,2,3) \left( S_{013} - S_{314} - S_{014} - S_{\widetilde{02}1\widetilde{32}} + S_{\widetilde{32}14} + S_{\widetilde{02}14} \right)
 \right. \right. } & & \nonumber \\
 & & \left. \left.
       + D_3^0(0,3,2) \left( S_{012} - S_{214} - S_{014} - S_{\widetilde{03}1\widetilde{23}} + S_{\widetilde{23}14} + S_{\widetilde{03}14} \right)
 \right. \right. \nonumber \\
 & & \left. \left.
       + D_3^0(4,3,2) \left( S_{214} - S_{012} - S_{014} - S_{\widetilde{23}1\widetilde{43}} + S_{01\widetilde{23}} + S_{01\widetilde{43}} \right)
 \right. \right. \nonumber \\
 & & \left. \left.
       + D_3^0(4,2,3) \left( S_{314} - S_{013} - S_{014} - S_{\widetilde{32}1\widetilde{42}} + S_{01\widetilde{32}} + S_{01\widetilde{42}} \right)
 \right. \right. \nonumber \\
 & & \left. \left.
       + A_3^0(0,2,4) \left( S_{013} + S_{314} - S_{014} + S_{\widetilde{02}1\widetilde{42}} - S_{\widetilde{02}13} - S_{\widetilde{24}13} \right)
 \right. \right. \nonumber \\
 & & \left. \left.
       + A_3^0(0,3,4) \left( S_{012} + S_{214} - S_{014} + S_{\widetilde{03}1\widetilde{43}} - S_{\widetilde{03}12} - S_{\widetilde{34}12} \right)
   \right]
 \right. \nonumber \\
 & &
 \left.
  + \frac{1}{4} 
    \left[
         A_3^0(0,2,4) \left( S_{013} + S_{314} - S_{014} + S_{\widetilde{02}1\widetilde{42}} - S_{\widetilde{02}13} - S_{\widetilde{24}13} \right)
 \right. \right. \nonumber \\
 & & \left. \left.
       + A_3^0(0,3,4) \left( S_{012} + S_{214} - S_{014} + S_{\widetilde{03}1\widetilde{43}} - S_{\widetilde{03}12} - S_{\widetilde{34}12} \right)
    \right]
 \right\}
 \circ \left| {\cal A}_3^{(0)} \right|^2 d\phi_5.
\eq
Collinear: $0_{\bar{q}} || 1_g$.
\bq
\lefteqn{
 \lim\limits_{0 || 1} \left( - d\alpha^{double}_3 \right)
 = 
 \frac{1}{6}  
 \left\{
  \frac{N_c^2}{4}
   \left[ 
          \left( - 2 D_3^0(01,2,3) - 2 D_3^0(01,3,2) + A_3^0(01,2,4) + A_3^0(01,3,4) \right) P_{01}
   \right]
 \right.
} & &
 \nonumber \\
& &
 \left.
  + \frac{1}{4} 
    \left[
          2 \left( A_3^0(01,2,4) + A_3^0(01,3,4) \right) P_{01}
    \right]
 \right. \nonumber \\
 & & \left.
  + \frac{1}{4N_c^2} 
    \left[
           - \left( A_3^0(01,2,4) + A_3^0(01,3,4) \right) P_{01}
    \right]
 \right\}
 \circ \left| {\cal A}_3^{(0)} \right|^2 d\phi_5,
 \nonumber \\
\lefteqn{
 \lim\limits_{0 || 1} \; d\alpha^{iterated,relevant}_3 
 = 
 \frac{1}{6}  
 \left\{
  \frac{N_c^2}{4}
   \left[
          \left( D_3^0(01,2,3) + D_3^0(01,3,2) + D_3^0(4,3,2) + D_3^0(4,2,3) \right) P_{01} 
   \right]
 \right.
} & &
 \nonumber \\
 & &
 \left.
  + \frac{1}{4} 
    \left[
           - \left( D_3^0(01,2,3) + D_3^0(01,3,2) + D_3^0(4,3,2) + D_3^0(4,2,3) 
                    + A_3^0(01,2,4) 
 \right. \right. \right. \nonumber \\
 & & \left. \left. \left.
                    + A_3^0(01,3,4) \right) P_{01}
    \right]
  + \frac{1}{4N_c^2} 
    \left[
           \left( A_3^0(01,2,4) + A_3^0(01,3,4) \right) P_{01}
    \right]
 \right\}
 \circ \left| {\cal A}_3^{(0)} \right|^2 d\phi_5,
 \nonumber \\
\lefteqn{
 \lim\limits_{0 || 1} \left( - d\alpha^{almost}_3 \right)
 = 
 \frac{1}{6}  
 \left\{
  \frac{N_c^2}{4}
   \left[ 
          \left( - D_3^0(4,3,2) - D_3^0(4,2,3) + D_3^0(01,2,3) + D_3^0(01,3,2) 
 \right. \right. \right.
} & &
 \nonumber \\
& &
 \left. \left. \left.
                 - A_3^0(01,2,4) - A_3^0(01,3,4) \right) P_{01}
   \right]
  + \frac{1}{4} 
    \left[
           \left( D_3^0(01,2,3) + D_3^0(01,3,2) + D_3^0(4,2,3) 
 \right. \right. \right. \nonumber \\
 & & \left. \left. \left.
                + D_3^0(4,3,2) 
                - A_3^0(01,2,4) - A_3^0(01,3,4) \right) P_{01}
    \right]
 \right\}
 \circ \left| {\cal A}_3^{(0)} \right|^2 d\phi_5,
 \hspace*{45mm}
 \nonumber \\
\lefteqn{
 \lim\limits_{0 || 1} \left( - d\alpha^{soft}_3 \right)
 = \mbox{integrable}.
}
\eq
Collinear: $1_g || 2_g$.
\bq
\lefteqn{
 \lim\limits_{1 || 2} \left( - d\alpha^{double}_3 \right)
 = 
 \frac{1}{6}  
 \left\{
  \frac{N_c^2}{4}
   \left[ 
          2 \left( - D_3^0(0,12,3) - D_3^0(0,3,12) - D_3^0(4,12,3) - D_3^0(4,3,12) \right) P_{12}
   \right]
 \right.
} & &
 \nonumber \\
& &
 \left.
  + \frac{1}{4} 
    \left[
           2 A_3^0(0,12,4) P_{12}
    \right]
 \right\}
 \circ \left| {\cal A}_3^{(0)} \right|^2 d\phi_5,
 \hspace*{80mm}
 \nonumber \\
\lefteqn{
 \lim\limits_{1 || 2} \; d\alpha^{iterated,relevant}_3 
 = 
 \frac{1}{6}  
 \left\{
  \frac{N_c^2}{4}
   \left[ 
          2 \left( D_3^0(0,12,3) + D_3^0(0,3,12) + D_3^0(4,3,12) + D_3^0(4,12,3) \right) P_{12} 
   \right]
 \right.
} & &
 \nonumber \\
 & & \left.
  + \frac{1}{4} 
    \left[
           - 2 \left( A_3^0(0,12,4) + A_3^0(0,3,4) \right) P_{12}
    \right]
 \right\}
 \circ \left| {\cal A}_3^{(0)} \right|^2 d\phi_5,
 \nonumber \\
\lefteqn{
 \lim\limits_{1 || 2} \left( - d\alpha^{almost}_3 \right)
 = 
 \frac{1}{6}  
 \left\{
  \frac{1}{4} 
    \left[
           2 A_3^0(0,3,4) P_{12}
    \right]
 \right\}
 \circ \left| {\cal A}_3^{(0)} \right|^2 d\phi_5,
} & &
 \nonumber \\
\lefteqn{
 \lim\limits_{1 || 2} \left( - d\alpha^{soft}_3 \right)
 = \mbox{integrable}.
} & &
\eq
In the double unresolved limit we have for the $0_{\bar{q}} 1_{\bar{q}'} 2_g 3_{q'} 4_q$-channel:
\\
\\
Soft-collinear: $2_g$ soft, $1_{\bar{q}'} || 3_{q'}$ collinear.
\bq
\lefteqn{
 \lim\limits_{2 \rightarrow 0, \; 1||3} d\sigma_5^{(0)}
 =
  \left( \frac{1}{4} + \frac{N_f-1}{2} \right)
 \left\{
 \frac{N_c}{4} 
    \left( SC_{02(31)} P_{13} + SC_{42(13)} P_{13} \right)
 \right. 
} & &
 \nonumber \\
 & & \left. 
 + \frac{1}{4N_c}
    \left( 
           - S_{024} P_{13} - 2 SC_{02(31)} P_{13} + 2 SC_{02(13)} P_{13} - 2 SC_{42(13)} P_{13} + 2 SC_{42(31)} P_{13} 
    \right)
\right\}
 \nonumber \\
 & &
 \circ \left| {\cal A}_3^{(0)} \right|^2 d\phi_5,
 \nonumber \\
\lefteqn{
 \lim\limits_{2 \rightarrow 0, \; 1||3} \left( - d\alpha^{double}_3 \right)
 =
  \left( \frac{1}{4} + \frac{N_f-1}{2} \right)
 \left\{
 \frac{N_c}{4} 
    \left( 
           - SC_{02(31)} P_{13} - SC_{42(13)} P_{13}
    \right)
\right\}
 \circ \left| {\cal A}_3^{(0)} \right|^2 d\phi_5,
} & &
 \nonumber \\
\lefteqn{
 \lim\limits_{2 \rightarrow 0, \; 1||3} \left( - d\alpha^{almost}_3 \right)
 =
  \left( \frac{1}{4} + \frac{N_f-1}{2} \right)
 \left\{
 \frac{N_c}{4} \frac{1}{4}
    \left( 
           S_{021} P_{13} - S_{023} P_{13} + S_{423} P_{13} - S_{421} P_{13}
    \right)
 \right.
} & &
 \nonumber \\
 & &
 \left.
 + \frac{1}{4N_c}
    \left( 
           S_{024} P_{13}
           - 2 S_{021} P_{13} + 2 S_{023} P_{13}  - 2 S_{324} P_{13} + 2 S_{124} P_{13}
    \right)
\right\}
 \circ \left| {\cal A}_3^{(0)} \right|^2 d\phi_5.
\eq
Triple collinear: $1_{\bar{q}'} || 2_g || 3_{q'}$.
\bq
& &
 \lim\limits_{1||2||3} d\sigma_5^{(0)}
 =
  \left( \frac{1}{4} + \frac{N_f-1}{2} \right)
 \left\{
 \frac{N_c}{4} 
    \left( P_{213} + P_{231} \right)
 + \frac{1}{4N_c}
    \left( 
           - P_{123}^{sc} 
    \right)
\right\}
 \circ \left| {\cal A}_3^{(0)} \right|^2 d\phi_5,
 \nonumber \\
& &
 \lim\limits_{1||2||3} \left( - d\alpha^{double}_3 \right)
 =
  \left( \frac{1}{4} + \frac{N_f-1}{2} \right)
 \left\{
 \frac{N_c}{4} 
    \left( 
           - P_{231} - P_{213}
    \right)
 + \frac{1}{4N_c}
           P_{123}^{sc}
\right\}
 \circ \left| {\cal A}_3^{(0)} \right|^2 d\phi_5,
 \nonumber \\
& &
 \lim\limits_{1||2||3} \left( - d\alpha^{almost}_3 \right)
 = 0.
\eq
Triple collinear: $0_{\bar{q}} || 1_{\bar{q}'} || 3_{q'}$.
\bq
\lefteqn{
 \lim\limits_{0||1||3} d\sigma_5^{(0)}
 =
 \left\{
 \frac{N_c}{4} 
  \left( \frac{1}{4} + \frac{N_f-1}{2} \right)
    P_{013}
 + \frac{1}{4N_c}
  \left( \frac{1}{4} + \frac{N_f-1}{2} \right)
    \left( 
           - P_{013}
    \right)
\right.
} & &
 \nonumber \\
 & & 
 \left.
 + \frac{1}{4} \left( - \frac{1}{2} P_{013}^{id} \right)
 + \frac{1}{4N_c^2} \left( \frac{1}{2} P_{013}^{id} \right)
\right\}
 \circ \left| {\cal A}_3^{(0)} \right|^2 d\phi_5,
 \hspace*{60mm}
 \nonumber \\
\lefteqn{
 \lim\limits_{0||1||3} \left( - d\alpha^{double}_3 \right)
 =
 \left\{
 \frac{N_c}{4} 
  \left( \frac{1}{4} + \frac{N_f-1}{2} \right)
    \left( 
           - P_{013}
    \right)
 + \frac{1}{4N_c}
  \left( \frac{1}{4} + \frac{N_f-1}{2} \right)
           P_{013} 
 \right.
} & &
 \nonumber \\
 & &
 \left.
 + \frac{1}{4} 
    \left( 
           \frac{1}{2} P_{013}^{id}
    \right)
 + \frac{1}{4N_c^2} 
    \left( 
           - \frac{1}{2} P_{013}^{id}
    \right)
\right\}
 \circ \left| {\cal A}_3^{(0)} \right|^2 d\phi_5
 \nonumber \\
\lefteqn{
 \lim\limits_{0||1||3} \left( - d\alpha^{almost}_3 \right)
 = 0.
} & &
\eq
Two collinear pairs: $0_{\bar{q}} || 2_g$ and $1_{\bar{q}'} || 3_{q'}$.
\bq
& &
 \lim\limits_{0||2, \; 1||3} d\sigma_5^{(0)}
 =
  \left( \frac{1}{4} + \frac{N_f-1}{2} \right)
 \left\{
 \frac{N_c}{4} 
    \left( P_{02} P_{13} \right)
 + \frac{1}{4N_c}
    \left( 
           - P_{02} P_{13} 
    \right)
\right\}
 \circ \left| {\cal A}_3^{(0)} \right|^2 d\phi_5,
 \nonumber \\
& &
 \lim\limits_{0||2, \; 1||3} \left( - d\alpha^{double}_3 \right)
 =
 \left\{
 \frac{N_c}{4} 
  \left( \frac{1}{4} + \frac{N_f-1}{2} \right)
    \left( 
           - P_{02} P_{13}
    \right)
\right\}
 \circ \left| {\cal A}_3^{(0)} \right|^2 d\phi_5,
 \nonumber \\
& &
 \lim\limits_{0||2, \; 1||3} \left( - d\alpha^{almost}_3 \right)
 =
  \left( \frac{1}{4} + \frac{N_f-1}{2} \right)
 \left\{
 \frac{1}{4N_c}
    \left( 
           P_{02} P_{13}
    \right)
\right\}
 \circ \left| {\cal A}_3^{(0)} \right|^2 d\phi_5.
\eq
In the single unresolved limit we have for the $0_{\bar{q}} 1_{\bar{q}'} 2_g 3_{q'} 4_q$-channel:
\\
\\
Soft: $2_g$ soft.
\bq
\lefteqn{
 \lim\limits_{2 \rightarrow 0} \left( - d\alpha^{double}_{3,\bar{q}\bar{q}'gq'q} \right) = 
 \left\{
  \frac{N_c}{4} 
  \left[ \left( \frac{1}{4} + \frac{N_f-1}{2} \right) 
         \left( 
                - 2 \left( E_3^0(0,1,3) + E_3^0(3,4,0) \right) S_{023} 
 \right. \right. \right.
} \nonumber \\
 & & 
 \left. \left. \left.
                - 2 \left( E_3^0(1,0,4) + E_3^0(4,3,1) \right) S_{124}  
         \right)
 \right. \right. \nonumber \\
 & & \left. \left.
         + \frac{1}{4} 
         \left( 
                - 2 \left( E_3^0(0,1,4) + E_3^0(4,3,0) \right) S_{024}  
                - 2 \left( E_3^0(1,0,3) + E_3^0(3,4,1) \right) S_{123}  
         \right)
  \right]
 \right. \nonumber \\
 & & \left.
 + \frac{1}{4N_c}
  \left[ \left( \frac{1}{4} + \frac{N_f-1}{2} \right) 
         \left( 
                \left( E_3^0(0,1,3) + E_3^0(4,3,1) \right) S_{123} 
              + \left( E_3^0(1,0,4) + E_3^0(3,4,0) \right) S_{024}  
         \right)
 \right. \right. \nonumber \\
 & & \left. \left.
         + \frac{1}{4} 
         \left( 
                \left( E_3^0(0,1,4) + E_3^0(3,4,1) \right) S_{124}  
              + \left( E_3^0(1,0,3) + E_3^0(4,3,0) \right) S_{023}  
         \right)
  \right]
 \right\}
 \circ \left| {\cal A}_3^{(0)} \right|^2 d\phi_5,
 \nonumber \\
\lefteqn{
 \lim\limits_{2 \rightarrow 0} \; d\alpha^{iterated}_{3,\bar{q}\bar{q}'gq'q\rightarrow \bar{q}\bar{q}'q'q\rightarrow \bar{q}gq} = 
} \nonumber \\
 & & 
 \left\{
  \frac{N_c}{4} 
  \left[ \left( \frac{1}{4} + \frac{N_f-1}{2} \right) 
         \left( E_3^0(0,1,3) + E_3^0(4,3,1) + E_3^0(1,0,4) + E_3^0(3,4,0) \right) \left( S_{124} + S_{023} \right)
 \right. \right. \nonumber \\
 & & \left. \left.
         + \frac{1}{4} 
         \left( E_3^0(1,0,3) + E_3^0(4,3,0) + E_3^0(0,1,4) + E_3^0(3,4,1) \right) \left( S_{024} + S_{123} \right)
  \right]
 \right. \nonumber \\
 & & \left.
 + \frac{1}{4N_c}
  \left[ \left( \frac{1}{4} + \frac{N_f-1}{2} \right) 
         \left( - E_3^0(0,1,3) - E_3^0(4,3,1) - E_3^0(1,0,4) - E_3^0(3,4,0) \right) 
 \right. \right. \nonumber \\
 & & \left. \left.
         \left( 2 S_{023} + 2 S_{124} - 2 S_{021} - 2 S_{324} + S_{024} + S_{123} \right)  
 \right. \right. \nonumber \\
 & & \left. \left.
         + \frac{1}{4} 
         \left( - E_3^0(1,0,3) - E_3^0(4,3,0) - E_3^0(0,1,4) - E_3^0(3,4,1) \right) 
 \right. \right. \nonumber \\
 & & \left. \left.
         \left( 2 S_{024} + 2 S_{123} - 2 S_{021} - 2 S_{324} + S_{023} + S_{124} \right) 
  \right]
 \right\}
 \circ \left| {\cal A}_3^{(0)} \right|^2 d\phi_5,
 \nonumber \\
\lefteqn{
 \lim\limits_{2 \rightarrow 0} \left( - d\alpha^{almost}_{3,\bar{q}\bar{q}'gq'q\rightarrow \bar{q}ggq\rightarrow \bar{q}gq} 
     - d\alpha^{almost}_{3,\bar{q}\bar{q}'gq'q\rightarrow \bar{q}\bar{q}'q'q\rightarrow \bar{q}gq} \right) = 
} \nonumber \\
 & & 
 \left\{
  \frac{N_c}{4} 
  \frac{1}{2}
  \left[ \left( \frac{1}{4} + \frac{N_f-1}{2} \right) 
  \left[
   E_3^0(0,1,3) \left( S_{023} - 2 S_{124} + S_{024} + S_{324} + S_{\widetilde{01}2\widetilde{31}} - S_{42\widetilde{31}} - S_{\widetilde{01}24} \right)
 \right. \right. \right. \nonumber \\
 & & \left. \left. \left.
 + E_3^0(1,0,4) \left( S_{124} - 2 S_{023} + S_{123} + S_{324} + S_{\widetilde{10}2\widetilde{40}} - S_{32\widetilde{40}} - S_{\widetilde{10}23} \right)
 \right. \right. \right. \nonumber \\
 & & \left. \left. \left.
 + E_3^0(4,3,1) \left( S_{124} - 2 S_{023} + S_{024} + S_{021} + S_{\widetilde{43}2\widetilde{13}} - S_{02\widetilde{13}} - S_{02\widetilde{43}} \right)
 \right. \right. \right. \nonumber \\
 & & \left. \left. \left.
 + E_3^0(3,4,0) \left( S_{023} - 2 S_{124} + S_{123} + S_{021} + S_{\widetilde{34}2\widetilde{04}} - S_{12\widetilde{04}} - S_{12\widetilde{34}} \right)
  \right]
 \right. \right. \nonumber \\
 & & \left. \left.
 + \frac{1}{4} 
  \left[
   E_3^0(0,1,4) \left( S_{024} - 2 S_{123} + S_{023} + S_{324} + S_{\widetilde{01}2\widetilde{41}} - S_{32\widetilde{41}} - S_{\widetilde{01}23} \right)
 \right. \right. \right. \nonumber \\
 & & \left. \left. \left.
 + E_3^0(1,0,3) \left( S_{123} - 2 S_{024} + S_{124} + S_{324} + S_{\widetilde{10}2\widetilde{30}} - S_{42\widetilde{30}} - S_{\widetilde{10}24} \right)
 \right. \right. \right. \nonumber \\
 & & \left. \left. \left.
 + E_3^0(3,4,1) \left( S_{123} - 2 S_{024} + S_{023} + S_{021} + S_{\widetilde{34}2\widetilde{14}} - S_{02\widetilde{14}} - S_{02\widetilde{34}} \right)
 \right. \right. \right. \nonumber \\
 & & \left. \left. \left.
 + E_3^0(4,3,0) \left( S_{024} - 2 S_{123} + S_{124} + S_{021} + S_{\widetilde{43}2\widetilde{03}} - S_{12\widetilde{03}} - S_{12\widetilde{43}} \right)
  \right] 
  \right]
 \right. \nonumber \\
 & & \left.
 + \frac{1}{4N_c}
  \left[ \left( \frac{1}{4} + \frac{N_f-1}{2} \right) 
  \left[
         \left( E_3^0(0,1,3) + E_3^0(4,3,1) \right) S_{024} 
         + \left( E_3^0(1,0,4) + E_3^0(3,4,0) \right) S_{123} 
 \right. \right. \right. \nonumber \\
 & & \left. \left. \left.
         +
         2 \left( E_3^0(0,1,3) + E_3^0(4,3,1) + E_3^0(1,0,4) + E_3^0(3,4,0) \right) 
         \left( S_{023} + S_{124} - S_{021} - S_{324} \right)  
  \right]
 \right. \right. \nonumber \\
 & & \left. \left.
  + \frac{1}{4} 
  \left[
         \left( E_3^0(0,1,4) + E_3^0(3,4,1) \right) S_{023} 
         + \left( E_3^0(1,0,3) + E_3^0(4,3,0) \right) S_{124} 
 \right. \right. \right. \nonumber \\
 & & \left. \left. \left.
         +
         2 \left( E_3^0(1,0,3) + E_3^0(4,3,0) + E_3^0(0,1,4) + E_3^0(3,4,1) \right) 
         \left( S_{024} + S_{123} - S_{021} - S_{324} \right) 
  \right]
  \right]
 \right\}
 \nonumber \\
 & & 
 \circ \left| {\cal A}_3^{(0)} \right|^2 d\phi_5,
 \nonumber \\
\lefteqn{
 \lim\limits_{2 \rightarrow 0} \left( - d\alpha^{soft}_{3,\bar{q}\bar{q}'gq'q} \right) = 
  \frac{N_c}{4} 
  \frac{1}{2}
  \left\{ \left( \frac{1}{4} + \frac{N_f-1}{2} \right) 
  \left[
   E_3^0(0,1,3) \left( S_{023} - S_{024} - S_{324} - S_{\widetilde{01}2\widetilde{31}} 
\right. \right. \right. } & & \nonumber \\
 & & \left. \left. \left.
+ S_{42\widetilde{31}} + S_{\widetilde{01}24} \right)
 + E_3^0(1,0,4) \left( S_{124} - S_{123} - S_{324} - S_{\widetilde{10}2\widetilde{40}} + S_{32\widetilde{40}} + S_{\widetilde{10}23} \right)
\right. \right. \nonumber \\
 & & \left. \left.
 + E_3^0(4,3,1) \left( S_{124} - S_{024} - S_{021} - S_{\widetilde{43}2\widetilde{13}} + S_{02\widetilde{13}} + S_{02\widetilde{43}} \right)
\right. \right. \nonumber \\
 & & \left. \left.
 + E_3^0(3,4,0) \left( S_{023} - S_{123} - S_{021} - S_{\widetilde{34}2\widetilde{04}} + S_{12\widetilde{04}} + S_{12\widetilde{34}} \right)
  \right]
\right. \nonumber \\
 & & \left. 
 + \frac{1}{4} 
  \left[
   E_3^0(0,1,4) \left( S_{024} - S_{023} - S_{324} - S_{\widetilde{01}2\widetilde{41}} + S_{32\widetilde{41}} + S_{\widetilde{01}23} \right)
\right. \right. \nonumber \\
 & & \left. \left.
 + E_3^0(1,0,3) \left( S_{123} - S_{124} - S_{324} - S_{\widetilde{10}2\widetilde{30}} + S_{42\widetilde{30}} + S_{\widetilde{10}24} \right)
\right. \right. \nonumber \\
 & & \left. \left.
 + E_3^0(3,4,1) \left( S_{123} - S_{023} - S_{021} - S_{\widetilde{34}2\widetilde{14}} + S_{02\widetilde{14}} + S_{02\widetilde{34}} \right)
\right. \right. \nonumber \\
 & & \left. \left.
 + E_3^0(4,3,0) \left( S_{024} - S_{124} - S_{021} - S_{\widetilde{43}2\widetilde{03}} + S_{12\widetilde{03}} + S_{12\widetilde{43}} \right)
  \right] 
 \right\}
 \circ \left| {\cal A}_3^{(0)} \right|^2 d\phi_5.
\eq
Collinear: $1_{\bar{q}'} || 3_{q'}$.
\\
In this case we have
\bq
 d\alpha^{single}_{4,\bar{q}\bar{q}'gq'q\rightarrow \bar{q}\bar{q}'q'q} 
 - d\alpha^{iterated}_{3,\bar{q}\bar{q}'gq'q\rightarrow \bar{q}\bar{q}'q'q\rightarrow \bar{q}gq} 
 & = & \mbox{integrable},
 \nonumber \\
 d\alpha^{soft}_{3,\bar{q}\bar{q}'gq'q}
 & = & \mbox{integrable}.
\eq
The non-trivial limits are:
\bq
\lefteqn{
 \lim\limits_{1 || 3} \left( - d\alpha^{double}_{3,\bar{q}\bar{q}'gq'q} \right) = 
 \left( \frac{1}{4} + \frac{N_f-1}{2} \right) 
 \left\{
  \frac{N_c}{4} 
  \left( - D_3^0(0,13,2) - D_3^0(0,2,13) - D_3^0(4,13,2) 
 \right. \right.
} \nonumber \\
 & & 
 \left. \left.
 - D_3^0(4,2,13) \right)
  P_{13}
 + \frac{1}{4N_c}
  A_3^0(0,13,4) P_{13}
 \right\} 
 \circ \left| {\cal A}_3^{(0)} \right|^2 d\phi_5,
 \nonumber \\
\lefteqn{
 \lim\limits_{1 || 3} \; d\alpha^{iterated}_{3,\bar{q}\bar{q}'gq'q\rightarrow \bar{q}ggq\rightarrow \bar{q}gq} = 
\left( \frac{1}{4} + \frac{N_f-1}{2} \right)
 \left\{
  \frac{N_c}{4}  
  \left( D_3^0(0,13,2) + D_3^0(0,2,13) + D_3^0(4,13,2) 
 \right. \right.
} \nonumber \\
 & & 
 \left. \left.
 + D_3^0(4,2,13) \right)
  P_{13}
 + \frac{1}{4N_c}
    \left( 
           - A_3^0(0,13,4) - A_3^0(0,2,4)
    \right) P_{13}
 \right\}
 \circ \left| {\cal A}_3^{(0)} \right|^2 d\phi_5,
 \hspace*{20mm}
 \nonumber \\
\lefteqn{
 \lim\limits_{1 || 3} \left( - d\alpha^{almost}_{3,\bar{q}\bar{q}'gq'q\rightarrow \bar{q}ggq\rightarrow \bar{q}gq} 
     - d\alpha^{almost}_{3,\bar{q}\bar{q}'gq'q\rightarrow \bar{q}\bar{q}'q'q\rightarrow \bar{q}gq} \right) 
=
  \left( \frac{1}{4} + \frac{N_f-1}{2} \right)
 \frac{1}{4N_c}
           A_3^0(0,2,4) P_{13}
 } & &
 \nonumber \\
 & &
 \circ \left| {\cal A}_3^{(0)} \right|^2 d\phi_5.
\eq
Collinear: $1_{\bar{q}'} || 2_g$.
\bq
 d\alpha^{single}_{4,\bar{q}\bar{q}'gq'q\rightarrow \bar{q}ggq} - d\alpha^{iterated}_{3,\bar{q}\bar{q}'gq'q\rightarrow \bar{q}ggq\rightarrow \bar{q}gq} & = & \mbox{integrable},
 \nonumber \\
 d\alpha^{soft}_{3,\bar{q}\bar{q}'gq'q}
 & = & \mbox{integrable}.
\eq
\bq
\lefteqn{
 \lim\limits_{1 || 2} \left( - d\alpha^{double}_{3,\bar{q}\bar{q}'gq'q\rightarrow \bar{q}\bar{q}'q'q\rightarrow \bar{q}gq} \right) = 
 \left\{
  \frac{N_c}{4} 
  \left[ \left( \frac{1}{4} + \frac{N_f-1}{2} \right) 
         \left( - 2 E_3^0(12,0,4) - 2 E_3^0(4,3,12) \right) P_{12} 
 \right. \right. 
} \nonumber \\
 & & 
\left. \left.
         + \frac{1}{4} 
         \left( -2 E_3^0(12,0,3) - 2 E_3^0(3,4,12) \right) P_{12} 
  \right]
 + \frac{1}{4N_c}
    \left[ 
           \left( \frac{1}{4} + \frac{N_f-1}{2} \right)
           \left( E_3^0(0,12,3) 
 \right. \right. \right. \nonumber \\
 & & \left. \left. \left.
+ E_3^0(4,3,12) \right) P_{12}
         + \frac{1}{4}
           \left( E_3^0(0,12,4) + E_3^0(3,4,12) \right) P_{12}
    \right]
 \right\}
 \circ \left| {\cal A}_3^{(0)} \right|^2 d\phi_5,
 \nonumber \\
\lefteqn{
 \lim\limits_{1 || 2} \; d\alpha^{iterated}_{3,\bar{q}\bar{q}'gq'q\rightarrow \bar{q}\bar{q}'q'q\rightarrow \bar{q}gq} = 
} \nonumber \\
 & & 
 \left\{
  \frac{N_c}{4} 
  \left[ \left( \frac{1}{4} + \frac{N_f-1}{2} \right) 
         \left( E_3^0(0,12,3) + E_3^0(4,3,12) + E_3^0(12,0,4) + E_3^0(3,4,0) \right) P_{12} 
 \right. \right. \nonumber \\
 & & \left. \left.
         + \frac{1}{4} 
         \left( E_3^0(12,0,3) + E_3^0(4,3,0) + E_3^0(0,12,4) + E_3^0(3,4,12) \right) P_{12} 
  \right]
 \right. \nonumber \\
 & & \left.
 + \frac{1}{4N_c}
  \left[ \left( \frac{1}{4} + \frac{N_f-1}{2} \right) 
         \left( - E_3^0(0,12,3) - E_3^0(4,3,12) - E_3^0(12,0,4) - E_3^0(3,4,0) \right) P_{12} 
 \right. \right. \nonumber \\
 & & \left. \left.
         + \frac{1}{4} 
         \left( - E_3^0(12,0,3) - E_3^0(4,3,0) - E_3^0(0,12,4) - E_3^0(3,4,12) \right) P_{12} 
  \right]
 \right\}
 \circ \left| {\cal A}_3^{(0)} \right|^2 d\phi_5,
 \nonumber \\
\lefteqn{
 \lim\limits_{1 || 2} \left( - d\alpha^{almost}_{3,\bar{q}\bar{q}'gq'q\rightarrow \bar{q}ggq\rightarrow \bar{q}gq}
- d\alpha^{almost}_{3,\bar{q}\bar{q}'gq'q\rightarrow \bar{q}\bar{q}'q'q\rightarrow \bar{q}gq} \right) = 
 \left\{
  \frac{N_c}{4} 
  \left[ \left( \frac{1}{4} + \frac{N_f-1}{2} \right) 
 \right. \right.
} \nonumber \\
 & & 
 \left. \left.
         \left( 
                 E_3^0(12,0,4) - E_3^0(3,4,0) 
                -E_3^0(0,12,3) + E_3^0(4,3,12) 
         \right) P_{12} 
 \right. \right. \nonumber \\
 & & \left. \left.
         + \frac{1}{4} 
         \left( 
                E_3^0(12,0,3) - E_3^0(4,3,0) 
              - E_3^0(0,12,4) + E_3^0(3,4,12) 
         \right) P_{12} 
  \right]
 \right. \nonumber \\
 & & \left.
 + \frac{1}{4N_c}
  \left[ \left( \frac{1}{4} + \frac{N_f-1}{2} \right) 
         \left( E_3^0(12,0,4) + E_3^0(3,4,0) \right) P_{12} 
 \right. \right. \nonumber \\
 & & \left. \left.
         + \frac{1}{4} 
         \left( E_3^0(12,0,3) + E_3^0(4,3,0) \right) P_{12} 
  \right]
 \right\}
 \circ \left| {\cal A}_3^{(0)} \right|^2 d\phi_5.
\eq

\end{appendix}

\bibliography{/home/stefanw/notes/biblio}
\bibliographystyle{/home/stefanw/latex-style/h-physrev3}

\end{document}